\documentclass[showpacs,aps,prd,nofootinbib,floatfix,amsmath,amssymb]{revtex4}
\usepackage{graphicx}
\usepackage[utf8x]{inputenc}
\usepackage{color}
\usepackage{subfig}
\usepackage{multirow}
\usepackage{slashed}

\begin{document}

\title{Copositivity criteria, scalar mass eigenstates, and custodial symmetry parameter $\Delta \rho$ for the $S_3 \otimes \mathbb{Z}_2$ model}

\author{G. De Conto}
\email{george.de.conto@gmail.com}
\affiliation{Departamento de F\'{i}sica - Centro de Ci\^{e}ncias Exatas,
	 Universidade Estadual de Londrina, Londrina - PR, 86051-990, Brazil.}

\author{A. C. B. Machado}
\email{a.c.b.machado1@gmail.com}
\affiliation{Centro de Ci\^{e}ncias Naturais e Humanas,
	 Universidade Federal do ABC, 09210-580, Santo Andr\'{e}-SP, Brazil.}

\begin{abstract}
   We present the copositivity criteria for the scalar sector of our $S_3 \otimes \mathbb{Z}_2$ model, which ensures that it is bounded from below. We also show the scalar mass eigenstates, and identify the Standard Model Higgs boson among these by imposing that its Yukawa couplings are the same in our model and the Standard Model. The $\Delta \rho$ parameter, related to the custodial symmetry, is calculated as well.
   
\end{abstract}

\maketitle

\section{Introduction}
\label{sec:intro}

The Standard Model of elementary particles (SM), governed by the local $SU(3)\times SU(2) \times U(1)$ gauge symmetry which allows us to obtain the most general Lagrangian that describes the dynamics of the fields, has only 19 free parameters. One of those parameters was finally measured by the LHC. The ATLAS~\cite{ATLASHiggs} and CMS~\cite{CMSHiggs} collaborations
showed the results that show the discovery of a new scalar boson, whose properties are comparable to those of the Standard Model Higgs boson~\cite{hhg} with mass of 125.3 GeV. However, there are still 18 parameters, namely: 6 quark masses, 3 masses of the charged leptons, 3 mixing angles in the quark sector and one CP violating phase (that give rise to the Cabibbo-Kobayashi-Maskawa,
CKM, matrix), 3 coupling constants, the vacuum expectation value and the $\theta$QCD. 

Now, after the discovery of the massive neutrinos, we have more free parameters and some open questions that the SM is not able to solve. Among these parameters we have: three more masses for the three neutrinos, three new mixing angles for the leptonic sector, and possibly a CP violating phase in the lepton mixing matrix. 

Another important evidence that the SM needs an extension is the observed dark matter (DM) from galaxy rotation curves, once there is no viable candidate in the SM as it is today. Within many possibilities, maybe the simplest DM models corresponds to coupling the DM sector to SM sector, with the SM Higgs scalar as the interaction mediator, the Higgs-portal. 

In the context of dark matter, the simplest models that provide good scalar DM candidate is the model (I(1+1)HDM), one Higgs SM-like and one inert. This doublet, is called inert because it does not couple to fermions, which provides a stable DM candidate, so this is an example of the Higgs-portal \cite{higgsportal} type of DM model, where the DM sector communicates with the SM sector through the Higgs boson exchange. As a result, the DM-Higgs coupling, $g_{DMh}$, governs the DM annihilation rate $\langle\sigma v\rangle$, the DM-nucleon scattering cross-section $\sigma_{DM-N}$ and the Higgs invisible decays.

Although the experimental constraints for these three types of processes when simultaneously considered is a difficult task to solve, one possible solution to this problem is to introduce coannihilation processes between DM  and other inert particles which are close in mass. Coannihilation processes lead to an increase or decrease of the effective annihilation cross-section, which in turn gives, respectively, smaller or larger DM relic density values. In the I(1+1)HDM, for example, the DM candidate could coannihilate with neutral and/or charged $Z_2$-odd particles. In models with a richer particle spectrum, more coannihilation processes could come into play \cite{Keus:2014jha}.

Therefore, despite the fact that the SM is an experimentally proven theory with a high predictive power, it is still incomplete. In this context, we have proposed an extension of the SM with three scalar doublets of $SU(2)$ with $S_3\otimes\mathbb{Z}_2$ symmetries\cite{Dias:2012bh,Fortes:2017ndr}. We had analyzed all the mass spectra in the scalar sectors and used the scotogenic mechanism for generating neutrino masses. Moreover, we had obtained the PMNS matrix once the unitary matrices which diagonalize the lepton masses were obtained.
The ultraviolet (UV) completion of the model is obtained when the energies are of the order of the $\zeta$ field masses, and we have to consider the production of such fields. Hence, this UV completion is as follows: we add at least one superheavy inert Higgs scalar doublet $D_{SH}$ which is singlet under $S_3$ and transforms as $\omega^2$ under $\mathbb{Z}_2$. Then the Yukawa interaction $\overline{L^c}N_sD_{SH}$ and the trilinear $D^\dagger_{SH}[D\zeta]_1$ are allowed and the effective operator ($LDN\zeta$) arises by the exchange ($s$-channel) of the $D_{SH}$.

The model has an accidental $\mathbb{Z}_2$ symmetry under which the inert doublets and right-handed neutrinos are odd, -1, and all the other particles even, +1. This implies that at least the lightest neutral inert scalar or right-handed neutrino is completely stable. This fact has implications such as the followings: the effect of the inert particles occurs only through loops, as the processes considered here. In addition, all the inert particles, mainly the lighter, are dark matter candidates. This topic will be considered somewhere.

Our analysis predicts a mass for the right-handed neutrino starting from $M_{\nu R}>295$ GeV, besides, from the obtained constraints for $\mu\to e\gamma$, we can predict that Br$(\mu\to ee\bar{e})< 10^{-14}$, this is two orders of magnitude below the current upper limit  and two orders above the upcoming expected upper limit.

We have $N_1$, $N_2$, $N_s$  and $\eta_2$ as DM candidates. 
For the set of selected parameters by us, we give examples of the annihilation channels of three DM candidates.
For $\eta_{2}$ being DM candidate we also have interesting annihilation channels which change flavor in the final states. This candidate can also be a self-interacting DM candidate according to the parameter choice and many studies in small scales can be performed considering this candidate. We can have other parameters that could  solve the core-cusp and the too-big-to-fail problem and also the missing satellite problem.
Like the other two candidate presented above, $N_{1}$ candidate can also present final states which change flavor. The $N_{2}$ DM candidate presents a similar behavior of $N_{1}$. We conclude that all the candidates offer well motivated options of additional studies and they are already being done by us.

In this work we will continue the study of this model starting with the analysis of the copositivity criteria for the potential, which ensures that it is bounded from below. Also, we will present the mass eigenstates for the scalar sector and, by imposing that its Yukawa couplings are the same as the SM, identify the SM Higgs boson among the available CP-even scalars of the model. Another study presented here is the $\Delta \rho$ parameter, related to the custodial symmetry. This symmetry relates the masses of the $W$ and $Z$ bosons and is violated in multi-Higgs models.

The outline of the paper is as follows: in Secs. \ref{sec:model} and \ref{sec:scalars} we present the model and its particle content, in Sec. \ref{sec:copositividade} we present the copositivity criteria of the scalar potential and in Secs. \ref{sec:massmatrices} and \ref{sec:CPpar3x3} we present the scalar mass eigenstates and identify the SM Higgs boson among those. Afterwards, we calculate $\Delta \rho$ in Sec. \ref{sec:custodialsymmetry}. Finally, in Sec. \ref{sec:conclusions}, we present our conclusions.

\section{The $S_3 \otimes \mathbb{Z}_2$ model}
\label{sec:model}

Here we will use the  $S_3$ discrete symmetry in order to obtain a model with 3 Higgs doublets, being  two of them inert.  The $S_3$ symmetry consists of all permutations among three objects. However, the representation of order 3 is reducible and is decomposed in two irreducible representations: $\textbf{3}=\textbf{1}\oplus \textbf{2}$. Here we will write only the multiplications involving two doublets and two singlets (which will be used here for obtaining the Yukawa interactions) and the scalar potential that is invariant under the full symmetry, $SU(2)_L\otimes U(1)_Y\otimes S_3\otimes \mathbb{Z}_2$. Let $[x_1,x_2]$ and $[y_1,y_2]$ be two doublets of $S_3$, the multiplication $\textbf{2}\otimes\textbf{2}$ is given by
\begin{equation}
\left[\begin{array}{c} x_1 \\ x_2\end{array}\right]_2 \otimes \left[\begin{array}{c} y_1 \\y_2\end{array}\right]_2 = [x_1 y_1 + x_2 y_2 ]_1 + [x_1 y_2 - x_2 y_1 ]_{1^\prime} + \left[\begin{array}{c} x_1 y_2 + x_1 y_2 \\ x_1 y_1 - x_2 y_2\end{array}\right]_{2^\prime}=\textbf{1}\oplus\textbf{1}^\prime\oplus \textbf{2}^\prime,
\end{equation}
being $\textbf{1}$ and $\textbf{1}^\prime$ singlets and $\textbf{2}^\prime$ a doublet. Besides we have that $\textbf{1}\otimes\textbf{1}=\textbf{1}$ and $\textbf{1}^\prime\otimes\textbf{1}\prime=\textbf{1}$. For more details about this and other discrete symmetries see Ref.~\cite{Ishimori:2010au}.  The $Z_N$ group, that is Abelian,  can be represented as discrete rotations, whose generators corresponds to a $2\pi/N$ rotation.

\begin{table}[ht]
\begin{tabular}{|c|c|c|c|c|c|c|c|}\hline
Symmetry & $Q_L,L_{i}$ &  $u_R,d_R,l_{jR}$ & $N_s$ & $N_d$ & $S$ & $D$  & $\zeta_d$  \\ \hline
$S_{3}$  & 1 &  1  & $1$ & 2 & 1
& 2  & 2  \\ \hline
$\mathbb{Z}_2$ & 1 & -1 & 1 & 1 & -1 & 1 & 1 \\ \hline
\end{tabular}
\caption{Transformation properties of the fermion and scalar fields under $S_{3}$ and $\mathbb{Z}_2$ symmetries. }
\label{table1}
\end{table}

\section{Scalar sector}
\label{sec:scalars}

The scalar sector of the model is presented as follows:
\begin{equation}
S=\left (\begin{array}{c}
S^+ \\
\frac{1}{\sqrt2}(v_{SM}+\textrm{Re}S^0+i\textrm{Im}S^0)
\end{array}\right),\quad D_{1,2}=\left( \begin{array}{c}
D^+_{1,2} \\
\frac{1}{\sqrt2}(\eta_{1,2}+i\chi_{1,2}).
\end{array}\right),
\label{notation}
\end{equation}
plus two real the singlets $\zeta_i=\frac{v_i+\xi_i}{\sqrt{2}}$, $i=1,2$.

The scalar potential invariant under the gauge and $S_3\otimes \mathbb{Z}_2$ symmetries is
\begin{eqnarray*}
V_{S_3}  &=& \mu^2_sS^\dagger S+\mu^2_d [D^\dagger\otimes  D]_1+ \mu^2_\zeta [\zeta_d\otimes  \zeta_d]_1 +\mu^2_{12}\zeta_1\zeta_2+a_1
([D^\dagger\otimes  D]_1)^2
+  a_2 [[D^\dagger\otimes D]_{1^\prime}[D^\dagger\otimes
D]_{1^\prime}]\nonumber\\
&&+a_3[(D^\dagger \otimes D)_{2^\prime}(D^\dagger\otimes D)_{2^\prime}]_1
+a_4(S^\dagger S)^2+
a_5[D^\dagger\otimes D]_1 S^\dagger  S + a_6 [[S ^\dagger D]_{2^\prime} [S^\dagger   D]_{2^\prime}]_1
\nonumber \\
&&+ H.c.]+
a_7 S^\dagger [ D \otimes D^\dagger]_1 S+
b_1 S^\dagger S [ \zeta_d \otimes \zeta_d]_1 + b_2  [D^\dagger\otimes  D]_1 [ \zeta_d \otimes \zeta_d]_1
+ b_3 [[D^\dagger\otimes  D]_{2^\prime} [ \zeta_d \otimes \zeta_d]_{2^\prime}]_{1} \nonumber\\&&+b_4 [[[D^\dagger\otimes D]_{1^\prime}[\zeta_d\otimes\zeta_d]_{1^\prime}]_1
+  c_1([\zeta_d\otimes  \zeta_d]_1)^2+
c_2 [[\zeta_d\otimes \zeta_d]_{2^\prime}[\zeta_d\otimes
\zeta_d]_{2^\prime}]_1,
\label{potential1}
\end{eqnarray*}
with $\mu^2_d>0$ since $\langle D^0_{1,2}\rangle=0$ is guaranteed by the $S_3$ symmetry. The parameter $a_6$ has been chosen real without loss of generality.

We can write Eq.~(\ref{potential1}) explicitly as
\begin{equation}
V(S,D,\zeta_d)=V^{(2)}+V^{(4a)}+V^{(4b)}+V^{(4c)},
\label{potential2}
\end{equation}
where
\begin{eqnarray*}
V^{(2)}&=&\mu^2_{SM}S^\dagger S+\mu^2_d (D^\dagger_1D_1+D^\dagger_2 D_2)+\mu^2_\zeta(\zeta^2_1+\zeta^2_2)+\mu^2_{12}\zeta_1\zeta_2,
\nonumber\\
V^{(4a)} &=&  a_1
(D^\dagger_1D_1+D^\dagger_2 D_2)^2
+  a_2 (D^\dagger_1D_2-D^\dagger_2D_1)^2
\nonumber\\
&&+ a_3[(D^\dagger_1D_2+D^\dagger_2D_1)^2+(D^\dagger_1D_1-D^\dagger_2D_2)^2]
+a_4(S^\dagger S)^2+
a_5  (D^\dagger_1D_1+D^\dagger_2 D_2)S^\dagger  S
\nonumber\\
&&+ a_6[(S^\dagger D_1)^2 +(S^\dagger D_2)^2 +H.c.]
+
a_7 [(S^\dagger D_1)(D^\dagger_1S)+(S^\dagger D_2)(D^\dagger_2S)],
\nonumber\\
V^{(4b)}&=& b_1S^\dagger S(\zeta^2_1+\zeta^2_2)+b_2(D^\dagger_1D_1+D^\dagger_2D_2)(\zeta^2_1+\zeta^2_2)+
b_3[(D^\dagger_1D_2+D^\dagger_2D_1)(\zeta_1\zeta_2+\zeta_1\zeta_2)
\nonumber \\
&&+ (D^\dagger_1D_1-D^\dagger_2D_2)(\zeta^2_1-\zeta^2_2)+H.c.] +  b_4[(D^\dagger_1D_2-D^\dagger_2D_1)(\zeta_1\zeta_2-\zeta_1\zeta_2)]
\nonumber\\
V^{(4c)}&=&c_1(\zeta^2_1+\zeta^2_2)^2
+ c_2[(\zeta_1\zeta_2+\zeta_2\zeta_1)^2+(\zeta^2_1-
\zeta^2_2)^2],
\label{potential3}
\end{eqnarray*}
where we have used $[\zeta_d\,\zeta_d]_{2^\prime}=(\zeta_1\zeta_2+\zeta_2\zeta_1,\zeta_1\zeta_1-\zeta_2\zeta_2)$. Here, we will consider all the couplings to be real parameters i.e., there is no $C\!P$ violation in the scalar sector.
The $S_3$ symmetry forbids linear terms with the doublets $D_1,D_2$ in the scalar potential and also some of the Yukawa interactions with charged leptons. This ensures the inert character of the these doublets after the $S_3$ symmetry is introduced. Notice that, although the term $\mu^2_{12}$ breaks softly the $S_3$ symmetry, it happens in the sector of the singlets $\zeta_{1,2}$ and does not spoil the inert character of the doublets.

From Eq.~(\ref{potential3}), we obtain the following stability conditions for the potential (i.e., setting its derivatives to zero):
\begin{eqnarray*}
\frac{1}{2} v_{SM} \left(2 a_4 v_{SM}^2+b_1 \left(v_1^2+v_2^2\right)+2 \mu_{SM}^2\right)&=&0,\nonumber \\
\frac{1}{2} \left(b_1 v_1 v_{SM}^2+2 v_1 (c_1+c_2) \left(v_1^2+v_2^2\right)+\mu_{12}^2 v_2\right)+\mu_\zeta^2 v_1&=&0,\nonumber\\
\frac{1}{2} b_1 v_2 v_{SM}^2+v_2 (c_1+c_2) \left(v_1^2+v_2^2\right)+\frac{\mu_{12}^2 v_1}{2}+\mu_\zeta^2 v_2&=&0,
\label{ce1}
\end{eqnarray*}
From Eq. \ref{ce1} we find three sets of solutions
\begin{equation}
    v_1=v_2=0, \quad \mu_{SM}=-a_4 v_{SM}^2
    \label{eq:solderivadas1}
\end{equation}
\begin{equation}
    v_2= -v_1,\quad \mu_\zeta^2= \frac{1}{2} \left(-b_1 v_{SM}^2-4 v_1^2 (c_1+c_2)+\mu_{12}^2\right), \quad \mu_{SM}=-a_4 v_{SM}^2-b_1 v_1^2
    \label{eq:solderivadas2}
\end{equation}
\begin{equation}
    v_2= v_1,\quad \mu_\zeta^2= \frac{1}{2} \left(-b_1 v_{SM}^2-4 v_1^2 (c_1+c_2)-\mu_{12}^2\right), \quad \mu_{SM}=-a_4 v_{SM}^2-b_1 v_1^2
    \label{eq:solderivadas3}
\end{equation}
These solutions will be used in the analyses presented throughout this work.

\section{Mass matrices and eigenstates}\label{sec:massmatrices}

The potential in Eq. \ref{potential1} gives us four mass matrices: one for the charged scalars, one for the CP-odd neutral scalars and two for the CP-even neutral scalars. In the sections below we shall show these matrices and their corresponding eigenvalues and eigenvectors. Also, when we have $\pm $ or $\mp$, the upper sign corresponds to the vacuum stability criteria from Eq. \ref{eq:solderivadas2} and the lower sign to Eq. \ref{eq:solderivadas3}.

\subsection{Charged scalars}

From Eq. \ref{potential2}, in the basis $(S^+, D_1^+, D_2^+) M_C (S^-,D_1^-,D_2^-)^T$, we find the mass matrix $M_C$ for the charged scalars to be
\begin{equation}
    M_C=
    \left(
\begin{array}{ccc}
 0 & 0 & 0 \\
 0 & b_2 v_1^2+\frac{a_5 v_{SM}^2}{2}+\mu_d^2 & \mp \, 2 b_3 v_1^2 \\
 0 & \mp \, 2 b_3 v_1^2 & b_2 v_1^2+\frac{a_5 v_{SM}^2}{2}+\mu_d^2 \\
\end{array}
\right).
\label{eq:matmassacarregados}
\end{equation}
The above matrix can be diagonalized as $R_C^T M_C R_C$, where $R_C$ is the orthogonal rotation matrix. For the matrix in Eq. \ref{eq:matmassacarregados} we find that the symmetry and mass eigenstates are related as
\begin{equation}
    \left(
    \begin{array}{c}
    S^+ \\ D_1^+ \\ D_2^+
    \end{array}
    \right)
    = R_C
    \left(
    \begin{array}{c}
    G^+ \\ H_1^+ \\ H_2^+
    \end{array}
    \right)
    =
    \left(
\begin{array}{ccc}
 1 & 0 & 0 \\
 0 & \pm \, \frac{1}{\sqrt{2}} & \mp \, \frac{1}{\sqrt{2}} \\
 0 & \frac{1}{\sqrt{2}} & \frac{1}{\sqrt{2}} \\
\end{array}
\right)
    \left(
    \begin{array}{c}
    G^+ \\ H_1^+ \\ H_2^+
    \end{array}
    \right).
\end{equation}
And their masses are:
\begin{equation}
    m_{G^+}^2=0, \quad m_{H_1^+}^2=\frac{a_5 v_{SM}^2}{2}+v_1^2 (b_2-2 b_3)+\mu_d^2, \quad m_{H_2^+}^2=\frac{a_5 v_{SM}^2}{2}+v_1^2 (b_2+2 b_3)+\mu_d^2.
\end{equation}

\subsection{CP-odd scalars}

From Eq. \ref{potential2}, in the basis $(Im[S^0],\chi_1,\chi_2) M_O (Im[S^0],\chi_1,\chi_2)^T$, we find the mass matrix $M_O$ for the CP-odd scalars to be
\begin{equation}
    M_O=
   \left(
\begin{array}{ccc}
 0 & 0 & 0 \\
 0 &  b_2   v_1 ^2+\frac{1}{2} ( a_5 -2  a_6 + a_7 )  v_{SM} ^2+ \mu_d^2  & \mp 2  b_3   v_1 ^2 \\
 0 & \mp 2  b_3   v_1 ^2 &  b_2   v_1 ^2+\frac{1}{2} ( a_5 -2  a_6 + a_7 )  v_{SM} ^2+ \mu_d^2  \\
\end{array}
\right).
\end{equation}
The diagonalization for this mass matrix is given by
\begin{equation}
    \left(
    \begin{array}{c}
         Im[S^0] \\ \chi_1 \\ \chi_2
    \end{array}
    \right)
    =
    R_O
    \left(
    \begin{array}{c}
         G_A^0  \\ A_1^0 \\ A_2^0
    \end{array}
    \right)
    =
    \left(
\begin{array}{ccc}
 1 & 0 & 0 \\
 0 & \pm \, \frac{1}{\sqrt{2}} & \mp \, \frac{1}{\sqrt{2}} \\
 0 & \frac{1}{\sqrt{2}} & \frac{1}{\sqrt{2}} \\
\end{array}
\right)
    \left(
    \begin{array}{c}
         G_A^0  \\ A_1^0 \\ A_2^0
    \end{array}
    \right).
\end{equation}
With masses:
\begin{align}
    m_{G_A^0}^2=0, \quad m_{A_1^0}^2=\frac{1}{2}  v_{SM} ^2 ( a_5 -2  a_6 + a_7 )+ v_1 ^2 ( b_2 -2  b_3 )+ \mu_d^2 ,\\ m_{A_2^0}^2=\frac{1}{2}  v_{SM} ^2 ( a_5 -2  a_6 + a_7 )+ v_1 ^2 ( b_2 +2  b_3 )+ \mu_d^2 . \nonumber
\end{align}

\subsection{CP-even scalars, $2\times 2$ matrix}

Eq. \ref{potential2} gives us for the CP-even sector two matrices, one $2\times 2$ and one $3 \times 3$. The $2\times 2$ matrix, in the basis $(\eta_1,\eta_2) M_{E1} (\eta_1,\eta_2)^T$, is
\begin{equation}
    M_{E1}=
    \left(
\begin{array}{cc}
  b_2   v_1 ^2+\frac{1}{2} ( a_5 +2  a_6 + a_7 )  v_{SM} ^2+ \mu_d^2  & \mp 2  b_3   v_1 ^2 \\
 \mp 2  b_3   v_1 ^2 &  b_2   v_1 ^2+\frac{1}{2} ( a_5 +2  a_6 + a_7 )  v_{SM} ^2+ \mu_d^2  \\
\end{array}
\right).
\end{equation}
The symmetry and mass eigenstates are related as
\begin{equation}
    \left(
    \begin{array}{c}
        \eta_1 \\ \eta_2
    \end{array}
    \right)
    =
    R_{E1}
    \left(
    \begin{array}{c}
        h^0_1 \\ h^0_2
    \end{array}
    \right)
    =
    \left(
\begin{array}{cc}
 \pm \, \frac{1}{\sqrt{2}} & \mp \, \frac{1}{\sqrt{2}} \\
 \frac{1}{\sqrt{2}} & \frac{1}{\sqrt{2}} \\
\end{array}
\right)
    \left(
    \begin{array}{c}
        h_1^0 \\ h_2^0
    \end{array}
    \right).
\end{equation}
The masses are:
\begin{align}
    m_{h_1^0}^2=\frac{1}{2}  v_{SM} ^2 ( a_5 +2  a_6 + a_7 )+ v_1 ^2 ( b_2 -2  b_3 )+ \mu_d^2 , \\ m_{h_2^0}^2=\frac{1}{2}  v_{SM} ^2 ( a_5 +2  a_6 + a_7 )+ v_1 ^2 ( b_2 +2  b_3 )+ \mu_d^2 . \nonumber
\end{align}

\section{The CP-even $3 \times 3$ matrix and the Higgs boson}\label{sec:CPpar3x3}

Again from Eq. \ref{potential2}, we obtain the $3 \times 3$ matrix for the CP-even scalars. Considering the basis $(Re[S^0], \xi_1,\xi_2) M_{E2} (Re[S^0], \xi_1, \xi_2)$ we find
\begin{equation}
    M_{E2}=
   \left(
\begin{array}{ccc}
 2  a_4   v_{SM} ^2 & b_1   v_1   v_{SM}  & \mp b_1   v_1   v_{SM}  \\
 b_1   v_1   v_{SM}  & \frac{1}{2} \left(4 ( c_1 + c_2 )  v_1 ^2 \pm \mu_{12}^2 \right) & \frac{1}{2} \left( \mu_{12}^2 \mp 4 ( c_1 + c_2 )  v_1 ^2\right) \\
 \mp b_1   v_1   v_{SM}  & \frac{1}{2} \left( \mu_{12}^2 \mp 4 ( c_1 + c_2 )  v_1 ^2\right) & \frac{1}{2} \left(4 ( c_1 + c_2 )  v_1 ^2 \pm \mu_{12}^2 \right) \\
\end{array}
\right),
\label{eq:MatMassaCPpar3x3}
\end{equation}
where, when we have $\pm$ or $\mp$, the upper sign corresponds to the vacuum stability criteria from Eq. \ref{eq:solderivadas2} and the lower sign to Eq. \ref{eq:solderivadas3}. To find the mass eigenstates of this sector, we will follow Ref. \cite{Higgsm331}, where we impose that the Yukawa couplings of the Higgs boson are the same in the SM and in the $S_3\otimes \mathbb{Z}_2$ model.

In the $S_3\otimes \mathbb{Z}_2$ model, the Yukawa sector for the leptons is given by
\begin{eqnarray*}
    - \mathcal { L } _ { Y u k a w a } ^ { l e p t o n s } &=& G _ { i j } ^ { l } \overline { L } _ { i } l _ { j R } S + G _ { i d } ^ { \nu } \overline { ( L ) ^ { c } } _ { i a } \epsilon _ { a b } \left[ N _ { d } D _ { b } \right] _ { 1 } + \frac { 1 } { \Lambda } G _ { i s } ^ { \nu } \overline { ( L ) ^ { c } } \epsilon _ { a b } \left[ N _ { s 1 ^ { \prime } } \left[ D _ { b } \zeta _ { d } \right] _ { 1 ^ { \prime } } \right] _ { 1 } \\&& + \frac { 1 } { 2 } M _ { s } \overline { N _ { s 1 ^ { \prime } } ^ { c } } N _ { s 1 ^ { \prime } } + \frac { 1 } { 2 } M _ { d } \left[ \overline { N _ { d } ^ { c } } N _ { d } \right] _ { 1 } + y \left[ \overline { N _ { s 1 ^ { \prime } } ^ { c } } \left[ N _ { d } \zeta d \right] _ { 1 ^ { \prime } } \right] _ { 1 } + H . c . ; \nonumber
    \label{eq:Yukawa}
\end{eqnarray*}
where $a$, $b$ are $SU(2)$ indices, $i, j = e, \mu, \tau$ (we omit summation symbols), $L_i(l_{iR})$ denote the usual left-handed lepton doublets (right-handed charged lepton singlets), $G's$ are the Yukawa couplings, and $N_{s,d}$ are the right-handed neutrinos; $\left[ D \zeta _ { d } \right] _ { 1 ^ { \prime } } = D _ { 1 } \zeta _ { 2 } - D _ { 2 } \zeta _ { 1 }$, $\left[ N _ { d } D \right] _ { 1 } = N _ { 2 R } D _ { 1 } + N _ { 3 R } D _ { 2 }$, according to the $S_3$ multiplication rules, and $\tilde{D}_{1,2}=i \tau_2 D_{1,2}$, where $\tau_2$ is the second Pauli matrix.

From Eq. \ref{eq:Yukawa}, $Re[S^0]$ is the only one that gives mass to the known fermions, therefore we will identify it as the SM Higgs. The matrix in Eq. \ref{eq:MatMassaCPpar3x3} can be diagonalized by an orthogonal $3\times3$ matrix, such that $R^T_{E2} M_{E2} R_{E2}=diag(m^2_H, m^2_{H^0_1}, m^2_{H^0_2})$, where $H$ is the SM Higgs boson and $H^0_{1,2}$ are the other CP-even scalars from the $3 \times 3$ CP-even mass matrix. This implies
\begin{align}
\left(\begin{array}{c}
H \\ H_1^0 \\ H_2^0
\end{array}\right)
&=R^T_{E2}
\left(
\begin{array}{c}
Re[S^0] \\ \xi_1 \\ \xi_2
\end{array}
\right)
\\&=
\left(
\begin{array}{ccc}
 cos \theta_2 & -cos \theta_3 sin \theta_2 & sin \theta_2 sin \theta_3 \\
 cos \theta_1 sin \theta_2 & cos \theta_1 cos \theta_2 cos \theta_3-sin \theta_1 sin \theta_3 & -cos \theta_3 sin \theta_1-cos \theta_1 cos \theta_2 sin \theta_3 \\
 sin \theta_1 sin \theta_2 & cos \theta_2 cos \theta_3 sin \theta_1+cos \theta_1 sin \theta_3 & cos \theta_1 cos \theta_3-cos \theta_2 sin \theta_1 sin \theta_3 \\
\end{array}
\right)
\left(
\begin{array}{c}
Re[S^0] \\ \xi_1 \\ \xi_2
\end{array}
\right). \nonumber
\end{align}
Since $Re[S^0]$ is the scalar we identify as the SM Higgs, we need that $(R^T_{E2})_{11}=1$ and all the other elements from the first row to be zero. To do so, we need $\theta_2=0$, which gives us $cos \theta_2=1$ and $sin \theta_2=0$, leaving $R_{E2}$ as
\begin{equation}
    R_{E2}=
    \left(
\begin{array}{ccc}
 1 & 0 & 0 \\
 0 & cos \theta_1 cos \theta_3-sin \theta_1 sin \theta_3 & cos \theta_3 sin \theta_1+cos \theta_1 sin \theta_3 \\
 0 & -cos \theta_3 sin \theta_1-cos \theta_1 sin \theta_3 & cos \theta_1 cos \theta_3-sin \theta_1 sin \theta_3 \\
\end{array}
\right)
=
\left(
\begin{array}{ccc}
	1 & 0 & 0 \\
	0 & cos \theta & sin \theta \\
	0 & -sin \theta & cos \theta \\
\end{array}
\right),
\label{eq:RotacaoCPpar3x3}
\end{equation}
where $\theta=\theta_1+\theta_3$. The matrix $R_{E2}$ from Eq. \ref{eq:RotacaoCPpar3x3} does not automatically diagonalize $M_{E2}$, i.e., it does not lead to $R^T_{E2} M_{E2} R_{E2}=diag(m^2_H, m^2_{H^0_1}, m^2_{H^0_2})$. To have that we need to impose conditions on the parameters that make up $M_{E2}$ and $R_{E2}$, so that we fulfill the equation $R^T_{E2} M_{E2} R_{E2}=diag(m^2_H, m^2_{H^0_1}, m^2_{H^0_2})$. The possible solutions depend on which solution for the vacuum stability conditions (Eqs. \ref{eq:solderivadas1}-\ref{eq:solderivadas3}) we choose. 
\begin{enumerate}
    \item $v_1=v_2=0$:
    \subitem This solution trivializes our mass matrix $M_{E2}$, making it diagonal from the very beginning, leaving one scalar with mass $m^2_{Re[S^0]}=a_4 v_{SM}^2$ and the other two scalars with mass $m^2_{\xi_1,\xi_2}=-\mu_{12}^2/4$. Therefore, there is no need for the diagonalization method presented in this section.
    
    \item $ v_2= -v_1$ and $\mu_\zeta^2= \frac{1}{2} \left(-b_1 v_{SM}^2-4 v_1^2 (c_1+c_2)+\mu_{12}^2\right)$:
    \subitem In this case we find 10 possible solutions for the parameters $a_4$, $b_1$, $\mu_{12}^2$ and the sines and cosines of $\theta$. Amongst all solutions, we either have all masses equal, $m^2=4 v_1^2 (c_1+c_2)$; or two equal masses, $m^2_1=2 a_4 v_{SM}^2$, and a third different mass $m^2_2=4 v_1^2(c_1+c_2)$. In both cases, $b_1=0$. When all masses are equal, $a_4=\frac{2 v_1^2 (c_1+c_2)}{v_{SM}^2}$ and $\mu_{12}^2=4 v_1^2 (c_1+c_2)$; when we have different masses, $\mu_{12}^2=2 a_4 v_{SM}^2$. As for the sines and cosines, they can either be functions of each other or have values $\pm \, 1/\sqrt{2}$.
    
    \item $ v_2= v_1,\quad \mu_\zeta^2= \frac{1}{2} \left(-b_1 v_{SM}^2-4 v_1^2 (c_1+c_2)-\mu_{12}^2\right)$:
    \subitem Once again we have 10 possible solutions for the parameters $a_4$, $b_1$, $\mu_{12}^2$ and the sines and cosines of $\theta$. Amongst all solutions, we either have all masses equal, $m^2=4 v_1^2 (c_1+c_2)$; or two equal masses, $m^2_1=2 a_4 v_{SM}^2$, and a third different mass $m^2_2=4 v_1^2 (c_1+c_2)$. In both cases, $b_1=0$. When all masses are equal, $a_4=\frac{2 v_1^2 (c_1+c_2)}{v_{SM}^2}$ and $\mu_{12}^2=-4 v_1^2 (c_1+c_2)$; when we have different masses, $\mu_{12}^2=-2 a_4 v_{SM}^2$. As for the sines and cosines, they can either be functions of each other or have values $\pm \, 1/\sqrt{2}$.
\end{enumerate}
The complete set of solutions are shown in Appendix \ref{sec:SolucoesCPpar3x3}. The mass degeneracy in this sector may look like a problem at first. However, having particles with the same mass helps to reduce the parameter $\Delta \rho$, which we address in the following section.

\section{Copositivity criteria}\label{sec:copositividade}

From Eq. \ref{potential2} all we found so far are the conditions for the potential stability (Eq. \ref{ce1}), which guarantees that the vacua are in a stable point, but do not guarantee that it is a minimum. To do so we will follow the method presented in Ref. \cite{Kannike:2012pe} also using Ref. \cite{ping}. A scalar potential with the form $\lambda_{ab} \phi_a^2 \phi_b^2$ is bounded from below if its matrix of quartic couplings is copositive, which means that its eignevalues are all non-negative. Therefore, we must build our matrix using the basis $(S^\dagger S, D_1^\dagger D_1, D_2^\dagger D_2, \zeta_1^2, \zeta_2^2)$. Although we have a $5\times 5$ matrix from our basis, two of its columns/rows are equal, which reduces to the following $4 \times 4$ matrix
\begin{equation}
\mathcal{A}=
\left(
\begin{array}{cccc}
a_4 & a_5+a_6+2 a_7 & a_5+a_6+2 a_7 & b_1 \\
a_5+a_6+2 a_7 & a_1+a_3 & 2 (a_1+a_3) & b_2+b_4 \\
a_5+a_6+2 a_7 & 2 (a_1+a_3) & a_1+a_3 & b_2-b_4 \\
b_1 & b_2+b_4 & b_2-b_4 & c_1+c_3 \\
\end{array}
\right).
\label{matriz44}
\end{equation}

Below are shown the different conditions under which the potential is no just bounded from below, but also guarantee that all scalar masses and vacuum expectation values are positive, and that all coupling constants are real. All indices $i,j,k,l,...$ are fixed and different from each other. Also, we want to remind the reader that all diagonal elements from matrix $\mathcal{A}$ given in Eq.~(\ref{matriz44}) should be positive for $\mathcal{A}$ to be copositive.

\newpage
\begin{center}
\fbox{{\bf {\textbf{Case 1.} All $\mathcal{A}_{ij}$ positive.}}}
\end{center}

\begin{eqnarray*}
&& b_2=b_1\text{ and } b_4=b_1\text{ and } a_4>b_1\text{ and } a_7>\frac{-3 a_5 v_{SM}^2-4 b_3 v_1^2-2 \mu _d^2}{5 v_{SM}^2}\text{ and } c_2>-c_1 \text{ and } v_{SM}>b_1\text{ and }  a_1\geq -a_3
 \\\nonumber&& 
\text{ and } \mu _d^2>\frac{1}{2} \left(-a_5 v_{SM}^2-4 b_3 v_1^2\right)\text{ and } v_1>b_1  
\text{ and } c_1\geq -c_3\text{ and } a_6<\frac{a_5 v_{SM}^2+a_7 v_{SM}^2+4 b_3 v_1^2+2 \mu _d^2}{2 v_{SM}^2}
\\\nonumber&&
\text{ and } -a_5-2 a_7\leq a_6\text{ and } b_3\leq b_1 
\\\nonumber&&
\textbf{or}
\\\nonumber&&
\left(a_3\left|a_5\right|c_3\right)\in \mathbb{R}\text{ and } b_2=b_1\text{ and } b_4=b_1\text{ and } a_4>b_1
\text{ and } a_7>\frac{-3 a_5 v_{SM}^2+4 b_3 v_1^2-2 \mu _d^2}{5 v_{SM}^2}\text{ and } b_3>b_1\text{ and } c_2>-c_1
\\\nonumber&&
\text{ and } \mu _d^2>\frac{1}{2} \left(4 b_3 v_1^2-a_5 v_{SM}^2\right)\text{ and } v_1>b_1
\text{ and } v_{SM}>b_1\text{ and } a_1\geq -a_3\text{ and } c_1\geq -c_3
\\\nonumber&&
\text{ and } a_6<\frac{a_5 v_{SM}^2+a_7 v_{SM}^2-4 b_3 v_1^2+2 \mu _d^2}{2 v_{SM}^2}\text{ and } -a_5-2 a_7\leq a_6 
\\\nonumber&&
\textbf{or}
\\\nonumber&&
\left(a_3\left|a_5\right|c_3\right)\in \mathbb{R}\text{ and } a_4>b_1\text{ and } a_7>\frac{-3 a_5 v_{SM}^2-2 b_2 v_1^2-4 b_3 v_1^2-2 \mu _d^2}{5 v_{SM}^2}
\text{ and } b_2>b_1\text{ and } c_2>-c_1
\\\nonumber&&
\text{ and } \mu _d^2>\frac{1}{2} \left(-a_5 v_{SM}^2-2 b_2 v_1^2-4 b_3 v_1^2\right)\text{ and } v_1>b_1
\text{ and } v_{SM}>b_1\text{ and } a_1\geq -a_3\text{ and } c_1\geq -c_3
\\\nonumber&&
\text{ and } a_6<\frac{a_5 v_{SM}^2+a_7 v_{SM}^2+2 b_2 v_1^2+4 b_3 v_1^2+2 \mu _d^2}{2 v_{SM}^2}
\text{ and } -a_5-2 a_7\leq a_6\text{ and } -b_2\leq b_4\text{ and } b_3\leq b_1\text{ and } b_4\leq b_2
\\\nonumber&&
\textbf{or}
\\\nonumber&&
\left(a_3\left|a_5\right|c_3\right)\in \mathbb{R}\text{ and } a_4>b_1\text{ and } a_7>\frac{-3 a_5 v_{SM}^2-2 b_2 v_1^2+4 b_3 v_1^2-2 \mu _d^2}{5 v_{SM}^2}
\text{ and } b_2>b_1\text{ and } b_3>b_1\text{ and } c_2>-c_1
\\\nonumber&&
\text{ and } \mu _d^2>\frac{1}{2} \left(-a_5 v_{SM}^2-2 b_2 v_1^2+4 b_3 v_1^2\right)
\text{ and } v_1>b_1\text{ and } v_{SM}>b_1 \text{ and } a_1\geq -a_3\text{ and } c_1\geq -c_3
\\\nonumber&&
\text{ and } a_6<\frac{a_5 v_{SM}^2+a_7 v_{SM}^2+2 b_2 v_1^2-4 b_3 v_1^2+2 \mu _d^2}{2 v_{SM}^2}\text{ and } -a_5-2 a_7\leq a_6
\text{ and } -b_2\leq b_4\text{ and } b_4\leq b_2
\end{eqnarray*}

\newpage
\begin{center}
\fbox{{\bf { \textbf{Case 2.} $\mathcal{A}_{ij} \leq 0$ and the other entries positive.}}}
\end{center}
\begin{center}
\fbox{{If $a5+a6+2 a7\leq 0$, then:  }}
\end{center}
\begin{eqnarray*}
&& a_1>\frac{a_5^2+2 a_6 a_5+4 a_7 a_5+a_6^2+4 a_7^2-a_3 a_4+4 a_6 a_7}{a_4}\text{ and } a_4>b_1
\text{ and } a_7>\frac{-3 a_5 v_{SM}^2-2 b_2 v_1^2-4 b_3 v_1^2-2 \mu _d^2}{5 v_{SM}^2}
\\\nonumber&& 
\text{ and } c_1>-c_2\text{ and } \mu _d^2>\frac{1}{2} \left(-a_5 v_{SM}^2-2 b_2 v_1^2-4 b_3 v_1^2\right)\text{ and } v_1>b_1\text{ and } v_{SM}>b_1 
\text{ and } a_6\leq -a_5-2 a_7\text{ and } b_3\leq b_1
\\\nonumber&& 
\textbf{or}  
\\\nonumber&& 
b_4\in \mathbb{R}\text{ and } c_3\in \mathbb{R}\text{ and } \left(a_3\left|a_5\right|b_2|c_2\right)\in \mathbb{R}
\text{ and } a_1>\frac{a_5^2+2 a_6 a_5+4 a_7 a_5+a_6^2+4 a_7^2-a_3 a_4+4 a_6 a_7}{a_4}\text{ and } a_4>b_1
\\\nonumber&&
\text{ and } a_7>\frac{-3 a_5 v_{SM}^2-2 b_2 v_1^2+4 b_3 v_1^2-2 \mu _d^2}{5 v_{SM}^2}\text{ and } b_3>b_1
\text{ and } c_1>-c_2\text{ and } \mu _d^2>\frac{1}{2} \left(-a_5 v_{SM}^2-2 b_2 v_1^2+4 b_3 v_1^2\right)
\\\nonumber&& 
\text{ and } v_{SM}>b_1\text{ and } a_6\leq -a_5-2 a_7 \text{ and } v_1>b_1
\\\nonumber&& 
\textbf{or}  
\\\nonumber&& 
a_1>\frac{a_5^2+2 a_6 a_5+4 a_7 a_5+a_6^2+4 a_7^2-a_3 a_4+4 a_6 a_7}{a_4}\text{ and } a_4>b_1\text{ and } b_3>b_1 \text{ and } v_1>b_1\text{ and } v_{SM}>b_1
\\\nonumber&& 
\text{ and } c_1>-c_2\text{ and } \mu _d^2>\frac{1}{2} \left(-a_5 v_{SM}^2-2 b_2 v_1^2+4 b_3 v_1^2\right)
\\\nonumber&& 
\text{ and } a_6<\frac{a_5 v_{SM}^2+a_7 v_{SM}^2+2 b_2 v_1^2-4 b_3 v_1^2+2 \mu _d^2}{2 v_{SM}^2} 
\text{ and } a_7\leq \frac{-3 a_5 v_{SM}^2-2 b_2 v_1^2+4 b_3 v_1^2-2 \mu _d^2}{5 v_{SM}^2}
\\\nonumber&& 
\textbf{or}  
\\\nonumber&& 
b_4\in \mathbb{R}\text{ and } c_3\in \mathbb{R}\text{ and } \left(a_3\left|a_5\right|b_2|c_2\right)\in \mathbb{R}
\text{ and } a_1>\frac{a_5^2+2 a_6 a_5+4 a_7 a_5+a_6^2+4 a_7^2-a_3 a_4+4 a_6 a_7}{a_4}
\\\nonumber&&
\text{ and } a_4>b_1\text{ and } c_1>-c_2
\text{ and } \mu _d^2>\frac{1}{2} \left(-a_5 v_{SM}^2-2 b_2 v_1^2-4 b_3 v_1^2\right)\text{ and } v_1>b_1\text{ and } v_{SM}>b_1
\\\nonumber&&
\text{ and } a_6<\frac{a_5 v_{SM}^2+a_7 v_{SM}^2+2 b_2 v_1^2+4 b_3 v_1^2+2 \mu _d^2}{2 v_{SM}^2}
\text{ and } a_7\leq \frac{-3 a_5 v_{SM}^2-2 b_2 v_1^2-4 b_3 v_1^2-2 \mu _d^2}{5 v_{SM}^2}\text{ and } b_3\leq b_1
\end{eqnarray*}

\begin{center}
\fbox{{ If $a5+a6+2 a7\leq 0$, then: }}
\end{center}

\begin{eqnarray*}
&& a_1>\frac{a_5^2+2 a_6 a_5+4 a_7 a_5+a_6^2+4 a_7^2-a_3 a_4+4 a_6 a_7}{a_4}\text{ and } a_4>b_1\text{ and } c_1>-c_2 \text{ and } a_6\leq -a_5-2 a_7\text{ and } b_3\leq b_1
\\\nonumber&&
\text{ and } a_7>\frac{-3 a_5 v_{SM}^2-2 b_2 v_1^2-4 b_3 v_1^2-2 \mu _d^2}{5 v_{SM}^2}
\text{ and } \mu _d^2>\frac{1}{2} \left(-a_5 v_{SM}^2-2 b_2 v_1^2-4 b_3 v_1^2\right)\text{ and } v_1>b_1\text{ and } v_{SM}>b_1
\\\nonumber&& 
\textbf{or}  
\\\nonumber&&
a_1>\frac{a_5^2+2 a_6 a_5+4 a_7 a_5+a_6^2+4 a_7^2-a_3 a_4+4 a_6 a_7}{a_4}\text{ and } a_4>b_1
\text{ and } a_7>\frac{-3 a_5 v_{SM}^2-2 b_2 v_1^2+4 b_3 v_1^2-2 \mu _d^2}{5 v_{SM}^2}
\\\nonumber&&
\text{ and } \mu _d^2>\frac{1}{2} \left(-a_5 v_{SM}^2-2 b_2 v_1^2+4 b_3 v_1^2\right)\text{ and } v_1>b_1\text{ and } v_{SM}>b_1\text{ and } a_6\leq -a_5-2 a_7 \text{ and } b_3>b_1\text{ and } c_1>-c_2
\\\nonumber&& 
\textbf{or}  
\\\nonumber&&
b_4\in \mathbb{R}\text{ and } c_3\in \mathbb{R}
\text{ and } \left(a_3\left|a_5\right|b_2|c_2\right)\in \mathbb{R}
\text{ and } a_1>\frac{a_5^2+2 a_6 a_5+4 a_7 a_5+a_6^2+4 a_7^2-a_3 a_4+4 a_6 a_7}{a_4} \text{ and } v_1>b_1
\\\nonumber&&
\text{ and } a_4>b_1\text{ and } b_3>b_1\text{ and } c_1>-c_2\text{ and } \mu _d^2>\frac{1}{2} \left(-a_5 v_{SM}^2-2 b_2 v_1^2+4 b_3 v_1^2\right)
\\\nonumber&&
\text{ and } v_{SM}>b_1\text{ and } a_6<\frac{a_5 v_{SM}^2+a_7 v_{SM}^2+2 b_2 v_1^2-4 b_3 v_1^2+2 \mu _d^2}{2 v_{SM}^2}
\text{ and } a_7\leq \frac{-3 a_5 v_{SM}^2-2 b_2 v_1^2+4 b_3 v_1^2-2 \mu _d^2}{5 v_{SM}^2}
\end{eqnarray*}

\begin{eqnarray*} 
\\\nonumber&& 
\textbf{or}  
\\\nonumber&&
b_4\in \mathbb{R}\text{ and } c_3\in \mathbb{R}\text{ and } \left(a_3\left|a_5\right|b_2|c_2\right)\in \mathbb{R}
\text{ and } a_1>\frac{a_5^2+2 a_6 a_5+4 a_7 a_5+a_6^2+4 a_7^2-a_3 a_4+4 a_6 a_7}{a_4}\text{ and } a_4>b_1
\\\nonumber&&
\text{ and } \mu _d^2>\frac{1}{2} \left(-a_5 v_{SM}^2-2 b_2 v_1^2-4 b_3 v_1^2\right)\text{ and } v_1>b_1\text{ and } v_{SM}>b_1 \text{ and } c_1>-c_2 
\\\nonumber&&
\text{ and } a_7\leq \frac{-3 a_5 v_{SM}^2-2 b_2 v_1^2-4 b_3 v_1^2-2 \mu _d^2}{5 v_{SM}^2}\text{ and } b_3\leq b_1
\text{ and } a_6<\frac{a_5 v_{SM}^2+a_7 v_{SM}^2+2 b_2 v_1^2+4 b_3 v_1^2+2 \mu _d^2}{2 v_{SM}^2}
\end{eqnarray*}

\begin{center}
\fbox{{ If $0\leq 0$, then:}}
\end{center}

\begin{eqnarray*}
&& a_4>b_1\text{ and } b_3>b_1\text{ and } c_1>-c_3\text{ and } c_2>-c_1
\text{ and } \mu _d^2>\frac{1}{2} \left(-a_5 v_{SM}^2-2 b_2 v_1^2+4 b_3 v_1^2\right)\text{ and } v_1>b_1\text{ and } v_{SM}>b_1
\\\nonumber&&
\text{ and } a_6<\frac{a_5 v_{SM}^2+a_7 v_{SM}^2+2 b_2 v_1^2-4 b_3 v_1^2+2 \mu _d^2}{2 v_{SM}^2}
\\\nonumber&& 
\textbf{or}  
\\\nonumber&&
a_4>b_1\text{ and } c_1>-c_3\text{ and } c_2>-c_1\text{ and } \mu _d^2>\frac{1}{2} \left(-a_5 v_{SM}^2-2 b_2 v_1^2-4 b_3 v_1^2\right)
\text{ and } v_1>b_1\text{ and } v_{SM}>b_1
\\\nonumber&&
\text{ and } a_6<\frac{a_5 v_{SM}^2+a_7 v_{SM}^2+2 b_2 v_1^2+4 b_3 v_1^2+2 \mu _d^2}{2 v_{SM}^2}
\text{ and } b_3\leq b_1
\end{eqnarray*}

\begin{center}
\fbox{{ If $2 a1+2 a3\leq 0$, then:}}
\end{center}
There is no solution

\begin{center}
\fbox{{ If $b2+b4\leq 0$, then:}}
\end{center}

\begin{eqnarray*}
&&  a_3>-a_1\text{ and } a_4>b_1\text{ and } b_2>\frac{-a_5 v_{SM}^2-4 b_3 v_1^2-2 \mu _d^2}{2 v_1^2}
\text{ and } c_2>-c_1\text{ and } c_3>\frac{-a_1 c_1-a_3 c_1+b_2^2+2 b_4 b_2+b_4^2}{a_1+a_3}
\\\nonumber&& 
\text{ and } v_1>b_1
\text{ and } v_{SM}>b_1\text{ and } a_6<\frac{a_5 v_{SM}^2+a_7 v_{SM}^2+2 b_2 v_1^2+4 b_3 v_1^2+2 \mu _d^2}{2 v_{SM}^2}
\text{ and } b_3<b_1\text{ and } b_4\leq -b_2
\\\nonumber&& 
\textbf{or}  
\\\nonumber&&
 a_3>-a_1\text{ and } a_4>b_1\text{ and } b_2>\frac{-a_5 v_{SM}^2+4 b_3 v_1^2-2 \mu _d^2}{2 v_1^2}\text{ and } c_2>-c_1
 \text{ and } c_3>\frac{-a_1 c_1-a_3 c_1+b_2^2+2 b_4 b_2+b_4^2}{a_1+a_3}
\\\nonumber&&
 \text{ and } v_1>b_1\text{ and } v_{SM}>b_1
 \text{ and } b_3\geq b_1\text{ and } a_6<\frac{a_5 v_{SM}^2+a_7 v_{SM}^2+2 b_2 v_1^2-4 b_3 v_1^2+2 \mu _d^2}{2 v_{SM}^2}\text{ and } b_4\leq -b_2 
\\\nonumber&& 
\textbf{or}  
\\\nonumber&&
  a_4>b_1\text{ and } b_2>\frac{-a_5 v_{SM}^2-4 b_3 v_1^2-2 \mu _d^2}{2 v_1^2}\text{ and } c_2>-c_1\text{ and } v_1>b_1\text{ and } v_{SM}>b_1
\text{ and } a_3<-a_1
\\\nonumber&&  
\text{ and } a_6<\frac{a_5 v_{SM}^2+a_7 v_{SM}^2+2 b_2 v_1^2+4 b_3 v_1^2+2 \mu _d^2}{2 v_{SM}^2}\text{ and } b_3<b_1
\text{ and } c_3<\frac{-a_1 c_1-a_3 c_1+b_2^2+2 b_4 b_2+b_4^2}{a_1+a_3}\text{ and } b_4\leq -b_2
\\\nonumber&& 
\textbf{or}  
\\\nonumber&&
a_4>b_1\text{ and } b_2>\frac{-a_5 v_{SM}^2+4 b_3 v_1^2-2 \mu _d^2}{2 v_1^2}\text{ and } c_2>-c_1\text{ and } v_1>b_1\text{ and } v_{SM}>b_1
\text{ and } b_3\geq b_1
\\\nonumber&&
\text{ and } a_3<-a_1\text{ and } a_6<\frac{a_5 v_{SM}^2+a_7 v_{SM}^2+2 b_2 v_1^2-4 b_3 v_1^2+2 \mu _d^2}{2 v_{SM}^2}
\text{ and } c_3<\frac{-a_1 c_1-a_3 c_1+b_2^2+2 b_4 b_2+b_4^2}{a_1+a_3}\text{ and } b_4\leq -b_2
\end{eqnarray*}

\begin{center}
\fbox{{ If $b2-b4\leq 0$, then:}}
\end{center}

\begin{eqnarray*}
\nonumber
&&  a_3>-a_1\text{ and } a_4>b_1\text{ and } b_2>\frac{-a_5 v_{SM}^2-4 b_3 v_1^2-2 \mu _d^2}{2 v_1^2}\text{ and } c_2>-c_1 \text{ and } v_{SM}>b_1\text{ and } b_4\geq b_2
\\\nonumber&& 
\text{ and } c_3>\frac{-a_1 c_1-a_3 c_1+b_2^2-2 b_4 b_2+b_4^2}{a_1+a_3}\text{ and } v_1>b_1 
\text{ and } a_6<\frac{a_5 v_{SM}^2+a_7 v_{SM}^2+2 b_2 v_1^2+4 b_3 v_1^2+2 \mu _d^2}{2 v_{SM}^2}\text{ and } b_3<b_1
\\\nonumber&& 
\textbf{or}  
\\\nonumber&&
 a_3>-a_1\text{ and } a_4>b_1\text{ and } b_2>\frac{-a_5 v_{SM}^2+4 b_3 v_1^2-2 \mu _d^2}{2 v_1^2}\text{ and } c_2>-c_1
 \text{ and } c_3>\frac{-a_1 c_1-a_3 c_1+b_2^2-2 b_4 b_2+b_4^2}{a_1+a_3}
  \\\nonumber&&
 \text{ and } v_1>b_1\text{ and } v_{SM}>b_1\text{ and } b_3\geq b_1
\text{ and } b_4\geq b_2\text{ and } a_6<\frac{a_5 v_{SM}^2+a_7 v_{SM}^2+2 b_2 v_1^2-4 b_3 v_1^2+2 \mu _d^2}{2 v_{SM}^2}
\\\nonumber&& 
\textbf{or}  
\\\nonumber&&
 a_4>b_1\text{ and } b_2>\frac{-a_5 v_{SM}^2-4 b_3 v_1^2-2 \mu _d^2}{2 v_1^2}\text{ and } c_2>-c_1\text{ and } v_1>b_1\text{ and } v_{SM}>b_1
 \text{ and } b_4\geq b_2\text{ and } a_3<-a_1
  \\\nonumber&&
 \text{ and } a_6<\frac{a_5 v_{SM}^2+a_7 v_{SM}^2+2 b_2 v_1^2+4 b_3 v_1^2+2 \mu _d^2}{2 v_{SM}^2}
 \text{ and } b_3<b_1\text{ and } c_3<\frac{-a_1 c_1-a_3 c_1+b_2^2-2 b_4 b_2+b_4^2}{a_1+a_3}
\\\nonumber&& 
\textbf{or}  
\\\nonumber&&
a_4>b_1\text{ and } b_2>\frac{-a_5 v_{SM}^2+4 b_3 v_1^2-2 \mu _d^2}{2 v_1^2}\text{ and } c_2>-c_1\text{ and } v_1>b_1\text{ and } v_{SM}>b_1 \text{ and } b_3\geq b_1\text{ and } b_4\geq b_2
\\\nonumber&&
\text{ and } a_6<\frac{a_5 v_{SM}^2+a_7 v_{SM}^2+2 b_2 v_1^2-4 b_3 v_1^2+2 \mu _d^2}{2 v_{SM}^2}
\text{ and } a_3<-a_1
\text{ and } c_3<\frac{-a_1 c_1-a_3 c_1+b_2^2-2 b_4 b_2+b_4^2}{a_1+a_3}
\end{eqnarray*}

\begin{center}
\fbox{{\bf { \textbf{Case 3.} $A_{ij}\leq0$, $A_{kl} \leq0$ and all other entries positive.}}}
\end{center}
\begin{center}
\fbox{{If $a5+a6+2 a7\leq 0$ and $2 a1+2 a3\leq 0$, then:}}
\end{center}
There is no solution.

\begin{center}
\fbox{{If $a5+a6+2 a7\leq 0$ and $b2+b4\leq 0$, then:}}
\end{center}
 
\begin{eqnarray*}
\nonumber
&& a_1>-a_3\text{ and } a_4>\frac{a_5^2+2 a_6 a_5+4 a_7 a_5+a_6^2+4 a_7^2+4 a_6 a_7}{a_1+a_3}
\text{ and } a_7>\frac{-3 a_5 v_{SM}^2-2 b_2 v_1^2-4 b_3 v_1^2-2 \mu _d^2}{5 v_{SM}^2}
\\\nonumber&& 
\text{ and } c_1>\frac{-a_1 c_3-a_3 c_3+b_2^2+2 b_4 b_2+b_4^2}{a_1+a_3} 
\text{ and } c_2>-c_1\text{ and } \mu _d^2>\frac{1}{2} \left(-a_5 v_{SM}^2-2 b_2 v_1^2-4 b_3 v_1^2\right)\text{ and } v_1>b_1
\\\nonumber&& 
\text{ and } v_{SM}>b_1\text{ and } a_6\leq -a_5-2 a_7\text{ and } b_3\leq b_1\text{ and } b_4\leq -b_2
\\\nonumber&& 
\textbf{or}  
\\\nonumber&&
a_1>-a_3\text{ and } a_4>\frac{a_5^2+2 a_6 a_5+4 a_7 a_5+a_6^2+4 a_7^2+4 a_6 a_7}{a_1+a_3}
\text{ and } a_7>\frac{-3 a_5 v_{SM}^2-2 b_2 v_1^2+4 b_3 v_1^2-2 \mu _d^2}{5 v_{SM}^2}\text{ and } b_3>b_1
\\\nonumber&&
\text{ and } c_1>\frac{-a_1 c_3-a_3 c_3+b_2^2+2 b_4 b_2+b_4^2}{a_1+a_3}\text{ and } c_2>-c_1
\text{ and } \mu _d^2>\frac{1}{2} \left(-a_5 v_{SM}^2-2 b_2 v_1^2+4 b_3 v_1^2\right)
\\\nonumber&&
\text{ and } v_1>b_1\text{ and } v_{SM}>b_1\text{ and } a_6\leq -a_5-2 a_7\text{ and } b_4\leq -b_2
\\\nonumber&& 
\textbf{or}  
\\\nonumber&&
a_1>-a_3\text{ and } a_4>\frac{a_5^2+2 a_6 a_5+4 a_7 a_5+a_6^2+4 a_7^2+4 a_6 a_7}{a_1+a_3}
\text{ and } b_3>b_1\text{ and } c_1>\frac{-a_1 c_3-a_3 c_3+b_2^2+2 b_4 b_2+b_4^2}{a_1+a_3}
\\\nonumber&&
\text{ and } c_2>-c_1\text{ and } \mu _d^2>\frac{1}{2} \left(-a_5 v_{SM}^2-2 b_2 v_1^2+4 b_3 v_1^2\right)\text{ and } v_1>b_1\text{ and } b_4\leq -b_2
\\\nonumber&&
\text{ and } v_{SM}>b_1\text{ and } a_6<\frac{a_5 v_{SM}^2+a_7 v_{SM}^2+2 b_2 v_1^2-4 b_3 v_1^2+2 \mu _d^2}{2 v_{SM}^2}
\text{ and } a_7\leq \frac{-3 a_5 v_{SM}^2-2 b_2 v_1^2+4 b_3 v_1^2-2 \mu _d^2}{5 v_{SM}^2}
\end{eqnarray*}
 
\begin{eqnarray*}
\\\nonumber&& 
\textbf{or}  
\\\nonumber&&
 a_1>-a_3\text{ and } a_4>\frac{a_5^2+2 a_6 a_5+4 a_7 a_5+a_6^2+4 a_7^2+4 a_6 a_7}{a_1+a_3}
 \text{ and } c_1>\frac{-a_1 c_3-a_3 c_3+b_2^2+2 b_4 b_2+b_4^2}{a_1+a_3}\text{ and } c_2>-c_1
 \\\nonumber&&
 \text{ and } \mu _d^2>\frac{1}{2} \left(-a_5 v_{SM}^2-2 b_2 v_1^2-4 b_3 v_1^2\right)\text{ and } v_1>b_1\text{ and } v_{SM}>b_1
 \text{ and } a_6<\frac{a_5 v_{SM}^2+a_7 v_{SM}^2+2 b_2 v_1^2+4 b_3 v_1^2+2 \mu _d^2}{2 v_{SM}^2}
 \\\nonumber&&
 \text{ and } a_7\leq \frac{-3 a_5 v_{SM}^2-2 b_2 v_1^2-4 b_3 v_1^2-2 \mu _d^2}{5 v_{SM}^2}\text{ and } b_3\leq b_1\text{ and } b_4\leq -b_2
 \end{eqnarray*}

\begin{center}
\fbox{{ If $a5+a6+2 a7\leq 0$ and $b2-b4\leq 0$, then:}}
\end{center}

\begin{eqnarray*}
\\\nonumber
&& a_1>-a_3\text{ and } a_4>\frac{a_5^2+2 a_6 a_5+4 a_7 a_5+a_6^2+4 a_7^2+4 a_6 a_7}{a_1+a_3} \text{ and } v_{SM}>b_1\text{ and } b_4\geq b_2\text{ and } a_6\leq -a_5-2 a_7\\\nonumber&& 
\text{ and } a_7>\frac{-3 a_5 v_{SM}^2-2 b_2 v_1^2-4 b_3 v_1^2-2 \mu _d^2}{5 v_{SM}^2}\text{ and } c_1>\frac{-a_1 c_3-a_3 c_3+b_2^2-2 b_4 b_2+b_4^2}{a_1+a_3} 
\text{ and } c_2>-c_1
\\\nonumber&& 
\\\nonumber&& 
\text{ and } b_3\leq b_1
\text{ and } \mu _d^2>\frac{1}{2} \left(-a_5 v_{SM}^2-2 b_2 v_1^2-4 b_3 v_1^2\right)\text{ and } v_1>b_1 
\textbf{or}  
\\\nonumber&&
 a_1>-a_3\text{ and } a_4>\frac{a_5^2+2 a_6 a_5+4 a_7 a_5+a_6^2+4 a_7^2+4 a_6 a_7}{a_1+a_3}
 \text{ and } a_7>\frac{-3 a_5 v_{SM}^2-2 b_2 v_1^2+4 b_3 v_1^2-2 \mu _d^2}{5 v_{SM}^2}\text{ and } b_3>b_1
 \\\nonumber&& 
 \text{ and } c_1>\frac{-a_1 c_3-a_3 c_3+b_2^2-2 b_4 b_2+b_4^2}{a_1+a_3}\text{ and } c_2>-c_1 
 \text{ and } \mu _d^2>\frac{1}{2} \left(-a_5 v_{SM}^2-2 b_2 v_1^2+4 b_3 v_1^2\right)
 \\\nonumber&& 
 \text{ and } v_1>b_1\text{ and } v_{SM}>b_1\text{ and } b_4\geq b_2\text{ and } a_6\leq -a_5-2 a_7
\\\nonumber&& 
\textbf{or}  
\\\nonumber&&
 a_1>-a_3\text{ and } a_4>\frac{a_5^2+2 a_6 a_5+4 a_7 a_5+a_6^2+4 a_7^2+4 a_6 a_7}{a_1+a_3}\text{ and } b_3>b_1 \text{ and } v_1>b_1\text{ and } v_{SM}>b_1\text{ and } b_4\geq b_2
 \\\nonumber&&
 \text{ and } c_1>\frac{-a_1 c_3-a_3 c_3+b_2^2-2 b_4 b_2+b_4^2}{a_1+a_3}\text{ and } c_2>-c_1\text{ and } \mu _d^2>\frac{1}{2} \left(-a_5 v_{SM}^2-2 b_2 v_1^2+4 b_3 v_1^2\right)
 \\\nonumber&&
 \text{ and } a_6<\frac{a_5 v_{SM}^2+a_7 v_{SM}^2+2 b_2 v_1^2-4 b_3 v_1^2+2 \mu _d^2}{2 v_{SM}^2}
 \text{ and } a_7\leq \frac{-3 a_5 v_{SM}^2-2 b_2 v_1^2+4 b_3 v_1^2-2 \mu _d^2}{5 v_{SM}^2}
 \\\nonumber&& 
\textbf{or}  
\\\nonumber&&
a_1>-a_3\text{ and } a_4>\frac{a_5^2+2 a_6 a_5+4 a_7 a_5+a_6^2+4 a_7^2+4 a_6 a_7}{a_1+a_3}
\text{ and } a_7\leq \frac{-3 a_5 v_{SM}^2-2 b_2 v_1^2-4 b_3 v_1^2-2 \mu _d^2}{5 v_{SM}^2}\text{ and } b_3\leq b_1
\\\nonumber&&
\text{ and } c_1>\frac{-a_1 c_3-a_3 c_3+b_2^2-2 b_4 b_2+b_4^2}{a_1+a_3}\text{ and } c_2>-c_1
\text{ and } \mu _d^2>\frac{1}{2} \left(-a_5 v_{SM}^2-2 b_2 v_1^2-4 b_3 v_1^2\right)\text{ and } v_1>b_1
\\\nonumber&&
\text{ and } b_4\geq b_2\text{ and } a_6<\frac{a_5 v_{SM}^2+a_7 v_{SM}^2+2 b_2 v_1^2+4 b_3 v_1^2+2 \mu _d^2}{2 v_{SM}^2} \text{ and } v_{SM}>b_1
\\\nonumber&&
\end{eqnarray*}

\begin{center}
\fbox{{ If $a5+a6+2 a7\leq 0$ and $b2+b4\leq 0$, then: }}
\end{center}

\begin{eqnarray*}
\nonumber
&& a_1>-a_3\text{ and } a_4>\frac{a_5^2+2 a_6 a_5+4 a_7 a_5+a_6^2+4 a_7^2+4 a_6 a_7}{a_1+a_3} \text{ and } c_2>-c_1\text{ and } \mu _d^2>\frac{1}{2} \left(-a_5 v_{SM}^2-2 b_2 v_1^2-4 b_3 v_1^2\right)
\\\nonumber&& 
\text{ and } a_7>\frac{-3 a_5 v_{SM}^2-2 b_2 v_1^2-4 b_3 v_1^2-2 \mu _d^2}{5 v_{SM}^2}\text{ and } c_1>\frac{-a_1 c_3-a_3 c_3+b_2^2+2 b_4 b_2+b_4^2}{a_1+a_3} \text{ and } v_1>b_1
\\\nonumber&& 
\text{ and } v_{SM}>b_1\text{ and } a_6\leq -a_5-2 a_7\text{ and } b_3\leq b_1\text{ and } b_4\leq -b_2
\end{eqnarray*} 

\begin{eqnarray*}
\\\nonumber&& 
\textbf{or}  
\\\nonumber&&
 a_1>-a_3\text{ and } a_4>\frac{a_5^2+2 a_6 a_5+4 a_7 a_5+a_6^2+4 a_7^2+4 a_6 a_7}{a_1+a_3}
 \text{ and } a_7>\frac{-3 a_5 v_{SM}^2-2 b_2 v_1^2+4 b_3 v_1^2-2 \mu _d^2}{5 v_{SM}^2}\text{ and } b_3>b_1
 \\\nonumber&&
 \text{ and } c_1>\frac{-a_1 c_3-a_3 c_3+b_2^2+2 b_4 b_2+b_4^2}{a_1+a_3}\text{ and } c_2>-c_1\text{ and } \mu _d^2>\frac{1}{2} \left(-a_5 v_{SM}^2-2 b_2 v_1^2+4 b_3 v_1^2\right)
 \\\nonumber&&
 \text{ and } v_1>b_1\text{ and } v_{SM}>b_1\text{ and } a_6\leq -a_5-2 a_7\text{ and } b_4\leq -b_2 
\\\nonumber&& 
\textbf{or}  
\\\nonumber&&
a_1>-a_3\text{ and } a_4>\frac{a_5^2+2 a_6 a_5+4 a_7 a_5+a_6^2+4 a_7^2+4 a_6 a_7}{a_1+a_3}
\text{ and } b_3>b_1\text{ and } c_1>\frac{-a_1 c_3-a_3 c_3+b_2^2+2 b_4 b_2+b_4^2}{a_1+a_3}
\\\nonumber&&
\text{ and } \mu _d^2>\frac{1}{2} \left(-a_5 v_{SM}^2-2 b_2 v_1^2+4 b_3 v_1^2\right)\text{ and } v_1>b_1\text{ and } v_{SM}>b_1 \text{ and } c_2>-c_1\text{ and } b_4\leq -b_2
\\\nonumber&&
\text{ and } a_6<\frac{a_5 v_{SM}^2+a_7 v_{SM}^2+2 b_2 v_1^2-4 b_3 v_1^2+2 \mu _d^2}{2 v_{SM}^2}
\text{ and } a_7\leq \frac{-3 a_5 v_{SM}^2-2 b_2 v_1^2+4 b_3 v_1^2-2 \mu _d^2}{5 v_{SM}^2}
\\\nonumber&& 
\textbf{or}  
\\\nonumber&&
a_1>-a_3\text{ and } a_4>\frac{a_5^2+2 a_6 a_5+4 a_7 a_5+a_6^2+4 a_7^2+4 a_6 a_7}{a_1+a_3}
\text{ and } c_1>\frac{-a_1 c_3-a_3 c_3+b_2^2+2 b_4 b_2+b_4^2}{a_1+a_3}\text{ and } c_2>-c_1
\\\nonumber&&
\text{ and } \mu _d^2>\frac{1}{2} \left(-a_5 v_{SM}^2-2 b_2 v_1^2-4 b_3 v_1^2\right)\text{ and } v_1>b_1\text{ and } v_{SM}>b_1\text{ and } b_4\leq -b_2
\\\nonumber&&
\text{ and } a_6<\frac{a_5 v_{SM}^2+a_7 v_{SM}^2+2 b_2 v_1^2+4 b_3 v_1^2+2 \mu _d^2}{2 v_{SM}^2}
\text{ and } a_7\leq \frac{-3 a_5 v_{SM}^2-2 b_2 v_1^2-4 b_3 v_1^2-2 \mu _d^2}{5 v_{SM}^2}\text{ and } b_3\leq b_1
\end{eqnarray*}

\begin{center}
\fbox{{If $a5+a6+2 a7\leq 0$ and $b2-b4\leq 0$, then:}}
\end{center}

\begin{eqnarray*}
\nonumber
&&a_1>-a_3\text{ and } a_4>\frac{a_5^2+2 a_6 a_5+4 a_7 a_5+a_6^2+4 a_7^2+4 a_6 a_7}{a_1+a_3} \text{ and } b_4\geq b_2\text{ and } a_6\leq -a_5-2 a_7\text{ and } b_3\leq b_1 \text{ and } c_2>-c_1
\\\nonumber&& 
\text{ and } a_7>\frac{-3 a_5 v_{SM}^2-2 b_2 v_1^2-4 b_3 v_1^2-2 \mu _d^2}{5 v_{SM}^2}\text{ and } c_1>\frac{-a_1 c_3-a_3 c_3+b_2^2-2 b_4 b_2+b_4^2}{a_1+a_3} \text{ and } v_1>b_1\text{ and } v_{SM}>b_1
\\\nonumber&& 
\text{ and } \mu _d^2>\frac{1}{2} \left(-a_5 v_{SM}^2-2 b_2 v_1^2-4 b_3 v_1^2\right)
\\\nonumber&& 
\textbf{or}  
\\\nonumber&&
a_1>-a_3\text{ and } a_4>\frac{a_5^2+2 a_6 a_5+4 a_7 a_5+a_6^2+4 a_7^2+4 a_6 a_7}{a_1+a_3} 
\text{ and } a_7>\frac{-3 a_5 v_{SM}^2-2 b_2 v_1^2+4 b_3 v_1^2-2 \mu _d^2}{5 v_{SM}^2}\text{ and } b_3>b_1
\\\nonumber&& 
\text{ and } c_1>\frac{-a_1 c_3-a_3 c_3+b_2^2-2 b_4 b_2+b_4^2}{a_1+a_3}\text{ and } c_2>-c_1 \text{ and } b_4\geq b_2\text{ and } a_6\leq -a_5-2 a_7
\\\nonumber&& 
\text{ and } \mu _d^2>\frac{1}{2} \left(-a_5 v_{SM}^2-2 b_2 v_1^2+4 b_3 v_1^2\right)\text{ and } v_1>b_1\text{ and } v_{SM}>b_1
\\\nonumber&& 
\textbf{or}  
\\\nonumber&&
a_1>-a_3\text{ and } a_4>\frac{a_5^2+2 a_6 a_5+4 a_7 a_5+a_6^2+4 a_7^2+4 a_6 a_7}{a_1+a_3}
\text{ and } b_3>b_1\text{ and } c_1>\frac{-a_1 c_3-a_3 c_3+b_2^2-2 b_4 b_2+b_4^2}{a_1+a_3}
\\\nonumber&& 
\text{ and } \mu _d^2>\frac{1}{2} \left(-a_5 v_{SM}^2-2 b_2 v_1^2+4 b_3 v_1^2\right)\text{ and } v_1>b_1\text{ and } v_{SM}>b_1 \text{ and } c_2>-c_1
\\\nonumber&& 
\text{ and } b_4\geq b_2\text{ and } a_6<\frac{a_5 v_{SM}^2+a_7 v_{SM}^2+2 b_2 v_1^2-4 b_3 v_1^2+2 \mu _d^2}{2 v_{SM}^2}
\text{ and } a_7\leq \frac{-3 a_5 v_{SM}^2-2 b_2 v_1^2+4 b_3 v_1^2-2 \mu _d^2}{5 v_{SM}^2}
\end{eqnarray*} 

\begin{eqnarray*}
\\\nonumber&& 
\textbf{or}  
\\\nonumber&&
 a_1>-a_3\text{ and } a_4>\frac{a_5^2+2 a_6 a_5+4 a_7 a_5+a_6^2+4 a_7^2+4 a_6 a_7}{a_1+a_3}
 \text{ and } c_1>\frac{-a_1 c_3-a_3 c_3+b_2^2-2 b_4 b_2+b_4^2}{a_1+a_3}\text{ and } c_2>-c_1
 \\\nonumber&& 
 \text{ and } \mu _d^2>\frac{1}{2} \left(-a_5 v_{SM}^2-2 b_2 v_1^2-4 b_3 v_1^2\right)\text{ and } v_1>b_1\text{ and } v_{SM}>b_1
 \\\nonumber&& 
 \text{ and } b_4\geq b_2\text{ and } a_6<\frac{a_5 v_{SM}^2+a_7 v_{SM}^2+2 b_2 v_1^2+4 b_3 v_1^2+2 \mu _d^2}{2 v_{SM}^2}
 \text{ and } a_7\leq \frac{-3 a_5 v_{SM}^2-2 b_2 v_1^2-4 b_3 v_1^2-2 \mu _d^2}{5 v_{SM}^2}\text{ and } b_3\leq b_1
\end{eqnarray*}

\begin{center}
\fbox{{If $2 a1+2 a3\leq 0$ and $0\leq 0$, then:}}
\end{center}
There is no solution

\begin{center}
\fbox{{IIf $2 a1+2 a3\leq 0$ and $b2-b4\leq 0$, then:}}
\end{center}
There is no solution

\newpage
\begin{center}
\fbox{{\bf { \textbf{Case 4.} $A_{ij}\leq0$, $A_{ik} \leq0$ and all other entries positive.
}}}
\end{center}
\begin{center}
\fbox{{ If $a5+a6+2 a7\leq 0$ and $a5+a6+2 a7\leq 0$, then:}}
\end{center}

\begin{eqnarray*}
\nonumber
&& a_1>\frac{-a_5^2-2 a_6 a_5-4 a_7 a_5-a_6^2-4 a_7^2-a_3 a_4-4 a_6 a_7}{a_4}\text{ and } a_4>b_1
\text{ and } a_7>\frac{-3 a_5 v_{SM}^2-2 b_2 v_1^2-4 b_3 v_1^2-2 \mu _d^2}{5 v_{SM}^2}
\\\nonumber&& 
\text{ and } \mu _d^2>\frac{1}{2} \left(-a_5 v_{SM}^2-2 b_2 v_1^2-4 b_3 v_1^2\right)\text{ and } v_1>b_1\text{ and } v_{SM}>b_1 \text{ and } c_1>-c_2
\text{ and } a_6\leq -a_5-2 a_7\text{ and } b_3\leq b_1
\\\nonumber&& 
\textbf{or}  
\\\nonumber&&
 a_1>\frac{-a_5^2-2 a_6 a_5-4 a_7 a_5-a_6^2-4 a_7^2-a_3 a_4-4 a_6 a_7}{a_4}\text{ and } a_4>b_1
 \text{ and } a_7>\frac{-3 a_5 v_{SM}^2-2 b_2 v_1^2+4 b_3 v_1^2-2 \mu _d^2}{5 v_{SM}^2}
 \\\nonumber&& 
 \text{ and } c_1>-c_2\text{ and } \mu _d^2>\frac{1}{2} \left(-a_5 v_{SM}^2-2 b_2 v_1^2+4 b_3 v_1^2\right)\text{ and } v_1>b_1 \text{ and } b_3>b_1
 \text{ and } v_{SM}>b_1\text{ and } a_6\leq -a_5-2 a_7
\\\nonumber&& 
\textbf{or}  
\\\nonumber&&
a_1>\frac{-a_5^2-2 a_6 a_5-4 a_7 a_5-a_6^2-4 a_7^2-a_3 a_4-4 a_6 a_7}{a_4}\text{ and } a_4>b_1
\text{ and } b_3>b_1\text{ and } c_1>-c_2 \text{ and } v_1>b_1\text{ and } v_{SM}>b_1
\\\nonumber&& 
\text{ and } \mu _d^2>\frac{1}{2} \left(-a_5 v_{SM}^2-2 b_2 v_1^2+4 b_3 v_1^2\right)
\text{ and } a_6<\frac{a_5 v_{SM}^2+a_7 v_{SM}^2+2 b_2 v_1^2-4 b_3 v_1^2+2 \mu _d^2}{2 v_{SM}^2}
\\\nonumber&& 
\text{ and } a_7\leq \frac{-3 a_5 v_{SM}^2-2 b_2 v_1^2+4 b_3 v_1^2-2 \mu _d^2}{5 v_{SM}^2}
\\\nonumber&& 
\textbf{or}  
\\\nonumber&&
 a_1>\frac{-a_5^2-2 a_6 a_5-4 a_7 a_5-a_6^2-4 a_7^2-a_3 a_4-4 a_6 a_7}{a_4}\text{ and } a_4>b_1\text{ and } c_1>-c_2
 \text{ and } \mu _d^2>\frac{1}{2} \left(-a_5 v_{SM}^2-2 b_2 v_1^2-4 b_3 v_1^2\right)
 \\\nonumber&& 
 \text{ and } a_6<\frac{a_5 v_{SM}^2+a_7 v_{SM}^2+2 b_2 v_1^2+4 b_3 v_1^2+2 \mu _d^2}{2 v_{SM}^2}
 \text{ and } a_7\leq \frac{-3 a_5 v_{SM}^2-2 b_2 v_1^2-4 b_3 v_1^2-2 \mu _d^2}{5 v_{SM}^2}\text{ and } b_3\leq b_1 
 \\\nonumber&&  
 \text{ and } v_1>b_1\text{ and } v_{SM}>b_1
\end{eqnarray*}

\begin{center}
\fbox{{ If $a5+a6+2 a7\leq 0$ and $0\leq 0$, then: }}
\end{center}

\begin{eqnarray*}
\nonumber
&& -c_3=c_1\text{ and } a_4>b_1\text{ and } a_7>\frac{-3 a_5 v_{SM}^2-2 b_2 v_1^2-4 b_3 v_1^2-2 \mu _d^2}{5 v_{SM}^2}
\text{ and } b_4>-b_2\text{ and } c_2>-c_1
\\\nonumber&& 
\text{ and } v_1>b_1\text{ and } v_{SM}>b_1\text{ and } a_6\leq -a_5-2 a_7\text{ and } b_3\leq b_1 \text{ and } \mu _d^2>\frac{1}{2} \left(-a_5 v_{SM}^2-2 b_2 v_1^2-4 b_3 v_1^2\right)
\\\nonumber&& 
\textbf{or}  
\\\nonumber&&
 -c_3=c_1\text{ and } a_4>b_1\text{ and } a_7>\frac{-3 a_5 v_{SM}^2-2 b_2 v_1^2+4 b_3 v_1^2-2 \mu _d^2}{5 v_{SM}^2} 
\text{ and } b_3>b_1\text{ and } b_4>-b_2\text{ and } c_2>-c_1
\\\nonumber&&
\text{ and } \mu _d^2>\frac{1}{2} \left(-a_5 v_{SM}^2-2 b_2 v_1^2+4 b_3 v_1^2\right)\text{ and } v_1>b_1\text{ and } v_{SM}>b_1\text{ and } a_6\leq -a_5-2 a_7
\\\nonumber&& 
\textbf{or}  
\\\nonumber&&
-c_3=c_1\text{ and } a_4>b_1\text{ and } a_7>\frac{-3 a_5 v_{SM}^2-2 b_2 v_1^2-4 b_3 v_1^2-2 \mu _d^2}{5 v_{SM}^2}\text{ and } b_4>-b_2
\text{ and } c_2>-c_1\text{ and } v_1>b_1
\\\nonumber&&
\text{ and } v_{SM}>b_1\text{ and } a_6\leq -a_5-2 a_7\text{ and } b_3\leq b_1 \text{ and } \mu _d^2>\frac{1}{2} \left(-a_5 v_{SM}^2-2 b_2 v_1^2-4 b_3 v_1^2\right)
\\\nonumber&& 
\textbf{or}  
\\\nonumber&&
-c_3=c_1\text{ and } a_4>b_1\text{ and } a_7>\frac{-3 a_5 v_{SM}^2-2 b_2 v_1^2+4 b_3 v_1^2-2 \mu _d^2}{5 v_{SM}^2}
\text{ and } b_3>b_1\text{ and } b_4>-b_2\text{ and } c_2>-c_1
\\\nonumber&&
 \text{ and } \mu _d^2>\frac{1}{2} \left(-a_5 v_{SM}^2-2 b_2 v_1^2+4 b_3 v_1^2\right) \text{ and } v_1>b_1\text{ and } v_{SM}>b_1\text{ and } a_6\leq -a_5-2 a_7
\end{eqnarray*}

\begin{eqnarray*}
\\\nonumber&& 
\textbf{or}  
\\\nonumber&&
 -c_3=c_1\text{ and } a_4>b_1\text{ and } b_3>b_1\text{ and } b_4>-b_2\text{ and } c_2>-c_1
 \text{ and } \mu _d^2>\frac{1}{2} \left(-a_5 v_{SM}^2-2 b_2 v_1^2+4 b_3 v_1^2\right)\text{ and } v_1>b_1
 \\\nonumber&&
 \text{ and } v_{SM}>b_1\text{ and } a_6<\frac{a_5 v_{SM}^2+a_7 v_{SM}^2+2 b_2 v_1^2-4 b_3 v_1^2+2 \mu _d^2}{2 v_{SM}^2}
 \text{ and } a_7\leq \frac{-3 a_5 v_{SM}^2-2 b_2 v_1^2+4 b_3 v_1^2-2 \mu _d^2}{5 v_{SM}^2}
\\\nonumber&& 
\textbf{or}  
\\\nonumber&&
 -c_3=c_1\text{ and } a_4>b_1\text{ and } b_4>-b_2\text{ and } c_2>-c_1
 \text{ and } \mu _d^2>\frac{1}{2} \left(-a_5 v_{SM}^2-2 b_2 v_1^2-4 b_3 v_1^2\right)\text{ and } v_1>b_1\text{ and } v_{SM}>b_1
 \\\nonumber&&
 \text{ and } a_6<\frac{a_5 v_{SM}^2+a_7 v_{SM}^2+2 b_2 v_1^2+4 b_3 v_1^2+2 \mu _d^2}{2 v_{SM}^2}
 \text{ and } a_7\leq \frac{-3 a_5 v_{SM}^2-2 b_2 v_1^2-4 b_3 v_1^2-2 \mu _d^2}{5 v_{SM}^2}\text{ and } b_3\leq b_1
 \\\nonumber&& 
\textbf{or}  
\\\nonumber&&
 a_1>\frac{a_4 b_2^2+a_4 b_4^2+2 a_4 b_2 b_4+a_5^2 c_1+a_5^2 c_3+2 a_6 a_5 c_1+4 a_7 a_5 c_1}{a_4 c_1+a_4 c_3}
\\\nonumber&+&
\frac{2 a_6 a_5 c_3+4 a_7 a_5 c_3+a_6^2 c_1+4 a_7^2 c_1-a_3 a_4 c_1+4 a_6 a_7 c_1+a_6^2 c_3+4 a_7^2 c_3-a_3 a_4 c_3+4 a_6 a_7 c_3}{a_4 c_1+a_4 c_3}
\\\nonumber&&
\text{ and } a_4>b_1\text{ and } a_7>\frac{-3 a_5 v_{SM}^2-2 b_2 v_1^2-4 b_3 v_1^2-2 \mu _d^2}{5 v_{SM}^2}
\text{ and } c_1>-c_3\text{ and } c_2>-c_1
\\\nonumber&&
\text{ and } \mu _d^2>\frac{1}{2} \left(-a_5 v_{SM}^2-2 b_2 v_1^2-4 b_3 v_1^2\right)\text{ and } v_1>b_1
\text{ and } v_{SM}>b_1\text{ and } a_6\leq -a_5-2 a_7\text{ and } b_3\leq b_1\text{ and } b_4\leq -b_2
\\\nonumber&&
\textbf{or}  
\\\nonumber&&
 a_1>\frac{a_4 b_2^2+a_4 b_4^2+2 a_4 b_2 b_4+a_5^2 c_1+a_5^2 c_3+2 a_6 a_5 c_1+4 a_7 a_5 c_1}{a_4 c_1+a_4 c_3}
\\\nonumber&+&
\frac{2 a_6 a_5 c_3+4 a_7 a_5 c_3+a_6^2 c_1+4 a_7^2 c_1-a_3 a_4 c_1+4 a_6 a_7 c_1+a_6^2 c_3+4 a_7^2 c_3-a_3 a_4 c_3+4 a_6 a_7 c_3}{a_4 c_1+a_4 c_3}
 \\\nonumber&&
 \text{ and } a_4>b_1\text{ and } a_7>\frac{-3 a_5 v_{SM}^2-2 b_2 v_1^2+4 b_3 v_1^2-2 \mu _d^2}{5 v_{SM}^2}\text{ and } b_3>b_1\text{ and } c_1>-c_3
 \\\nonumber&&
 \text{ and } c_2>-c_1\text{ and } \mu _d^2>\frac{1}{2} \left(-a_5 v_{SM}^2-2 b_2 v_1^2+4 b_3 v_1^2\right)
 \text{ and } v_1>b_1\text{ and } v_{SM}>b_1\text{ and } a_6\leq -a_5-2 a_7\text{ and } b_4\leq -b_2
\\\nonumber&& 
\textbf{or}  
\\\nonumber&&
 a_1>\frac{a_4 b_2^2+a_4 b_4^2+2 a_4 b_2 b_4+a_5^2 c_1+a_5^2 c_3+2 a_6 a_5 c_1+4 a_7 a_5 c_1}{a_4 c_1+a_4 c_3}
 \\\nonumber&+&
 \frac{2 a_6 a_5 c_3+4 a_7 a_5 c_3+a_6^2 c_1+4 a_7^2 c_1-a_3 a_4 c_1+4 a_6 a_7 c_1+a_6^2 c_3+4 a_7^2 c_3-a_3 a_4 c_3+4 a_6 a_7 c_3}{a_4 c_1+a_4 c_3}
 \\\nonumber&&
 \text{ and } a_4>b_1\text{ and } b_3>b_1\text{ and } c_1>-c_3\text{ and } c_2>-c_1\text{ and } \mu _d^2>\frac{1}{2} \left(-a_5 v_{SM}^2-2 b_2 v_1^2+4 b_3 v_1^2\right)\text{ and } v_1>b_1
 \\\nonumber&&
 \text{ and } v_{SM}>b_1
  \\\nonumber&&
 \text{ and } a_6<\frac{a_5 v_{SM}^2+a_7 v_{SM}^2+2 b_2 v_1^2-4 b_3 v_1^2+2 \mu _d^2}{2 v_{SM}^2}
 \text{ and } a_7\leq \frac{-3 a_5 v_{SM}^2-2 b_2 v_1^2+4 b_3 v_1^2-2 \mu _d^2}{5 v_{SM}^2}\text{ and } b_4\leq -b_2
\\\nonumber&& 
\textbf{or}  
\\\nonumber&&
a_1>\frac{a_4 b_2^2+a_4 b_4^2+2 a_4 b_2 b_4+a_5^2 c_1+a_5^2 c_3+2 a_6 a_5 c_1+4 a_7 a_5 c_1+2 a_6 a_5 c_3}{a_4 c_1+a_4 c_3}
 \\\nonumber&+&
 \frac{4 a_7 a_5 c_3+a_6^2 c_1+4 a_7^2 c_1-a_3 a_4 c_1+4 a_6 a_7 c_1+a_6^2 c_3+4 a_7^2 c_3-a_3 a_4 c_3+4 a_6 a_7 c_3}{a_4 c_1+a_4 c_3} \text{ and } b_3\leq b_1\text{ and } b_4\leq -b_2
 \\\nonumber&&
 \text{ and } a_4>b_1\text{ and } c_1>-c_3\text{ and } c_2>-c_1\text{ and } \mu _d^2>\frac{1}{2} \left(-a_5 v_{SM}^2-2 b_2 v_1^2-4 b_3 v_1^2\right)
 \text{ and } v_1>b_1\text{ and } v_{SM}>b_1 
  \\\nonumber&&
 \text{ and } a_6<\frac{a_5 v_{SM}^2+a_7 v_{SM}^2+2 b_2 v_1^2+4 b_3 v_1^2+2 \mu _d^2}{2 v_{SM}^2}
 \text{ and } a_7\leq \frac{-3 a_5 v_{SM}^2-2 b_2 v_1^2-4 b_3 v_1^2-2 \mu _d^2}{5 v_{SM}^2} \end{eqnarray*}

\begin{eqnarray*}
\\\nonumber&& 
\textbf{or}  
\\\nonumber&&
a_4>b_1\text{ and } a_7>\frac{-3 a_5 v_{SM}^2-2 b_2 v_1^2-4 b_3 v_1^2-2 \mu _d^2}{5 v_{SM}^2}\text{ and } b_4>-b_2\text{ and } c_1>-c_3 \text{ and } v_1>b_1\text{ and } v_{SM}>b_1
\\\nonumber&&
\text{ and } c_2>-c_1\text{ and } \mu _d^2>\frac{1}{2} \left(-a_5 v_{SM}^2-2 b_2 v_1^2-4 b_3 v_1^2\right)
\text{ and } a_1\geq \frac{a_5^2+2 a_6 a_5+4 a_7 a_5+a_6^2+4 a_7^2-a_3 a_4+4 a_6 a_7}{a_4}
\\\nonumber&&
\text{ and } a_6\leq -a_5-2 a_7\text{ and } b_3\leq b_1
\\\nonumber&& 
\textbf{or}  
\\\nonumber&&
 a_4>b_1\text{ and } a_7>\frac{-3 a_5 v_{SM}^2-2 b_2 v_1^2-4 b_3 v_1^2-2 \mu _d^2}{5 v_{SM}^2}\text{ and } b_4>-b_2\text{ and } c_2>-c_1\text{ and } v_1>b_1\text{ and } v_{SM}>b_1
 \\\nonumber&&
 \text{ and } a_1\leq \frac{a_5^2+2 a_6 a_5+4 a_7 a_5+a_6^2+4 a_7^2-a_3 a_4+4 a_6 a_7}{a_4}\text{ and } a_6\leq -a_5-2 a_7\text{ and } b_3\leq b_1 
 \\\nonumber&& 
\text{ and } \mu _d^2>\frac{1}{2} \left(-a_5 v_{SM}^2-2 b_2 v_1^2-4 b_3 v_1^2\right) \text{ and } c_1<-c_3
\\\nonumber&& 
\textbf{or}  
\\\nonumber&&
 a_4>b_1\text{ and } a_7>\frac{-3 a_5 v_{SM}^2-2 b_2 v_1^2-4 b_3 v_1^2-2 \mu _d^2}{5 v_{SM}^2}\text{ and } c_2>-c_1
\text{ and } \mu _d^2>\frac{1}{2} \left(-a_5 v_{SM}^2-2 b_2 v_1^2-4 b_3 v_1^2\right)
\\\nonumber&& 
\text{ and }  v_1>b_1\text{ and } v_{SM}>b_1
\text{ and } a_1<\frac{a_4 b_2^2+a_4 b_4^2+2 a_4 b_2 b_4+a_5^2 c_1+a_5^2 c_3+2 a_6 a_5 c_1+4 a_7 a_5 c_1}{a_4 c_1+a_4 c_3}
\\\nonumber&+&
\frac{2 a_6 a_5 c_3+4 a_7 a_5 c_3+a_6^2 c_1+4 a_7^2 c_1-a_3 a_4 c_1+4 a_6 a_7 c_1+a_6^2 c_3+4 a_7^2 c_3-a_3 a_4 c_3+4 a_6 a_7 c_3}{a_4 c_1+a_4 c_3}
\\\nonumber&&
\text{ and } c_1<-c_3\text{ and } a_6\leq -a_5-2 a_7\text{ and } b_3\leq b_1\text{ and } b_4\leq -b_2
\\\nonumber&& 
\textbf{or}  
\\\nonumber&&
a_4>b_1\text{ and } a_7>\frac{-3 a_5 v_{SM}^2-2 b_2 v_1^2+4 b_3 v_1^2-2 \mu _d^2}{5 v_{SM}^2}\text{ and } b_3>b_1\text{ and } b_4>-b_2 \text{ and } v_1>b_1\text{ and } v_{SM}>b_1
\\\nonumber&&
\text{ and } c_1>-c_3\text{ and } c_2>-c_1\text{ and } \mu _d^2>\frac{1}{2} \left(-a_5 v_{SM}^2-2 b_2 v_1^2+4 b_3 v_1^2\right)
\\\nonumber&&
\text{ and } a_1\geq \frac{a_5^2+2 a_6 a_5+4 a_7 a_5+a_6^2+4 a_7^2-a_3 a_4+4 a_6 a_7}{a_4}\text{ and } a_6\leq -a_5-2 a_7
\\\nonumber&&
\textbf{or}  
\\\nonumber&&
a_4>b_1\text{ and } a_7>\frac{-3 a_5 v_{SM}^2-2 b_2 v_1^2+4 b_3 v_1^2-2 \mu _d^2}{5 v_{SM}^2}\text{ and } b_3>b_1
\text{ and } b_4>-b_2\text{ and } c_2>-c_1
\\\nonumber&&
\text{ and } \mu _d^2>\frac{1}{2} \left(-a_5 v_{SM}^2-2 b_2 v_1^2+4 b_3 v_1^2\right)
\text{ and } v_1>b_1\text{ and } v_{SM}>b_1\text{ and } c_1<-c_3
\\\nonumber&&
\text{ and } a_1\leq \frac{a_5^2+2 a_6 a_5+4 a_7 a_5+a_6^2+4 a_7^2-a_3 a_4+4 a_6 a_7}{a_4}\text{ and } a_6\leq -a_5-2 a_7
\\\nonumber&& 
\textbf{or}  
\\\nonumber&&
a_4>b_1\text{ and } a_7>\frac{-3 a_5 v_{SM}^2-2 b_2 v_1^2+4 b_3 v_1^2-2 \mu _d^2}{5 v_{SM}^2}\text{ and } b_3>b_1\text{ and } c_2>-c_1
\\\nonumber&&
\text{ and } \mu _d^2>\frac{1}{2} \left(-a_5 v_{SM}^2-2 b_2 v_1^2+4 b_3 v_1^2\right)\text{ and } v_1>b_1
\\\nonumber&&
\text{ and } v_{SM}>b_1\text{ and } a_1<\frac{a_4 b_2^2+a_4 b_4^2+2 a_4 b_2 b_4+a_5^2 c_1+a_5^2 c_3+2 a_6 a_5 c_1+4 a_7 a_5 c_1}{a_4 c_1+a_4 c_3}
\\\nonumber&+&
\frac{2 a_6 a_5 c_3+4 a_7 a_5 c_3+a_6^2 c_1+4 a_7^2 c_1-a_3 a_4 c_1+4 a_6 a_7 c_1+a_6^2 c_3+4 a_7^2 c_3-a_3 a_4 c_3+4 a_6 a_7 c_3}{a_4 c_1+a_4 c_3}
\\\nonumber&&
\text{ and } c_1<-c_3\text{ and } a_6\leq -a_5-2 a_7\text{ and } b_4\leq -b_2
\\\nonumber&& 
\textbf{or}  
\\\nonumber&&
a_4>b_1\text{ and } b_3>b_1\text{ and } b_4>-b_2\text{ and } c_1>-c_3\text{ and } c_2>-c_1
\text{ and } \mu _d^2>\frac{1}{2} \left(-a_5 v_{SM}^2-2 b_2 v_1^2+4 b_3 v_1^2\right)\text{ and } v_1>b_1
\\\nonumber&&
\text{ and } v_{SM}>b_1
\text{ and } a_1\geq \frac{a_5^2+2 a_6 a_5+4 a_7 a_5+a_6^2+4 a_7^2-a_3 a_4+4 a_6 a_7}{a_4}
\\\nonumber&&
\text{ and } a_6<\frac{a_5 v_{SM}^2+a_7 v_{SM}^2+2 b_2 v_1^2-4 b_3 v_1^2+2 \mu _d^2}{2 v_{SM}^2}
\text{ and } a_7\leq \frac{-3 a_5 v_{SM}^2-2 b_2 v_1^2+4 b_3 v_1^2-2 \mu _d^2}{5 v_{SM}^2}
\end{eqnarray*}

\begin{eqnarray*}
\\\nonumber&& 
\textbf{or}  
\\\nonumber&&
a_4>b_1\text{ and } b_3>b_1\text{ and } b_4>-b_2\text{ and } c_2>-c_1
\text{ and } \mu _d^2>\frac{1}{2} \left(-a_5 v_{SM}^2-2 b_2 v_1^2+4 b_3 v_1^2 \right)\text{ and } v_1>b_1\text{ and } v_{SM}>b_1
\\\nonumber&&
\text{ and } a_6<\frac{a_5 v_{SM}^2+a_7 v_{SM}^2+2 b_2 v_1^2-4 b_3 v_1^2+2 \mu _d^2}{2 v_{SM}^2}\text{ and } c_1<-c_3 \text{ and } a_7\leq \frac{-3 a_5 v_{SM}^2-2 b_2 v_1^2+4 b_3 v_1^2-2 \mu _d^2}{5 v_{SM}^2}
\\\nonumber&&
\text{ and } a_1\leq \frac{a_5^2+2 a_6 a_5+4 a_7 a_5+a_6^2+4 a_7^2-a_3 a_4+4 a_6 a_7}{a_4}
\\\nonumber&&
\textbf{or}  
\\\nonumber&&
a_4>b_1\text{ and } b_3>b_1\text{ and } c_2>-c_1\text{ and } \mu _d^2>\frac{1}{2} \left(-a_5 v_{SM}^2-2 b_2 v_1^2+4 b_3 v_1^2\right)
\text{ and } v_1>b_1\text{ and } v_{SM}>b_1
\\\nonumber&& 
\text{ and } a_1<\frac{a_4 b_2^2+a_4 b_4^2+2 a_4 b_2 b_4+a_5^2 c_1+a_5^2 c_3+2 a_6 a_5 c_1+4 a_7 a_5 c_1}{a_4 c_1+a_4 c_3}
\\\nonumber&+& 
\frac{2 a_6 a_5 c_3+4 a_7 a_5 c_3+a_6^2 c_1+4 a_7^2 c_1-a_3 a_4 c_1+4 a_6 a_7 c_1+a_6^2 c_3+4 a_7^2 c_3-a_3 a_4 c_3+4 a_6 a_7 c_3}{a_4 c_1+a_4 c_3} \text{ and } b_4\leq -b_2
\\\nonumber&& 
\text{ and } a_6<\frac{a_5 v_{SM}^2+a_7 v_{SM}^2+2 b_2 v_1^2-4 b_3 v_1^2+2 \mu _d^2}{2 v_{SM}^2}\text{ and } c_1<-c_3
\text{ and } a_7\leq \frac{-3 a_5 v_{SM}^2-2 b_2 v_1^2+4 b_3 v_1^2-2 \mu _d^2}{5 v_{SM}^2}
\\\nonumber&& 
\textbf{or}  
\\\nonumber&&
a_4>b_1\text{ and } b_4>-b_2\text{ and } c_1>-c_3\text{ and } c_2>-c_1
\text{ and } \mu _d^2>\frac{1}{2} \left(-a_5 v_{SM}^2-2 b_2 v_1^2-4 b_3 v_1^2\right)\text{ and } v_1>b_1\text{ and } v_{SM}>b_1
\\\nonumber&& 
\text{ and } a_1\geq \frac{a_5^2+2 a_6 a_5+4 a_7 a_5+a_6^2+4 a_7^2-a_3 a_4+4 a_6 a_7}{a_4}
\text{ and } a_6<\frac{a_5 v_{SM}^2+a_7 v_{SM}^2+2 b_2 v_1^2+4 b_3 v_1^2+2 \mu _d^2}{2 v_{SM}^2}
\\\nonumber&& 
\text{ and } a_7\leq \frac{-3 a_5 v_{SM}^2-2 b_2 v_1^2-4 b_3 v_1^2-2 \mu _d^2}{5 v_{SM}^2}\text{ and } b_3\leq b_1
\\\nonumber&& 
\textbf{or}  
\\\nonumber&&
a_4>b_1\text{ and } b_4>-b_2\text{ and } c_2>-c_1\text{ and } \mu _d^2>\frac{1}{2} \left(-a_5 v_{SM}^2-2 b_2 v_1^2-4 b_3 v_1^2\right)
\text{ and } v_1>b_1 \text{ and } b_3\leq b_1
\\\nonumber&&
\text{ and } v_{SM}>b_1\text{ and } a_6<\frac{a_5 v_{SM}^2+a_7 v_{SM}^2+2 b_2 v_1^2+4 b_3 v_1^2+2 \mu _d^2}{2 v_{SM}^2} \text{ and } a_7\leq \frac{-3 a_5 v_{SM}^2-2 b_2 v_1^2-4 b_3 v_1^2-2 \mu _d^2}{5 v_{SM}^2}
\\\nonumber&&
\text{ and } c_1<-c_3\text{ and } a_1\leq \frac{a_5^2+2 a_6 a_5+4 a_7 a_5+a_6^2+4 a_7^2-a_3 a_4+4 a_6 a_7}{a_4}
\\\nonumber&&
\\\nonumber&&
\textbf{or}  
\\\nonumber&&
 a_4>b_1\text{ and } c_2> -c_1\text{ and } \mu _d^2>\frac{1}{2} \left(-a_5 v_{SM}^2-2 b_2 v_1^2-4 b_3 v_1^2\right)
 \text{ and }v_1>b_1\text{ and } v_{SM}>b_1 \text{ and } b_3\leq b_1\text{ and } b_4\leq - b_2
 \\\nonumber&&
 \text{ and } a_1<\frac{a_4 b_2^2+a_4 b_4^2+2 a_4 b_2 b_4+a_5^2 c_1+a_5^2 c_3+2 a_6 a_5 c_1+4 a_7 a_5 c_1+2 a_6 a_5 c_3}{a_4 c_1+a_4 c_3}
 \\\nonumber&+&
 \frac{4 a_7 a_5 c_3+a_6^2 c_1+4 a_7^2 c_1-a_3 a_4 c_1+4 a_6 a_7 c_1+a_6^2 c_3+4 a_7^2 c_3-a_3 a_4 c_3+4 a_6 a_7 c_3}{a_4 c_1+a_4 c_3} \text{ and } c_1<-c_3
 \\\nonumber&&
 \text{ and } a_6<\frac{a_5 v_{SM}^2+a_7 v_{SM}^2+2 b_2 v_1^2+4 b_3 v_1^2+2 \mu _d^2}{2 v_{SM}^2}
 \text{ and } a_7\leq \frac{-3 a_5 v_{SM}^2-2 b_2 v_1^2-4 b_3 v_1^2-2 \mu _d^2}{5 v_{SM}^2}
\end{eqnarray*}

\begin{center}
\fbox{{ If $a5+a6+2 a7\leq 0$ and $0\leq 0$, then: }}
\end{center}

\begin{eqnarray*}
\nonumber
&& -c_3=c_1\text{ and } a_4>b_1\text{ and } a_7>\frac{-3 a_5 v_{SM}^2-2 b_2 v_1^2-4 b_3 v_1^2-2 \mu _d^2}{5 v_{SM}^2}
\text{ and } c_2>-c_1\text{ and } \mu _d^2>\frac{1}{2} \left(-a_5 v_{SM}^2-2 b_2 v_1^2-4 b_3 v_1^2\right)
\\\nonumber&& 
\text{ and } v_{SM}>b_1\text{ and } b_3<b_1\text{ and } b_4<b_2\text{ and } a_6\leq -a_5-2 a_7 \text{ and } v_1>b_1
\\\nonumber&& 
\textbf{or}  
\\\nonumber&&
 -c_3=c_1\text{ and } a_4>b_1\text{ and } a_7>\frac{-3 a_5 v_{SM}^2-2 b_2 v_1^2+4 b_3 v_1^2-2 \mu _d^2}{5 v_{SM}^2}
 \text{ and } c_2>-c_1\text{ and } \mu _d^2>\frac{1}{2} \left(-a_5 v_{SM}^2-2 b_2 v_1^2+4 b_3 v_1^2\right)
 \\\nonumber&& 
 \text{ and } v_{SM}>b_1\text{ and } b_3\geq b_1\text{ and } b_4<b_2\text{ and } a_6\leq -a_5-2 a_7 \text{ and } v_1>b_1
 \end{eqnarray*}
 
\begin{eqnarray*} 
\\\nonumber&& 
\textbf{or}  
\\\nonumber&&
-c_3=c_1\text{ and } a_4>b_1\text{ and } c_2>-c_1\text{ and } \mu _d^2>\frac{1}{2} \left(-a_5 v_{SM}^2-2 b_2 v_1^2-4 b_3 v_1^2\right)\text{ and } v_1>b_1
\text{ and } v_{SM}>b_1
\\\nonumber&& 
\text{ and } a_7<\frac{-3 a_5 v_{SM}^2-2 b_2 v_1^2+4 b_3 v_1^2-2 \mu _d^2}{5 v_{SM}^2}\text{ and } b_3<b_1\text{ and } b_4<b_2 \text{ and } a_6<\frac{a_5 v_{SM}^2+a_7 v_{SM}^2+2 b_2 v_1^2+4 b_3 v_1^2+2 \mu _d^2}{2 v_{SM}^2}
\\\nonumber&& 
\textbf{or}  
\\\nonumber&&
-c_3=c_1\text{ and } a_4>b_1\text{ and } c_2>-c_1\text{ and } \mu _d^2>\frac{1}{2} \left(-a_5 v_{SM}^2-2 b_2 v_1^2+4 b_3 v_1^2\right) \text{ and } a_7\leq \frac{-3 a_5 v_{SM}^2-2 b_2 v_1^2+4 b_3 v_1^2-2 \mu _d^2}{5 v_{SM}^2} 
\\\nonumber&&
\text{ and } a_6<\frac{a_5 v_{SM}^2+a_7 v_{SM}^2+2 b_2 v_1^2-4 b_3 v_1^2+2 \mu _d^2}{2 v_{SM}^2}\text{ and } b_4<b_2 \text{ and } v_1>b_1\text{ and } v_{SM}>b_1\text{ and } b_3\geq b_1
\\\nonumber&&
\textbf{or}  
\\\nonumber&&
 b_4=b_2\text{ and } a_4>b_1\text{ and } c_2>-c_1\text{ and } \mu _d^2>\frac{1}{2} \left(-a_5 v_{SM}^2-2 b_2 v_1^2-4 b_3 v_1^2\right) 
 \text{ and } v_1>b_1\text{ and } v_{SM}>b_1
 \\\nonumber&&
 \text{ and } a_6<\frac{a_5 v_{SM}^2+a_7 v_{SM}^2+2 b_2 v_1^2+4 b_3 v_1^2+2 \mu _d^2}{2 v_{SM}^2} \text{ and } a_1<\frac{a_5^2+2 a_6 a_5+4 a_7 a_5+a_6^2+4 a_7^2-a_3 a_4+4 a_6 a_7}{a_4}
 \\\nonumber&&
 \text{ and } a_7<\frac{-3 a_5 v_{SM}^2-2 b_2 v_1^2+4 b_3 v_1^2-2 \mu _d^2}{5 v_{SM}^2}  \text{ and } b_3<b_1\text{ and } c_1<-c_3
\\\nonumber&& 
\textbf{or}  
\\\nonumber&&
 -c_3=c_1\text{ and } a_4>b_1\text{ and } c_2>-c_1\text{ and } \mu _d^2>\frac{1}{2} \left(-a_5 v_{SM}^2-2 b_2 v_1^2-4 b_3 v_1^2\right)
 \\\nonumber&&
 \text{ and } v_1>b_1\text{ and } v_{SM}>b_1
 \text{ and } a_6<\frac{a_5 v_{SM}^2+a_7 v_{SM}^2+2 b_2 v_1^2+4 b_3 v_1^2+2 \mu _d^2}{2 v_{SM}^2}
 \text{ and } b_3<b_1\text{ and } b_4<b_2
   \\\nonumber&&
 \text{ and } a_7\leq \frac{-3 a_5 v_{SM}^2-2 b_2 v_1^2-4 b_3 v_1^2-2 \mu _d^2}{5 v_{SM}^2}
 \text{ and } \frac{-3 a_5 v_{SM}^2-2 b_2 v_1^2+4 b_3 v_1^2-2 \mu _d^2}{5 v_{SM}^2}\leq a_7
\\\nonumber&& 
\textbf{or}  
\\\nonumber&&
 a_1>\frac{a_4 b_2^2+a_4 b_4^2-2 a_4 b_2 b_4+a_5^2 c_1+a_5^2 c_3+2 a_6 a_5 c_1+4 a_7 a_5 c_1+2 a_6 a_5 c_3+4 a_7 a_5 c_3+a_6^2 c_1}{a_4 c_1+a_4 c_3}
 \\\nonumber&+&
 \frac{4 a_7^2 c_1-a_3 a_4 c_1+4 a_6 a_7 c_1+a_6^2 c_3+4 a_7^2 c_3-a_3 a_4 c_3+4 a_6 a_7 c_3}{a_4 c_1+a_4 c_3}\text{ and } a_4>b_1
 \text{ and } a_7>\frac{-3 a_5 v_{SM}^2-2 b_2 v_1^2-4 b_3 v_1^2-2 \mu _d^2}{5 v_{SM}^2}
  \\\nonumber&&
 \text{ and } c_1>-c_3
 \text{ and } c_2>-c_1\text{ and } \mu _d^2>\frac{1}{2} \left(-a_5 v_{SM}^2-2 b_2 v_1^2-4 b_3 v_1^2\right)\text{ and } v_1>b_1
 \\\nonumber&&
 \text{ and } v_{SM}>b_1\text{ and } b_4\geq b_2\text{ and } b_3<b_1\text{ and } a_6\leq -a_5-2 a_7
\\\nonumber&& 
\textbf{or}  
\\\nonumber&&
 a_1>\frac{a_4 b_2^2+a_4 b_4^2-2 a_4 b_2 b_4+a_5^2 c_1+a_5^2 c_3+2 a_6 a_5 c_1+4 a_7 a_5 c_1+2 a_6 a_5 c_3}{a_4 c_1+a_4 c_3}
 \\\nonumber&+& 
 \frac{4 a_7 a_5 c_3+a_6^2 c_1+4 a_7^2 c_1-a_3 a_4 c_1+4 a_6 a_7 c_1+a_6^2 c_3+4 a_7^2 c_3-a_3 a_4 c_3+4 a_6 a_7 c_3}{a_4 c_1+a_4 c_3}
 \text{ and } a_4>b_1
  \\\nonumber&&
 \text{ and } a_7>\frac{-3 a_5 v_{SM}^2-2 b_2 v_1^2+4 b_3 v_1^2-2 \mu _d^2}{5 v_{SM}^2}
 \text{ and } c_1>-c_3\text{ and } c_2>-c_1\text{ and } \mu _d^2>\frac{1}{2} \left(-a_5 v_{SM}^2-2 b_2 v_1^2+4 b_3 v_1^2\right)
 \\\nonumber&&
 \text{ and } v_1>b_1\text{ and } v_{SM}>b_1\text{ and } b_3\geq b_1\text{ and } b_4\geq b_2\text{ and } a_6\leq -a_5-2 a_7
\\\nonumber&& 
\textbf{or}  
\\\nonumber&&
a_1>\frac{a_4 b_2^2+a_4 b_4^2-2 a_4 b_2 b_4+a_5^2 c_1+a_5^2 c_3+2 a_6 a_5 c_1+4 a_7 a_5 c_1+2 a_6 a_5 c_3+4 a_7 a_5 c_3}{a_4 c_1+a_4 c_3}
\\\nonumber&+&
\frac{a_6^2 c_1+4 a_7^2 c_1-a_3 a_4 c_1+4 a_6 a_7 c_1+a_6^2 c_3+4 a_7^2 c_3-a_3 a_4 c_3+4 a_6 a_7 c_3}{a_4 c_1+a_4 c_3} \text{ and } a_7<\frac{-3 a_5 v_{SM}^2-2 b_2 v_1^2+4 b_3 v_1^2-2 \mu _d^2}{5 v_{SM}^2}
\\\nonumber&&
\text{ and } a_4>b_1\text{ and } c_1>-c_3\text{ and } c_2>-c_1
\text{ and } \mu _d^2>\frac{1}{2} \left(-a_5 v_{SM}^2-2 b_2 v_1^2-4 b_3 v_1^2\right) 
\\\nonumber&&
\text{ and } v_1>b_1\text{ and } v_{SM}>b_1\text{ and } b_4\geq b_2
\text{ and } a_6<\frac{a_5 v_{SM}^2+a_7 v_{SM}^2+2 b_2 v_1^2+4 b_3 v_1^2+2 \mu _d^2}{2 v_{SM}^2} \text{ and } b_3<b_1
\\\nonumber&&
\end{eqnarray*}
 
\begin{eqnarray*} 
\\\nonumber&& 
\textbf{or}  
\\\nonumber&&
a_1>\frac{a_4 b_2^2+a_4 b_4^2-2 a_4 b_2 b_4+a_5^2 c_1+a_5^2 c_3+2 a_6 a_5 c_1+4 a_7 a_5 c_1+2 a_6 a_5 c_3+4 a_7 a_5 c_3}{a_4 c_1+a_4 c_3}
\\\nonumber&+&
\frac{a_6^2 c_1+4 a_7^2 c_1-a_3 a_4 c_1+4 a_6 a_7 c_1+a_6^2 c_3+4 a_7^2 c_3-a_3 a_4 c_3+4 a_6 a_7 c_3}{a_4 c_1+a_4 c_3}
\text{ and } a_4>b_1\text{ and } c_1>-c_3\text{ and } c_2>-c_1
\\\nonumber&&
\text{ and } \mu _d^2>\frac{1}{2} \left(-a_5 v_{SM}^2-2 b_2 v_1^2+4 b_3 v_1^2\right)
\text{ and } v_1>b_1\text{ and } v_{SM}>b_1\text{ and } b_3\geq b_1\text{ and } b_4\geq b_2
\\\nonumber&&
\text{ and } a_6<\frac{a_5 v_{SM}^2+a_7 v_{SM}^2+2 b_2 v_1^2-4 b_3 v_1^2+2 \mu _d^2}{2 v_{SM}^2}
\text{ and } a_7\leq \frac{-3 a_5 v_{SM}^2-2 b_2 v_1^2+4 b_3 v_1^2-2 \mu _d^2}{5 v_{SM}^2}
\\\nonumber&& 
\textbf{or}  
\\\nonumber&&
a_1>\frac{a_4 b_2^2+a_4 b_4^2-2 a_4 b_2 b_4+a_5^2 c_1+a_5^2 c_3+2 a_6 a_5 c_1+4 a_7 a_5 c_1+2 a_6 a_5 c_3}{a_4 c_1+a_4 c_3}
\\\nonumber&+&
\frac{4 a_7 a_5 c_3+a_6^2 c_1+4 a_7^2 c_1-a_3 a_4 c_1+4 a_6 a_7 c_1+a_6^2 c_3+4 a_7^2 c_3-a_3 a_4 c_3+4 a_6 a_7 c_3}{a_4 c_1+a_4 c_3}
\text{ and } a_4>b_1\text{ and } c_1>-c_3\text{ and } c_2>-c_1
\\\nonumber&&
\text{ and } \mu _d^2>\frac{1}{2} \left(-a_5 v_{SM}^2-2 b_2 v_1^2+4 b_3 v_1^2\right)\text{ and } v_1>b_1\text{ and } v_{SM}>b_1 \text{ and } b_3\geq b_1
\\\nonumber&&
\text{ and } b_4\geq b_2\text{ and } a_6<\frac{a_5 v_{SM}^2+a_7 v_{SM}^2+2 b_2 v_1^2-4 b_3 v_1^2+2 \mu _d^2}{2 v_{SM}^2}
\text{ and } a_7\leq \frac{-3 a_5 v_{SM}^2-2 b_2 v_1^2+4 b_3 v_1^2-2 \mu _d^2}{5 v_{SM}^2}
\\\nonumber&& 
\textbf{or}  
\\\nonumber&&
 a_4>b_1\text{ and } a_7>\frac{-3 a_5 v_{SM}^2-2 b_2 v_1^2-4 b_3 v_1^2-2 \mu _d^2}{5 v_{SM}^2}
 \text{ and } c_1>-c_3\text{ and } c_2>-c_1
 \text{ and } \mu _d^2>\frac{1}{2} \left(-a_5 v_{SM}^2-2 b_2 v_1^2-4 b_3 v_1^2\right)
 \\\nonumber&&
 \text{ and } a_1\geq \frac{a_5^2+2 a_6 a_5+4 a_7 a_5+a_6^2+4 a_7^2-a_3 a_4+4 a_6 a_7}{a_4}\text{ and } b_3<b_1
 \\\nonumber&&
  \text{ and } v_1>b_1\text{ and } v_{SM}>b_1
 \text{ and } b_4<b_2\text{ and } a_6\leq -a_5-2 a_7
\\\nonumber&& 
\textbf{or}  
\\\nonumber&&
a_4>b_1\text{ and } a_7>\frac{-3 a_5 v_{SM}^2-2 b_2 v_1^2-4 b_3 v_1^2-2 \mu _d^2}{5 v_{SM}^2}
\\\nonumber&&
\text{ and } c_2>-c_1\text{ and } \mu _d^2>\frac{1}{2} \left(-a_5 v_{SM}^2-2 b_2 v_1^2-4 b_3 v_1^2\right)\text{ and } v_1>b_1\text{ and } v_{SM}>b_1
\\\nonumber&&
\text{ and } b_4\geq b_2\text{ and } a_1<\frac{a_4 b_2^2+a_4 b_4^2-2 a_4 b_2 b_4+a_5^2 c_1+a_5^2 c_3+2 a_6 a_5 c_1+4 a_7 a_5 c_1}{a_4 c_1+a_4 c_3}
\\\nonumber&+&
\frac{2 a_6 a_5 c_3+4 a_7 a_5 c_3+a_6^2 c_1+4 a_7^2 c_1-a_3 a_4 c_1+4 a_6 a_7 c_1+a_6^2 c_3+4 a_7^2 c_3-a_3 a_4 c_3+4 a_6 a_7 c_3}{a_4 c_1+a_4 c_3}
\\\nonumber&&
\text{ and } b_3<b_1\text{ and } c_1<-c_3\text{ and } a_6\leq -a_5-2 a_7 
\\\nonumber&& 
\textbf{or}  
\\\nonumber&&
 a_4>b_1\text{ and } a_7>\frac{-3 a_5 v_{SM}^2-2 b_2 v_1^2-4 b_3 v_1^2-2 \mu _d^2}{5 v_{SM}^2}
 \\\nonumber&&
 \text{ and } c_2>-c_1\text{ and } \mu _d^2>\frac{1}{2} \left(-a_5 v_{SM}^2-2 b_2 v_1^2-4 b_3 v_1^2\right)
 \\\nonumber&&
 \text{ and } v_1>b_1\text{ and } v_{SM}>b_1\text{ and } b_3<b_1\text{ and } b_4<b_2
 \\\nonumber&&
 \text{ and } c_1<-c_3\text{ and } a_1\leq \frac{a_5^2+2 a_6 a_5+4 a_7 a_5+a_6^2+4 a_7^2-a_3 a_4+4 a_6 a_7}{a_4}\text{ and } a_6\leq -a_5-2 a_7
 \\\nonumber&& 
\textbf{or}  
\\\nonumber&&
 a_4>b_1\text{ and } a_7>\frac{-3 a_5 v_{SM}^2-2 b_2 v_1^2+4 b_3 v_1^2-2 \mu _d^2}{5 v_{SM}^2}\text{ and } c_1>-c_3
 \\\nonumber&&
 \text{ and } c_2>-c_1\text{ and } \mu _d^2>\frac{1}{2} \left(-a_5 v_{SM}^2-2 b_2 v_1^2+4 b_3 v_1^2\right)\text{ and } v_1>b_1
 \\\nonumber&&
 \text{ and } v_{SM}>b_1\text{ and } a_1\geq \frac{a_5^2+2 a_6 a_5+4 a_7 a_5+a_6^2+4 a_7^2-a_3 a_4+4 a_6 a_7}{a_4}\text{ and } b_3\geq b_1
 \\\nonumber&&
 \text{ and } b_4<b_2\text{ and } a_6\leq -a_5-2 a_7
\end{eqnarray*}

\begin{eqnarray*} 
\nonumber&& 
\textbf{or}  
\\\nonumber&&
 a_4>b_1\text{ and } a_7>\frac{-3 a_5 v_{SM}^2-2 b_2 v_1^2+4 b_3 v_1^2-2 \mu _d^2}{5 v_{SM}^2}\text{ and } c_2>-c_1 \text{ and } \mu _d^2>\frac{1}{2} \left(-a_5 v_{SM}^2-2 b_2 v_1^2+4 b_3 v_1^2\right)
 \\\nonumber&&
 \text{ and } b_3\geq b_1\text{ and } b_4\geq b_2 \text{ and } v_1>b_1\text{ and } v_{SM}>b_1
 \\\nonumber&&
 \text{ and } a_1<\frac{a_4 b_2^2+a_4 b_4^2-2 a_4 b_2 b_4+a_5^2 c_1+a_5^2 c_3+2 a_6 a_5 c_1+4 a_7 a_5 c_1+2 a_6 a_5 c_3+4 a_7 a_5 c_3}{a_4 c_1+a_4 c_3}
 \\\nonumber&+&
 \frac{a_6^2 c_1+4 a_7^2 c_1-a_3 a_4 c_1+4 a_6 a_7 c_1+a_6^2 c_3+4 a_7^2 c_3-a_3 a_4 c_3+4 a_6 a_7 c_3}{a_4 c_1+a_4 c_3}\text{ and } c_1<-c_3\text{ and } a_6\leq -a_5-2 a_7
\\\nonumber&& 
\textbf{or}  
\\\nonumber&&
\text{ and } \mu _d^2>\frac{1}{2} \left(-a_5 v_{SM}^2-2 b_2 v_1^2+4 b_3 v_1^2\right)\text{ and } v_1>b_1\text{ and } v_{SM}>b_1
\text{ and } b_3\geq b_1\text{ and } b_4<b_2\text{ and } c_1<-c_3
\\\nonumber&&
a_4>b_1\text{ and } a_7>\frac{-3 a_5 v_{SM}^2-2 b_2 v_1^2+4 b_3 v_1^2-2 \mu _d^2}{5 v_{SM}^2}\text{ and } c_2>-c_1
\\\nonumber&&
\text{ and } a_1\leq \frac{a_5^2+2 a_6 a_5+4 a_7 a_5+a_6^2+4 a_7^2-a_3 a_4+4 a_6 a_7}{a_4}\text{ and } a_6\leq -a_5-2 a_7
\\\nonumber&& 
\textbf{or}  
\\\nonumber&&
a_4>b_1\text{ and } b_4>b_2\text{ and } c_2>-c_1\text{ and } \mu _d^2>\frac{1}{2} \left(-a_5 v_{SM}^2-2 b_2 v_1^2-4 b_3 v_1^2\right)
\\\nonumber&&
\text{ and } v_1>b_1\text{ and } v_{SM}>b_1
\\\nonumber&&
\text{ and } a_1<\frac{a_4 b_2^2+a_4 b_4^2-2 a_4 b_2 b_4+a_5^2 c_1+a_5^2 c_3+2 a_6 a_5 c_1+4 a_7 a_5 c_1+2 a_6 a_5 c_3+4 a_7 a_5 c_3}{a_4 c_1+a_4 c_3}
\\\nonumber&+&
\frac{a_6^2 c_1+4 a_7^2 c_1-a_3 a_4 c_1+4 a_6 a_7 c_1+a_6^2 c_3+4 a_7^2 c_3-a_3 a_4 c_3+4 a_6 a_7 c_3}{a_4 c_1+a_4 c_3}
\\\nonumber&&
\text{ and } a_6<\frac{a_5 v_{SM}^2+a_7 v_{SM}^2+2 b_2 v_1^2+4 b_3 v_1^2+2 \mu _d^2}{2 v_{SM}^2}
\\\nonumber&&
\text{ and } a_7<\frac{-3 a_5 v_{SM}^2-2 b_2 v_1^2+4 b_3 v_1^2-2 \mu _d^2}{5 v_{SM}^2}\text{ and } b_3<b_1\text{ and } c_1<-c_3 
\\\nonumber&& 
\textbf{or}  
\\\nonumber&&
a_4>b_1\text{ and } c_1>-c_3\text{ and } c_2>-c_1\text{ and } \mu _d^2>\frac{1}{2} \left(-a_5 v_{SM}^2-2 b_2 v_1^2-4 b_3 v_1^2\right)
\\\nonumber&&
\text{ and } v_1>b_1\text{ and } v_{SM}>b_1\text{ and } a_1\geq \frac{a_5^2+2 a_6 a_5+4 a_7 a_5+a_6^2+4 a_7^2-a_3 a_4+4 a_6 a_7}{a_4}
\\\nonumber&&
\text{ and } a_6<\frac{a_5 v_{SM}^2+a_7 v_{SM}^2+2 b_2 v_1^2+4 b_3 v_1^2+2 \mu _d^2}{2 v_{SM}^2}
\\\nonumber&&
\text{ and } a_7<\frac{-3 a_5 v_{SM}^2-2 b_2 v_1^2+4 b_3 v_1^2-2 \mu _d^2}{5 v_{SM}^2}\text{ and } b_3<b_1\text{ and } b_4<b_2
\\\nonumber&& 
\textbf{or}  
\\\nonumber&&
a_4>b_1\text{ and } c_1>-c_3\text{ and } c_2>-c_1\text{ and } \mu _d^2>\frac{1}{2} \left(-a_5 v_{SM}^2-2 b_2 v_1^2+4 b_3 v_1^2\right)
\\\nonumber&&
\text{ and } v_1>b_1\text{ and } v_{SM}>b_1\text{ and } a_1\geq \frac{a_5^2+2 a_6 a_5+4 a_7 a_5+a_6^2+4 a_7^2-a_3 a_4+4 a_6 a_7}{a_4} \text{ and } b_3\geq b_1
\\\nonumber&&
\text{ and } a_6<\frac{a_5 v_{SM}^2+a_7 v_{SM}^2+2 b_2 v_1^2-4 b_3 v_1^2+2 \mu _d^2}{2 v_{SM}^2}
\text{ and } b_4<b_2\text{ and } a_7\leq \frac{-3 a_5 v_{SM}^2-2 b_2 v_1^2+4 b_3 v_1^2-2 \mu _d^2}{5 v_{SM}^2}
\end{eqnarray*}

\begin{eqnarray*}
\\\nonumber&& 
\textbf{or}  
\\\nonumber&&
 a_4>b_1\text{ and } c_2>-c_1\text{ and } \mu _d^2>\frac{1}{2} \left(-a_5 v_{SM}^2-2 b_2 v_1^2-4 b_3 v_1^2\right)
 \text{ and } v_1>b_1\text{ and } v_{SM}>b_1
  \\\nonumber&&
 \text{ and } a_6<\frac{a_5 v_{SM}^2+a_7 v_{SM}^2+2 b_2 v_1^2+4 b_3 v_1^2+2 \mu _d^2}{2 v_{SM}^2}
 \text{ and } a_7<\frac{-3 a_5 v_{SM}^2-2 b_2 v_1^2+4 b_3 v_1^2-2 \mu _d^2}{5 v_{SM}^2}
  \\\nonumber&&
 \text{ and } b_3<b_1
 \text{ and } b_4<b_2\text{ and } c_1<-c_3\text{ and } a_1\leq \frac{a_5^2+2 a_6 a_5+4 a_7 a_5+a_6^2+4 a_7^2-a_3 a_4+4 a_6 a_7}{a_4}
 \\\nonumber&& 
\textbf{or}  
\\\nonumber&&
 a_4>b_1\text{ and } c_2>-c_1\text{ and } \mu _d^2>\frac{1}{2} \left(-a_5 v_{SM}^2-2 b_2 v_1^2+4 b_3 v_1^2\right)\text{ and } v_1>b_1  \text{ and } a_7\leq \frac{-3 a_5 v_{SM}^2-2 b_2 v_1^2+4 b_3 v_1^2-2 \mu _d^2}{5 v_{SM}^2}
 \\\nonumber&&
 \text{ and } v_{SM}>b_1\text{ and } b_3\geq b_1\text{ and } b_4\geq b_2
 \\\nonumber&&
 \text{ and } a_1<\frac{a_4 b_2^2+a_4 b_4^2-2 a_4 b_2 b_4+a_5^2 c_1+a_5^2 c_3+2 a_6 a_5 c_1+4 a_7 a_5 c_1+2 a_6 a_5 c_3}{a_4 c_1+a_4 c_3}
 \\\nonumber&+&
 \frac{4 a_7 a_5 c_3+a_6^2 c_1+4 a_7^2 c_1-a_3 a_4 c_1+4 a_6 a_7 c_1+a_6^2 c_3+4 a_7^2 c_3-a_3 a_4 c_3+4 a_6 a_7 c_3}{a_4 c_1+a_4 c_3}
 \\\nonumber&&
 \text{ and } a_6<\frac{a_5 v_{SM}^2+a_7 v_{SM}^2+2 b_2 v_1^2-4 b_3 v_1^2+2 \mu _d^2}{2 v_{SM}^2}\text{ and } c_1<-c_3\
 \\\nonumber&&
\textbf{or}  
\\\nonumber&&
a_4>b_1\text{ and } c_2>-c_1\text{ and } \mu _d^2>\frac{1}{2} \left(-a_5 v_{SM}^2-2 b_2 v_1^2+4 b_3 v_1^2\right)\text{ and } v_1>b_1
\\\nonumber&&
\text{ and } v_{SM}>b_1\text{ and } b_3\geq b_1\text{ and } a_6<\frac{a_5 v_{SM}^2+a_7 v_{SM}^2+2 b_2 v_1^2-4 b_3 v_1^2+2 \mu _d^2}{2 v_{SM}^2}
\text{ and } b_4<b_2\text{ and } c_1<-c_3
\\\nonumber&&
\text{ and } a_1\leq \frac{a_5^2+2 a_6 a_5+4 a_7 a_5+a_6^2+4 a_7^2-a_3 a_4+4 a_6 a_7}{a_4}
\text{ and } a_7\leq \frac{-3 a_5 v_{SM}^2-2 b_2 v_1^2+4 b_3 v_1^2-2 \mu _d^2}{5 v_{SM}^2}
\\\nonumber&& 
\textbf{or}  
\\\nonumber&&
a_1>\frac{a_4 b_2^2+a_4 b_4^2-2 a_4 b_2 b_4+a_5^2 c_1+a_5^2 c_3+2 a_6 a_5 c_1+4 a_7 a_5 c_1+2 a_6 a_5 c_3+4 a_7 a_5 c_3}{a_4 c_1+a_4 c_3}
\\\nonumber&+&
\frac{a_6^2 c_1+4 a_7^2 c_1-a_3 a_4 c_1+4 a_6 a_7 c_1+a_6^2 c_3+4 a_7^2 c_3-a_3 a_4 c_3+4 a_6 a_7 c_3}{a_4 c_1+a_4 c_3} \text{ and } v_1>b_1\text{ and } v_{SM}>b_1\text{ and } b_4\geq b_2
\text{ and } a_4>b_1\text{ and } c_1>-c_3\text{ and } c_2>-c_1
\\\nonumber&&
\text{ and } a_6<\frac{a_5 v_{SM}^2+a_7 v_{SM}^2+2 b_2 v_1^2+4 b_3 v_1^2+2 \mu _d^2}{2 v_{SM}^2}\text{ and } b_3<b_1 \text{ and } \mu _d^2>\frac{1}{2} \left(-a_5 v_{SM}^2-2 b_2 v_1^2-4 b_3 v_1^2\right)
\\\nonumber&&
\text{ and } a_7\leq \frac{-3 a_5 v_{SM}^2-2 b_2 v_1^2-4 b_3 v_1^2-2 \mu _d^2}{5 v_{SM}^2}\text{ and } \frac{-3 a_5 v_{SM}^2-2 b_2 v_1^2+4 b_3 v_1^2-2 \mu _d^2}{5 v_{SM}^2}\leq a_7 
\\\nonumber&& 
\textbf{or}  
\\\nonumber&&
a_4>b_1\text{ and } c_1>-c_3\text{ and } c_2>-c_1\text{ and } \mu _d^2>\frac{1}{2} \left(-a_5 v_{SM}^2-2 b_2 v_1^2-4 b_3 v_1^2\right)
\text{ and } v_1>b_1\text{ and } v_{SM}>b_1
\\\nonumber&&
\text{ and } a_1\geq \frac{a_5^2+2 a_6 a_5+4 a_7 a_5+a_6^2+4 a_7^2-a_3 a_4+4 a_6 a_7}{a_4}
\text{ and } a_6<\frac{a_5 v_{SM}^2+a_7 v_{SM}^2+2 b_2 v_1^2+4 b_3 v_1^2+2 \mu _d^2}{2 v_{SM}^2}
\\\nonumber&&
\text{ and } a_7\leq \frac{-3 a_5 v_{SM}^2-2 b_2 v_1^2-4 b_3 v_1^2-2 \mu _d^2}{5 v_{SM}^2}
\text{ and } \frac{-3 a_5 v_{SM}^2-2 b_2 v_1^2+4 b_3 v_1^2-2 \mu _d^2}{5 v_{SM}^2}\leq a_7 \text{ and } b_3<b_1\text{ and } b_4<b_2
\\\nonumber&& 
\textbf{or}  
\\\nonumber&&
a_4>b_1\text{ and } c_2>-c_1\text{ and } \mu _d^2>\frac{1}{2} \left(-a_5 v_{SM}^2-2 b_2 v_1^2-4 b_3 v_1^2\right)\text{ and } v_1>b_1
\text{ and } v_{SM}>b_1 \text{ and } b_4\geq b_2
\\\nonumber&&
\text{ and } a_1<\frac{a_4 b_2^2+a_4 b_4^2-2 a_4 b_2 b_4+a_5^2 c_1+a_5^2 c_3+2 a_6 a_5 c_1+4 a_7 a_5 c_1+2 a_6 a_5 c_3+4 a_7 a_5 c_3}{a_4 c_1+a_4 c_3}
\\\nonumber&+&
\frac{a_6^2 c_1+4 a_7^2 c_1-a_3 a_4 c_1+4 a_6 a_7 c_1+a_6^2 c_3+4 a_7^2 c_3-a_3 a_4 c_3+4 a_6 a_7 c_3}{a_4 c_1+a_4 c_3}
\text{ and } a_6<\frac{a_5 v_{SM}^2+a_7 v_{SM}^2+2 b_2 v_1^2+4 b_3 v_1^2+2 \mu _d^2}{2 v_{SM}^2} 
\\\nonumber&&
\text{ and } a_7\leq \frac{-3 a_5 v_{SM}^2-2 b_2 v_1^2-4 b_3 v_1^2-2 \mu _d^2}{5 v_{SM}^2}\text{ and } \frac{-3 a_5 v_{SM}^2-2 b_2 v_1^2+4 b_3 v_1^2-2 \mu _d^2}{5 v_{SM}^2}\leq a_7 \text{ and } c_1<-c_3 \text{ and } b_3<b_1
\end{eqnarray*}

\begin{eqnarray*}
\\\nonumber&& 
\textbf{or}  
\\\nonumber&&
a_4>b_1\text{ and } c_2>-c_1\text{ and } \mu _d^2>\frac{1}{2} \left(-a_5 v_{SM}^2-2 b_2 v_1^2-4 b_3 v_1^2\right)\text{ and } v_1>b_1\text{ and } v_{SM}>b_1 \text{ and } b_3<b_1\text{ and } b_4<b_2
\\\nonumber&&
\text{ and } a_6<\frac{a_5 v_{SM}^2+a_7 v_{SM}^2+2 b_2 v_1^2+4 b_3 v_1^2+2 \mu _d^2}{2 v_{SM}^2}
\text{ and } a_1\leq \frac{a_5^2+2 a_6 a_5+4 a_7 a_5+a_6^2+4 a_7^2-a_3 a_4+4 a_6 a_7}{a_4}
\\\nonumber&&
\text{ and } a_7\leq \frac{-3 a_5 v_{SM}^2-2 b_2 v_1^2-4 b_3 v_1^2-2 \mu _d^2}{5 v_{SM}^2}\text{ and } \frac{-3 a_5 v_{SM}^2-2 b_2 v_1^2+4 b_3 v_1^2-2 \mu _d^2}{5 v_{SM}^2}\leq a_7 \text{ and } c_1<-c_3
\end{eqnarray*}

\begin{center}
\fbox{{ If $2 a1+2 a3\leq 0$ and $b2+b4\leq 0$, then: }}
\end{center} 

\begin{eqnarray*}
\nonumber
&&b_2=b_1\text{ and } b_4=b_1\text{ and } a_4>b_1\text{ and } b_3>b_1 
\text{ and } c_1>\frac{-3 a_1 c_3-3 a_3 c_3}{3 a_1+3 a_3}\text{ and } c_2>-c_1\text{ and } \mu _d^2>\frac{1}{2} \left(4 b_3 v_1^2-a_5 v_{SM}^2\right)
\\\nonumber&& 
\text{ and } v_1>b_1\text{ and } v_{SM}>b_1\text{ and } a_1<-a_3\text{ and } a_6<\frac{a_5 v_{SM}^2+a_7 v_{SM}^2-4 b_3 v_1^2+2 \mu _d^2}{2 v_{SM}^2}
\\\nonumber&& 
\textbf{or}  
\\\nonumber&&
b_2=b_1\text{ and } b_4=b_1\text{ and } a_4>b_1\text{ and } c_1>\frac{-3 a_1 c_3-3 a_3 c_3}{3 a_1+3 a_3}\text{ and } c_2>-c_1
\text{ and } \mu _d^2>\frac{1}{2} \left(-a_5 v_{SM}^2-4 b_3 v_1^2\right)\text{ and } v_1>b_1
\\\nonumber&& 
\text{ and } a_1<-a_3\text{ and } a_6<\frac{a_5 v_{SM}^2+a_7 v_{SM}^2+4 b_3 v_1^2+2 \mu _d^2}{2 v_{SM}^2}\text{ and } b_3\leq b_1 \text{ and } v_{SM}>b_1
\\\nonumber&& 
\textbf{or}  
\\\nonumber&&
b_4=b_1\text{ and } a_4>b_1\text{ and } b_3>b_1\text{ and } c_2>-c_1\text{ and } \mu _d^2>\frac{1}{2} \left(-a_5 v_{SM}^2-2 b_2 v_1^2+4 b_3 v_1^2\right)
\text{ and } v_1>b_1\text{ and } v_{SM}>b_1 
\\\nonumber&& 
\text{ and } a_6<\frac{a_5 v_{SM}^2+a_7 v_{SM}^2+2 b_2 v_1^2-4 b_3 v_1^2+2 \mu _d^2}{2 v_{SM}^2}\text{ and } b_2<b_1 text{ and } c_1\geq \frac{-a_1 c_3-a_3 c_3+b_2^2}{a_1+a_3} \text{ and } a_1<-a_3
\\\nonumber&& 
\textbf{or}  
\\\nonumber&&
b_4=b_1\text{ and } a_4>b_1\text{ and } c_2>-c_1\text{ and } \mu _d^2>\frac{1}{2} \left(-a_5 v_{SM}^2-2 b_2 v_1^2-4 b_3 v_1^2\right)
\text{ and } v_1>b_1\text{ and } v_{SM}>b_1\text{ and } a_1<-a_3
\\\nonumber&&
\text{ and } a_6<\frac{a_5 v_{SM}^2+a_7 v_{SM}^2+2 b_2 v_1^2+4 b_3 v_1^2+2 \mu _d^2}{2 v_{SM}^2}\text{ and } b_2<b_1\text{ and } b_3\leq b_1 \text{ and } c_1\geq \frac{-a_1 c_3-a_3 c_3+b_2^2}{a_1+a_3}
\\\nonumber&& 
\textbf{or}  
\\\nonumber&&
b_4=b_2\text{ and } a_4>b_1\text{ and } b_3>b_1\text{ and } c_1>\frac{-a_1 c_3-a_3 c_3+b_2^2+2 b_4 b_2+b_4^2}{a_1+a_3}\text{ and } c_2>-c_1 
\text{ and } b_4<b_1 \text{ and } v_1>b_1
\\\nonumber&&
\text{ and } a_6<\frac{a_5 v_{SM}^2+a_7 v_{SM}^2+2 b_2 v_1^2-4 b_3 v_1^2+2 \mu _d^2}{2 v_{SM}^2}
\text{ and } \mu _d^2>\frac{1}{2} \left(-a_5 v_{SM}^2-2 b_2 v_1^2+4 b_3 v_1^2\right)\text{ and } a_1<-a_3 \text{ and } v_{SM}>b_1
\\\nonumber&& 
\textbf{or}  
\\\nonumber&&
b_4=b_2\text{ and } a_4>b_1\text{ and } c_1>\frac{-a_1 c_3-a_3 c_3+b_2^2+2 b_4 b_2+b_4^2}{a_1+a_3}\text{ and } c_2>-c_1
\\\nonumber&&
\text{ and } \mu _d^2>\frac{1}{2} \left(-a_5 v_{SM}^2-2 b_2 v_1^2-4 b_3 v_1^2\right)\text{ and } v_1>b_1\text{ and } v_{SM}>b_1
\\\nonumber&&
\text{ and } a_1<-a_3\text{ and } a_6<\frac{a_5 v_{SM}^2+a_7 v_{SM}^2+2 b_2 v_1^2+4 b_3 v_1^2+2 \mu _d^2}{2 v_{SM}^2}\text{ and } b_4<b_1\text{ and } b_3\leq b_1
\\\nonumber&&
\textbf{or}  
\\\nonumber&&
a_4>b_1\text{ and } b_3>b_1\text{ and } b_4>b_1\text{ and } c_2>-c_1\text{ and } \mu _d^2>\frac{1}{2} \left(-a_5 v_{SM}^2-2 b_2 v_1^2+4 b_3 v_1^2\right)
\text{ and } v_1>b_1\text{ and } v_{SM}>b_1\text{ and } c_1\geq \frac{-a_1 c_3-a_3 c_3+b_2^2+2 b_4 b_2+b_4^2}{a_1+a_3}\text{ and } a_1<-a_3
\\\nonumber&&
\text{ and } a_6<\frac{a_5 v_{SM}^2+a_7 v_{SM}^2+2 b_2 v_1^2-4 b_3 v_1^2+2 \mu _d^2}{2 v_{SM}^2}\text{ and } b_2\leq -b_4
\end{eqnarray*}

\begin{eqnarray*}
\\\nonumber&& 
\textbf{or}  
\\\nonumber&&
a_4>b_1\text{ and } b_3>b_1\text{ and } c_2>-c_1\text{ and } \mu _d^2>\frac{1}{2} \left(-a_5 v_{SM}^2-2 b_2 v_1^2+4 b_3 v_1^2\right) \text{ and } b_2<b_4\text{ and } b_4<b_1
\\\nonumber&&
\text{ and } v_1>b_1\text{ and } v_{SM}>b_1\text{ and } c_1\geq \frac{-a_1 c_3-a_3 c_3+b_2^2+2 b_4 b_2+b_4^2}{a_1+a_3}
\text{ and } a_1<-a_3
\\\nonumber&&
\text{ and } a_6<\frac{a_5 v_{SM}^2+a_7 v_{SM}^2+2 b_2 v_1^2-4 b_3 v_1^2+2 \mu _d^2}{2 v_{SM}^2}
\\\nonumber&& 
\textbf{or}  
\\\nonumber&&
a_4>b_1\text{ and } b_4>b_1\text{ and } c_2>-c_1\text{ and } \mu _d^2>\frac{1}{2} \left(-a_5 v_{SM}^2-2 b_2 v_1^2-4 b_3 v_1^2\right) \text{ and } b_2\leq -b_4\text{ and } b_3\leq b_1
\\\nonumber&&
\text{ and } v_1>b_1\text{ and } v_{SM}>b_1\text{ and } c_1\geq \frac{-a_1 c_3-a_3 c_3+b_2^2+2 b_4 b_2+b_4^2}{a_1+a_3}
\\\nonumber&&
\text{ and } a_1<-a_3\text{ and } a_6<\frac{a_5 v_{SM}^2+a_7 v_{SM}^2+2 b_2 v_1^2+4 b_3 v_1^2+2 \mu _d^2}{2 v_{SM}^2}
\\\nonumber&& 
\textbf{or}  
\\\nonumber&&
a_4>b_1\text{ and } c_2>-c_1\text{ and } \mu _d^2>\frac{1}{2} \left(-a_5 v_{SM}^2-2 b_2 v_1^2-4 b_3 v_1^2\right)\text{ and } v_1>b_1 \text{ and } b_4<b_1\text{ and } b_3\leq b_1 \text{ and } b_2<b_4
\\\nonumber&&
\text{ and } v_{SM}>b_1\text{ and } c_1\geq \frac{-a_1 c_3-a_3 c_3+b_2^2+2 b_4 b_2+b_4^2}{a_1+a_3}\text{ and } a_1<-a_3
\text{ and } a_6<\frac{a_5 v_{SM}^2+a_7 v_{SM}^2+2 b_2 v_1^2+4 b_3 v_1^2+2 \mu _d^2}{2 v_{SM}^2}
\\\nonumber&&
\textbf{or}  
\\\nonumber&&
a_4>b_1\text{ and } b_3>b_1 \text{ and } c_2>-c_1\text{ and } \mu _d^2>\frac{1}{2} \left(-a_5 v_{SM}^2-2 b_2 v_1^2+4 b_3 v_1^2\right)\text{ and } v_1>b_1 \text{ and } v_{SM}>b_1 \text{ and } b_4<b_1
\\\nonumber&&
\text{ and } c_1>\frac{-3 a_1 c_3-3 a_3 c_3+2 b_2^2+8 b_4 b_2+2 b_4^2}{3 a_1+3 a_3}
\text{ and } a_1<-a_3
\\\nonumber&&
\text{ and } a_6<\frac{a_5 v_{SM}^2+a_7 v_{SM}^2+2 b_2 v_1^2-4 b_3 v_1^2+2 \mu _d^2}{2 v_{SM}^2} \text{ and } b_4<b_2\text{ and } b_2\leq -b_4
\\\nonumber&& 
\textbf{or}  
\\\nonumber&&
a_4>b_1\text{ and } c_1>\frac{-3 a_1 c_3-3 a_3 c_3+2 b_2^2+8 b_4 b_2+2 b_4^2}{3 a_1+3 a_3}\text{ and } c_2>-c_1
\\\nonumber&&
\text{ and } \mu _d^2>\frac{1}{2} \left(-a_5 v_{SM}^2-2 b_2 v_1^2-4 b_3 v_1^2\right)\text{ and } v_1>b_1\text{ and } v_{SM}>b_1
\\\nonumber&&
\text{ and } a_1<-a_3\text{ and } a_6<\frac{a_5 v_{SM}^2+a_7 v_{SM}^2+2 b_2 v_1^2+4 b_3 v_1^2+2 \mu _d^2}{2 v_{SM}^2}
\text{ and } b_4<b_1\text{ and } b_4<b_2\text{ and } b_2\leq -b_4\text{ and } b_3\leq b_1
\end{eqnarray*}

\newpage
\begin{center}
\fbox{{\bf { \textbf{Case 5.} $A_{ij}\leq0$, $A_{jk}\leq0$, $A_{ik}\leq0$ and the other entries positive.
}}}
\end{center}
\begin{center}
\fbox{{ If $a5+a6+2 a7\leq 0$, $2 a1+2 a3\leq 0$ and $a5+a6+2 a7\leq 0$, then:}}
\end{center}
There is no solution.
\begin{center}
\fbox{{If $a5+a6+2 a7\leq 0$, $b2+b4\leq 0$ and $0\leq 0$, then: }}
\end{center}

\begin{eqnarray*} 
\nonumber
&& a_1>\frac{-a_3 c_1-a_3 c_3+b_2^2+2 b_4 b_2+b_4^2}{c_1+c_3}
\\\nonumber&& 
\text{ and } a_4>\frac{a_5^2 c_1+a_5^2 c_3+2 a_6 a_5 c_1+4 a_7 a_5 c_1+2 a_6 a_5 c_3+4 a_7 a_5 c_3+a_6^2 c_1+4 a_7^2 c_1+4 a_6 a_7 c_1}{a_1 c_1+a_3 c_1+a_1 c_3+a_3 c_3-b_2^2-2 b_4 b_2-b_4^2}
\\\nonumber&+& 
\frac{a_6^2 c_3+4 a_7^2 c_3+4 a_6 a_7 c_3}{a_1 c_1+a_3 c_1+a_1 c_3+a_3 c_3-b_2^2-2 b_4 b_2-b_4^2}
\text{ and } a_7>\frac{-3 a_5 v_{SM}^2-2 b_2 v_1^2-4 b_3 v_1^2-2 \mu _d^2}{5 v_{SM}^2}\text{ and } c_1>-c_3\text{ and } c_2>-c_1
\\\nonumber&& 
\text{ and } \mu _d^2>\frac{1}{2} \left(-a_5 v_{SM}^2-2 b_2 v_1^2-4 b_3 v_1^2\right)\text{ and } v_1>b_1\text{ and } v_{SM}>b_1
\text{ and } a_6\leq -a_5-2 a_7\text{ and } b_3\leq b_1\text{ and } b_4\leq -b_2
\\\nonumber&& 
\textbf{or}  
\\\nonumber&&
a_1>\frac{-a_3 c_1-a_3 c_3+b_2^2+2 b_4 b_2+b_4^2}{c_1+c_3} 
\text{ and } a_4>\frac{a_5^2 c_1+a_5^2 c_3+2 a_6 a_5 c_1+4 a_7 a_5 c_1+2 a_6 a_5 c_3+4 a_7 a_5 c_3+a_6^2 c_1+4 a_7^2 c_1}{a_1 c_1+a_3 c_1+a_1 c_3+a_3 c_3-b_2^2-2 b_4 b_2-b_4^2}
\\\nonumber&+& 
\frac{4 a_6 a_7 c_1+a_6^2 c_3+4 a_7^2 c_3+4 a_6 a_7 c_3}{a_1 c_1+a_3 c_1+a_1 c_3+a_3 c_3-b_2^2-2 b_4 b_2-b_4^2} \text{ and } \mu _d^2>\frac{1}{2} \left(-a_5 v_{SM}^2-2 b_2 v_1^2+4 b_3 v_1^2\right)
\\\nonumber&& 
\text{ and } a_7>\frac{-3 a_5 v_{SM}^2-2 b_2 v_1^2+4 b_3 v_1^2-2 \mu _d^2}{5 v_{SM}^2}\text{ and } b_3>b_1
\text{ and } c_1>-c_3\text{ and } c_2>-c_1
\\\nonumber&& 
\text{ and } v_1>b_1\text{ and } v_{SM}>b_1\text{ and } a_6\leq -a_5-2 a_7\text{ and } b_4\leq -b_2
\\\nonumber&& 
\textbf{or}  
\\\nonumber&&
a_1>\frac{-a_3 c_1-a_3 c_3+b_2^2+2 b_4 b_2+b_4^2}{c_1+c_3} \text{ and } a_7\leq \frac{-3 a_5 v_{SM}^2-2 b_2 v_1^2+4 b_3 v_1^2-2 \mu _d^2}{5 v_{SM}^2}\text{ and } b_4\leq -b_2
\\\nonumber&& 
\text{ and } a_4>\frac{a_5^2 c_1+a_5^2 c_3+2 a_6 a_5 c_1+4 a_7 a_5 c_1+2 a_6 a_5 c_3+4 a_7 a_5 c_3+a_6^2 c_1+4 a_7^2 c_1+4 a_6 a_7 c_1}{a_1 c_1+a_3 c_1+a_1 c_3+a_3 c_3-b_2^2-2 b_4 b_2-b_4^2}
\\\nonumber&+& 
\frac{a_6^2 c_3+4 a_7^2 c_3+4 a_6 a_7 c_3}{a_1 c_1+a_3 c_1+a_1 c_3+a_3 c_3-b_2^2-2 b_4 b_2-b_4^2}\text{ and } b_3>b_1\text{ and } c_1>-c_3 \text{ and } v_1>b_1
\text{ and } c_2>-c_1
\\\nonumber&& 
\text{ and } v_{SM}>b_1\text{ and } a_6<\frac{a_5 v_{SM}^2+a_7 v_{SM}^2+2 b_2 v_1^2-4 b_3 v_1^2+2 \mu _d^2}{2 v_{SM}^2} \text{ and } \mu _d^2>\frac{1}{2} \left(-a_5 v_{SM}^2-2 b_2 v_1^2+4 b_3 v_1^2\right)
\\\nonumber&& 
\textbf{or}  
\\\nonumber&&
a_1>\frac{-a_3 c_1-a_3 c_3+b_2^2+2 b_4 b_2+b_4^2}{c_1+c_3}
\\\nonumber&&
\text{ and } a_4>\frac{a_5^2 c_1+a_5^2 c_3+2 a_6 a_5 c_1+4 a_7 a_5 c_1+2 a_6 a_5 c_3+4 a_7 a_5 c_3+a_6^2 c_1+4 a_7^2 c_1}{a_1 c_1+a_3 c_1+a_1 c_3+a_3 c_3-b_2^2-2 b_4 b_2-b_4^2}
\\\nonumber&+& 
\frac{4 a_6 a_7 c_1+a_6^2 c_3+4 a_7^2 c_3+4 a_6 a_7 c_3}{a_1 c_1+a_3 c_1+a_1 c_3+a_3 c_3-b_2^2-2 b_4 b_2-b_4^2}\text{ and } c_1>-c_3\text{ and } c_2>-c_1
\\\nonumber&& 
\text{ and } \mu _d^2>\frac{1}{2} \left(-a_5 v_{SM}^2-2 b_2 v_1^2-4 b_3 v_1^2\right)\text{ and } v_1>b_1\text{ and } v_{SM}>b_1 \text{ and } b_3\leq b_1\text{ and } b_4\leq -b_2
\\\nonumber&& 
\text{ and } a_6<\frac{a_5 v_{SM}^2+a_7 v_{SM}^2+2 b_2 v_1^2+4 b_3 v_1^2+2 \mu _d^2}{2 v_{SM}^2}
\text{ and } a_7\leq \frac{-3 a_5 v_{SM}^2-2 b_2 v_1^2-4 b_3 v_1^2-2 \mu _d^2}{5 v_{SM}^2}
\end{eqnarray*}

\newpage
\begin{center}
\fbox{{ If $a5+a6+2 a7\leq 0$, $b2-b4\leq 0$ and $0\leq 0$, then: }}
\end{center}

\begin{eqnarray*}
\nonumber
&& a_1>\frac{-a_3 c_1-a_3 c_3+b_2^2-2 b_4 b_2+b_4^2}{c_1+c_3}
\\\nonumber&& 
\text{ and } a_4>\frac{a_5^2 c_1+a_5^2 c_3+2 a_6 a_5 c_1+4 a_7 a_5 c_1+2 a_6 a_5 c_3+4 a_7 a_5 c_3+a_6^2 c_1+4 a_7^2 c_1}{a_1 c_1+a_3 c_1+a_1 c_3+a_3 c_3-b_2^2+2 b_4 b_2-b_4^2}
\\\nonumber&+& 
\frac{4 a_6 a_7 c_1+a_6^2 c_3+4 a_7^2 c_3+4 a_6 a_7 c_3}{a_1 c_1+a_3 c_1+a_1 c_3+a_3 c_3-b_2^2+2 b_4 b_2-b_4^2}
\text{ and } a_7>\frac{-3 a_5 v_{SM}^2-2 b_2 v_1^2-4 b_3 v_1^2-2 \mu _d^2}{5 v_{SM}^2}\text{ and } c_1>-c_3\text{ and } c_2>-c_1
\\\nonumber&& 
\text{ and } \mu _d^2>\frac{1}{2} \left(-a_5 v_{SM}^2-2 b_2 v_1^2-4 b_3 v_1^2\right)\text{ and } v_1>b_1\text{ and } v_{SM}>b_1
\text{ and } b_4\geq b_2\text{ and } a_6\leq -a_5-2 a_7\text{ and } b_3\leq b_1
\\\nonumber&& 
\textbf{or}  
\\\nonumber&&
 a_1>\frac{-a_3 c_1-a_3 c_3+b_2^2-2 b_4 b_2+b_4^2}{c_1+c_3}
 \\\nonumber&& 
 \text{ and } a_4>\frac{a_5^2 c_1+a_5^2 c_3+2 a_6 a_5 c_1+4 a_7 a_5 c_1+2 a_6 a_5 c_3+4 a_7 a_5 c_3+a_6^2 c_1+4 a_7^2 c_1}{a_1 c_1+a_3 c_1+a_1 c_3+a_3 c_3-b_2^2+2 b_4 b_2-b_4^2}
 \\\nonumber&+& 
 \frac{4 a_6 a_7 c_1+a_6^2 c_3+4 a_7^2 c_3+4 a_6 a_7 c_3}{a_1 c_1+a_3 c_1+a_1 c_3+a_3 c_3-b_2^2+2 b_4 b_2-b_4^2} \text{ and } a_6\leq -a_5-2 a_7
 \\\nonumber&& 
 \text{ and } a_7>\frac{-3 a_5 v_{SM}^2-2 b_2 v_1^2+4 b_3 v_1^2-2 \mu _d^2}{5 v_{SM}^2}\text{ and } b_3>b_1
 \\\nonumber&& 
 \text{ and } c_1>-c_3\text{ and } c_2>-c_1\text{ and } \mu _d^2>\frac{1}{2} \left(-a_5 v_{SM}^2-2 b_2 v_1^2+4 b_3 v_1^2\right)
 \text{ and } v_1>b_1\text{ and } v_{SM}>b_1\text{ and } b_4\geq b_2
 \\\nonumber&& 
\textbf{or}  
\\\nonumber&&
 a_1>\frac{-a_3 c_1-a_3 c_3+b_2^2-2 b_4 b_2+b_4^2}{c_1+c_3}
 \\\nonumber&& 
 \text{ and } a_4>\frac{a_5^2 c_1+a_5^2 c_3+2 a_6 a_5 c_1+4 a_7 a_5 c_1+2 a_6 a_5 c_3+4 a_7 a_5 c_3+a_6^2 c_1+4 a_7^2 c_1}{a_1 c_1+a_3 c_1+a_1 c_3+a_3 c_3-b_2^2+2 b_4 b_2-b_4^2}
 \\\nonumber&+& 
 \frac{4 a_6 a_7 c_1+a_6^2 c_3+4 a_7^2 c_3+4 a_6 a_7 c_3}{a_1 c_1+a_3 c_1+a_1 c_3+a_3 c_3-b_2^2+2 b_4 b_2-b_4^2}
 \\\nonumber&& 
 \text{ and } b_3>b_1\text{ and } c_1>-c_3\text{ and } c_2>-c_1\text{ and } \mu _d^2>\frac{1}{2} \left(-a_5 v_{SM}^2-2 b_2 v_1^2+4 b_3 v_1^2\right)
 \\\nonumber&& 
 \text{ and } v_1>b_1\text{ and } v_{SM}>b_1\text{ and } b_4\geq b_2\text{ and } a_6<\frac{a_5 v_{SM}^2+a_7 v_{SM}^2+2 b_2 v_1^2-4 b_3 v_1^2+2 \mu _d^2}{2 v_{SM}^2}
 \\\nonumber&& 
 \text{ and } a_7\leq \frac{-3 a_5 v_{SM}^2-2 b_2 v_1^2+4 b_3 v_1^2-2 \mu _d^2}{5 v_{SM}^2}
\\\nonumber&& 
\textbf{or}  
\\\nonumber&&
 a_1>\frac{-a_3 c_1-a_3 c_3+b_2^2-2 b_4 b_2+b_4^2}{c_1+c_3}
 \\\nonumber&& 
 \text{ and } a_4>\frac{a_5^2 c_1+a_5^2 c_3+2 a_6 a_5 c_1+4 a_7 a_5 c_1+2 a_6 a_5 c_3+4 a_7 a_5 c_3+a_6^2 c_1+4 a_7^2 c_1+4 a_6 a_7 c_1}{a_1 c_1+a_3 c_1+a_1 c_3+a_3 c_3-b_2^2+2 b_4 b_2-b_4^2}
 \\\nonumber&+& 
 \frac{a_6^2 c_3+4 a_7^2 c_3+4 a_6 a_7 c_3}{a_1 c_1+a_3 c_1+a_1 c_3+a_3 c_3-b_2^2+2 b_4 b_2-b_4^2}\text{ and } c_1>-c_3
 \\\nonumber&& 
 \text{ and } c_2>-c_1\text{ and } \mu _d^2>\frac{1}{2} \left(-a_5 v_{SM}^2-2 b_2 v_1^2-4 b_3 v_1^2\right)\text{ and } v_1>b_1
 \\\nonumber&& 
 \text{ and } v_{SM}>b_1\text{ and } b_4\geq b_2\text{ and } a_6<\frac{a_5 v_{SM}^2+a_7 v_{SM}^2+2 b_2 v_1^2+4 b_3 v_1^2+2 \mu _d^2}{2 v_{SM}^2}
 \\\nonumber&& 
 \text{ and } a_7\leq \frac{-3 a_5 v_{SM}^2-2 b_2 v_1^2-4 b_3 v_1^2-2 \mu _d^2}{5 v_{SM}^2}\text{ and } b_3\leq b_1
\end{eqnarray*}

\begin{center}
\fbox{{ If $2 a1+2 a3\leq 0$, $b2-b4\leq 0$ and $b2+b4\leq 0$, then: }}
\end{center}
There is no solution.

\newpage
\begin{center}
\fbox{{\bf { \textbf{Case 6.} $A_{ij}\leq0$, $A_{ik}\leq0$, $A_{il}\leq0$, and the other entries positive.
}}}
\end{center}
\begin{center}
\fbox{{ If $a5+a6+2 a7\leq 0$, $a5+a6+2 a7\leq 0$ and $0\leq 0$, then:}}
\end{center}

\begin{eqnarray*}
\nonumber
&& -a_6-2 a_7=a_5\text{ and } b_4=b_1\text{ and } a_1>-a_3\text{ and } a_4>b_1\text{ and } b_2>b_1\text{ and } b_3>b_1 \text{ and } c_1<\frac{A}{B}
\text{ and } -c_3<c_1
\\\nonumber&& 
\text{ and } c_2>-c_1\text{ and } \mu _d^2>\frac{1}{2} \left(-a_5 v_{SM}^2-2 b_2 v_1^2+4 b_3 v_1^2\right)\text{ and } v_1>b_1 
\text{ and } v_{SM}>b_1\text{ and } a_6<\frac{a_7}{2}
\\\nonumber&& 
\textbf{or}  
\\\nonumber&&
 -a_6-2 a_7=a_5\text{ and } b_4=b_1\text{ and } a_1>-a_3\text{ and } a_4>b_1\text{ and } b_2>b_1\text{ and } b_3>b_1\text{ and } c_2>-c_1  \text{ and } -c_3<c_1
 \\\nonumber&& 
 \text{ and } \mu _d^2>\frac{1}{2} \left(-a_5 v_{SM}^2+2 a_6 v_{SM}^2-a_7 v_{SM}^2-2 b_2 v_1^2+4 b_3 v_1^2\right)\text{ and } v_1>b_1 
 \text{ and } v_{SM}>b_1\text{ and } a_6\geq \frac{a_7}{2}  \text{ and } c_1<\frac{A}{B}
\\\nonumber&& 
\textbf{or}  
\\\nonumber&&
-a_6-2 a_7=a_5\text{ and } b_4=b_1\text{ and } a_1>-a_3\text{ and } a_4>b_1\text{ and } b_2>b_1\text{ and } c_2>-c_1 \text{ and } c_1<\frac{A}{B}
\text{ and } -c_3<c_1
\\\nonumber&& 
\text{ and } \mu _d^2>\frac{1}{2} \left(-a_5 v_{SM}^2-2 b_2 v_1^2-4 b_3 v_1^2\right)
\text{ and } v_1>b_1\text{ and } v_{SM}>b_1\text{ and } a_6<\frac{a_7}{2} \text{ and } b_3\leq b_1 
\\\nonumber&& 
\textbf{or}  
\\\nonumber&& 
-a_6-2 a_7=a_5\text{ and } b_4=b_1\text{ and } a_1>-a_3\text{ and } a_4>b_1\text{ and } b_2>b_1\text{ and } c_2>-c_1 \text{ and } -c_3<c_1\text{ and } b_3\leq b_1
\\\nonumber&& 
\text{ and } \mu _d^2>\frac{1}{2} \left(-a_5 v_{SM}^2+2 a_6 v_{SM}^2-a_7 v_{SM}^2-2 b_2 v_1^2-4 b_3 v_1^2\right)
\text{ and } v_1>b_1\text{ and } v_{SM}>b_1\text{ and } a_6\geq \frac{a_7}{2}
\text{ and } c_1<\frac{A}{B}
\\\nonumber&&
\textbf{or}  
\\\nonumber&&
-a_6-2 a_7=a_5\text{ and } a_1>-a_3\text{ and } a_4>b_1\text{ and } b_2>\sqrt{3} \sqrt{b_4^2}\text{ and } b_3>b_1\text{ and } b_4>b_1 \text{ and } c_1<\frac{C}{B}
\text{ and } -c_3<c_1 
\\\nonumber&& 
\text{ and } c_2>-c_1\text{ and } \mu _d^2>\frac{1}{2} \left(-a_5 v_{SM}^2-2 b_2 v_1^2+4 b_3 v_1^2\right)
\text{ and } v_1>b_1\text{ and } v_{SM}>b_1\text{ and } a_6<\frac{a_7}{2}
\\\nonumber&& 
\textbf{or}  
\\\nonumber&& 
-a_6-2 a_7=a_5\text{ and } a_1>-a_3\text{ and } a_4>b_1\text{ and } b_2>\sqrt{3} \sqrt{b_4^2}\text{ and } b_3>b_1 \text{ and } c_1<\frac{C}{B}
\text{ and } -c_3<c_1 \text{ and } a_6\geq \frac{a_7}{2}
\\\nonumber&& 
\text{ and } b_4>b_1\text{ and } c_2>-c_1\text{ and } \mu _d^2>\frac{1}{2} \left(-a_5 v_{SM}^2+2 a_6 v_{SM}^2-a_7 v_{SM}^2-2 b_2 v_1^2+4 b_3 v_1^2\right)
\text{ and } v_1>b_1\text{ and } v_{SM}>b_1
\\\nonumber&& 
\textbf{or}  
\\\nonumber&& 
-a_6-2 a_7=a_5\text{ and } a_1>-a_3\text{ and } a_4>b_1\text{ and } b_2>\sqrt{3} \sqrt{b_4^2}
\text{ and } b_3>b_1\text{ and } c_2>-c_1 \text{ and } c_1<\frac{C}{B}
\text{ and } -c_3<c_1
\\\nonumber&&
\text{ and } \mu _d^2>\frac{1}{2} \left(-a_5 v_{SM}^2-2 b_2 v_1^2+4 b_3 v_1^2\right)\text{ and } v_1>b_1 
\text{ and } v_{SM}>b_1\text{ and } a_6<\frac{a_7}{2}
\text{ and } b_4<b_1
\\\nonumber&&
\textbf{or}  
\\\nonumber&& 
 -a_6-2 a_7=a_5\text{ and } a_1>-a_3\text{ and } a_4>b_1\text{ and } b_2>\sqrt{3} \sqrt{b_4^2}\text{ and } b_3>b_1 \text{ and } c_1<\frac{C}{B}
\text{ and } -c_3<c_1 \text{ and } b_4<b_1 
\\\nonumber&& 
\text{ and } \mu _d^2>\frac{1}{2} \left(-a_5 v_{SM}^2+2 a_6 v_{SM}^2-a_7 v_{SM}^2-2 b_2 v_1^2+4 b_3 v_1^2\right)
\text{ and } v_1>b_1
\text{ and } v_{SM}>b_1\text{ and } a_6\geq \frac{a_7}{2} \text{ and } c_2>-c_1
\\\nonumber&& 
\textbf{or}  
\\\nonumber&& 
-a_6-2 a_7=a_5\text{ and } a_1>-a_3\text{ and } a_4>b_1\text{ and } b_2>\sqrt{3} \sqrt{b_4^2}\mu _d^2>\frac{1}{2} \left(-a_5 v_{SM}^2-2 b_2 v_1^2-4 b_3 v_1^2\right)
\\\nonumber&& 
\text{ and } v_1>b_1\text{ and } v_{SM}>b_1\text{ and } a_6<\frac{a_7}{2}
\text{ and } c_1<\frac{C}{B}
\text{ and } -c_3<c_1\text{ and } b_3\leq b_1 \text{ and } b_4>b_1 
\text{ and } c_2>-c_1\text{ and } 
\\\nonumber&& 
\textbf{or}  
\\\nonumber&& 
 -a_6-2 a_7=a_5\text{ and } a_1>-a_3\text{ and } a_4>b_1
\text{ and } \mu _d^2>\frac{1}{2} \left(-a_5 v_{SM}^2+2 a_6 v_{SM}^2-a_7 v_{SM}^2-2 b_2 v_1^2-4 b_3 v_1^2\right)  \text{ and } c_1<\frac{C}{B}
\\\nonumber&& 
\text{ and } v_1>b_1\text{ and } v_{SM}>b_1\text{ and } a_6\geq \frac{a_7}{2}
\text{ and } -c_3<c_1\text{ and } b_3\leq b_1
\text{ and } b_2>\sqrt{3} \sqrt{b_4^2}
\text{ and } b_4>b_1
\text{ and } c_2>-c_1 
\\\nonumber&& 
\textbf{or}  
\\\nonumber&& 
-a_6-2 a_7=a_5\text{ and } a_1>-a_3\text{ and } a_4>b_1\text{ and } b_2>\sqrt{3} \sqrt{b_4^2}
\text{ and } c_2>-c_1\text{ and } \mu _d^2>\frac{1}{2} \left(-a_5 v_{SM}^2-2 b_2 v_1^2-4 b_3 v_1^2\right)\\\nonumber&& 
\text{ and } v_{SM}>b_1\text{ and } a_6<\frac{a_7}{2}\text{ and } b_4<b_1
\text{ and } c_1<\frac{C}{B}
\text{ and } -c_3<c_1\text{ and } b_3\leq b_1 \text{ and } v_1>b_1
\end{eqnarray*}

\begin{eqnarray*}
\\\nonumber&& 
\textbf{or}  
\\\nonumber&& a_4>b_1\text{ and } b_2>\sqrt{3} \sqrt{b_4^2}\text{ and } c_2>-c_1 
\text{ and } \mu _d^2>\frac{1}{2} \left(-a_5 v_{SM}^2+2 a_6 v_{SM}^2-a_7 v_{SM}^2-2 b_2 v_1^2-4 b_3 v_1^2\right) \text{ and } v_1>b_1
\\\nonumber&& 
\text{ and } a_6\geq \frac{a_7}{2}\text{ and } b_4<b_1 \text{ and } v_{SM}>b_1
\text{ and } c_1<\frac{C}{B}
\text{ and } -c_3<c_1\text{ and } b_3\leq b_1  \text{ and }-a_6-2 a_7=a_5\text{ and } a_1>-a_3
 \\\nonumber&& 
\textbf{or}  
\\\nonumber&& 
b_4=b_1\text{ and } a_1>-a_3
\text{ and } b_2>b_1 \text{ and } c_1<\frac{A}{B}
\text{ and } -c_3<c_1  \text{ and } b_3>b_1\text{ and } c_2>-c_1\text{ and } a_5<-a_6-2 a_7\text{ and } a_6\!<\!\frac{a_7}{2} 
\\\nonumber&& 
\text{ and } \mu _d^2>\frac{1}{2} \left(-a_5 v_{SM}^2-2 b_2 v_1^2+4 b_3 v_1^2\right)\text{ and } v_1>b_1
\text{ and } a_4>\frac{a_5^2+2 a_6 a_5+4 a_7 a_5+a_6^2+4 a_7^2+4 a_6 a_7}{a_1+a_3}  \text{ and } v_{SM}>b_1
\\\nonumber&& 
\textbf{or}  
\\\nonumber&&
b_4=b_1\text{ and } a_1>-a_3\text{ and } a_4>\frac{a_5^2+2 a_6 a_5+4 a_7 a_5+a_6^2+4 a_7^2+4 a_6 a_7}{a_1+a_3}\text{ and } b_2>b_1
\\\nonumber&& 
\text{ and } b_3>b_1\text{ and } c_2>-c_1
\text{ and } \mu _d^2>\frac{1}{2} \left(-a_5 v_{SM}^2+2 a_6 v_{SM}^2-a_7 v_{SM}^2-2 b_2 v_1^2+4 b_3 v_1^2\right)
\text{ and } v_1>b_1\text{ and } v_{SM}>b_1
\\\nonumber&& 
\text{ and } -c_3<c_1 \text{ and } a_6\geq \frac{a_7}{2}\text{ and } a_5<-a_6-2 a_7
\text{ and } c_1<\frac{A}{B}
\\\nonumber&& 
\textbf{or}  
\\\nonumber&& 
b_4=b_1\text{ and } a_1>-a_3\text{ and } a_4>\frac{a_5^2+2 a_6 a_5+4 a_7 a_5+a_6^2+4 a_7^2+4 a_6 a_7}{a_1+a_3} 
\\\nonumber&& 
\text{ and } c_1<\frac{A}{B}
\text{ and } -c_3<c_1\text{ and } b_3\leq b_1 \text{ and } a_6<\frac{a_7}{2} 
\\\nonumber&& 
\text{ and } b_2>b_1\text{ and } c_2>-c_1\text{ and } \mu _d^2>\frac{1}{2} \left(-a_5 v_{SM}^2-2 b_2 v_1^2-4 b_3 v_1^2\right)\text{ and } v_1>b_1 
\text{ and } v_{SM}>b_1\text{ and } a_5<-a_6-2 a_7
 \\\nonumber&& 
\textbf{or}  
\\\nonumber&& 
 b_4=b_1\text{ and } a_1>-a_3\text{ and } a_4>\frac{a_5^2+2 a_6 a_5+4 a_7 a_5+a_6^2+4 a_7^2+4 a_6 a_7}{a_1+a_3}
\\\nonumber&& 
\text{ and } b_2>b_1\text{ and } c_2>-c_1\text{ and } \mu _d^2>\frac{1}{2} \left(-a_5 v_{SM}^2+2 a_6 v_{SM}^2-a_7 v_{SM}^2-2 b_2 v_1^2-4 b_3 v_1^2\right)
\\\nonumber&& 
\text{ and } v_1>b_1\text{ and } v_{SM}>b_1 
\text{ and } a_6\geq \frac{a_7}{2}\text{ and } a_5<-a_6-2 a_7
\text{ and } c_1<\frac{A}{B}
\text{ and } -c_3<c_1\text{ and } b_3\leq b_1
\\\nonumber&&
\textbf{or}  
\\\nonumber&& 
\sqrt{3} \sqrt{b_4^2}=b_2\text{ and } a_1>-a_3\text{ and } a_4>\frac{a_5^2+2 a_6 a_5+4 a_7 a_5+a_6^2+4 a_7^2+4 a_6 a_7}{a_1+a_3} \text{ and } c_1<\frac{C}{B} 
\text{ and } -c_3<c_1
\\\nonumber&& 
\text{ and } a_6>\frac{a_7}{2}\text{ and } b_3>b_1\text{ and } b_4>b_1\text{ and } c_2>-c_1
\\\nonumber&& 
\text{ and } \mu _d^2>\frac{1}{2} \left(-a_5 v_{SM}^2+2 a_6 v_{SM}^2-a_7 v_{SM}^2-2 b_2 v_1^2+4 b_3 v_1^2\right)
\text{ and } v_1>b_1\text{ and } v_{SM}>b_1\text{ and } a_5<-a_6-2 a_7 
\\\nonumber&& 
\textbf{or}  
\\\nonumber&& 
\sqrt{3} \sqrt{b_4^2}=b_2\text{ and } a_1>-a_3\text{ and } a_4>\frac{a_5^2+2 a_6 a_5+4 a_7 a_5+a_6^2+4 a_7^2+4 a_6 a_7}{a_1+a_3}
\\\nonumber&& 
\text{ and } a_6>\frac{a_7}{2}\text{ and } b_3>b_1
\text{ and } c_2>-c_1\text{ and } \mu _d^2>\frac{1}{2} \left(-a_5 v_{SM}^2+2 a_6 v_{SM}^2-a_7 v_{SM}^2-2 b_2 v_1^2+4 b_3 v_1^2\right)
\\\nonumber&& 
\text{ and } v_1>b_1\text{ and } v_{SM}>b_1\text{ and } a_5<-a_6-2 a_7\text{ and } b_4<b_1
\text{ and } c_1<\frac{C}{B}
\text{ and } -c_3<c_1
 \\\nonumber&& 
\textbf{or}  
\\\nonumber&& 
\sqrt{3} \sqrt{b_4^2}=b_2\text{ and } a_1>-a_3\text{ and } a_4>\frac{a_5^2+2 a_6 a_5+4 a_7 a_5+a_6^2+4 a_7^2+4 a_6 a_7}{a_1+a_3}
\\\nonumber&& 
\text{ and } a_6>\frac{a_7}{2}\text{ and } b_4>b_1\text{ and } c_2>-c_1
\\\nonumber&& 
\text{ and } \mu _d^2>\frac{1}{2} \left(-a_5 v_{SM}^2+2 a_6 v_{SM}^2-a_7 v_{SM}^2-2 b_2 v_1^2-4 b_3 v_1^2\right)\text{ and } v_1>b_1
\\\nonumber&& 
\text{ and } v_{SM}>b_1\text{ and } a_5<-a_6-2 a_7
\\\nonumber&& 
\text{ and } c_1<\frac{C}{B}
\\\nonumber&& 
\text{ and } -c_3<c_1\text{ and } b_3\leq b_1
\end{eqnarray*}

\begin{eqnarray*}
&& 
\textbf{or}  
\\\nonumber&& 
\sqrt{3} \sqrt{b_4^2}=b_2\text{ and } a_1>-a_3\text{ and } a_4>\frac{a_5^2+2 a_6 a_5+4 a_7 a_5+a_6^2+4 a_7^2+4 a_6 a_7}{a_1+a_3} \text{ and } c_1<\frac{C}{B}
\text{ and } -c_3<c_1\text{ and } b_3\leq b_1
\\\nonumber&& 
\text{ and } a_6>\frac{a_7}{2}\text{ and } c_2>-c_1 
\text{ and } \mu _d^2>\frac{1}{2} \left(-a_5 v_{SM}^2+2 a_6 v_{SM}^2-a_7 v_{SM}^2-2 b_2 v_1^2-4 b_3 v_1^2\right)
\\\nonumber&& 
\text{ and } v_1>b_1\text{ and } v_{SM}>b_1
\text{ and } a_5<-a_6-2 a_7\text{ and } b_4<b_1
\\\nonumber&& 
\textbf{or}  
\\\nonumber&& 
\sqrt{3} \sqrt{b_4^2}=b_2\text{ and } a_1>-a_3
\text{ and } a_4>\frac{a_5^2+2 a_6 a_5+4 a_7 a_5+a_6^2+4 a_7^2+4 a_6 a_7}{a_1+a_3} \text{ and } c_1<\frac{C}{B}
\text{ and } -c_3<c_1\text{ and } a_6\leq \frac{a_7}{2}
\\\nonumber&& 
\text{ and } b_3>b_1\text{ and } b_4>b_1\text{ and } c_2>-c_1\text{ and } \mu _d^2>\frac{1}{2} \left(-a_5 v_{SM}^2-2 b_2 v_1^2+4 b_3 v_1^2\right)
\\\nonumber&& 
\text{ and } v_1>b_1\text{ and } v_{SM}>b_1\text{ and } a_5<-a_6-2 a_7
\\\nonumber&&
\textbf{or}  
\\\nonumber&&
 \sqrt{3} \sqrt{b_4^2}=b_2\text{ and } a_1>-a_3\text{ and } a_4>\frac{a_5^2+2 a_6 a_5+4 a_7 a_5+a_6^2+4 a_7^2+4 a_6 a_7}{a_1+a_3}
 \\\nonumber&& 
 \text{ and } b_3>b_1\text{ and } c_2>-c_1
 \text{ and } \mu _d^2>\frac{1}{2} \left(-a_5 v_{SM}^2-2 b_2 v_1^2+4 b_3 v_1^2\right)\text{ and } v_1>b_1\text{ and } v_{SM}>b_1
 \\\nonumber&& 
 \text{ and } a_5<-a_6-2 a_7\text{ and } b_4<b_1
 \text{ and } c_1<\frac{C}{B}
 \text{ and } -c_3<c_1\text{ and } a_6\leq \frac{a_7}{2}
\\\nonumber&& 
\textbf{or}  
\\\nonumber&& 
\sqrt{3} \sqrt{b_4^2}=b_2\text{ and } a_1>-a_3
\text{ and } a_4>\frac{a_5^2+2 a_6 a_5+4 a_7 a_5+a_6^2+4 a_7^2+4 a_6 a_7}{a_1+a_3} \text{ and } v_{SM}>b_1\text{ and } a_5<-a_6-2 a_7 
\\\nonumber&& 
\text{ and } b_4>b_1\text{ and } c_2>-c_1\text{ and } \mu _d^2>\frac{1}{2} \left(-a_5 v_{SM}^2-2 b_2 v_1^2-4 b_3 v_1^2\right)\text{ and } v_1>b_1
\text{ and } -c_3<c_1\text{ and } a_6\leq \frac{a_7}{2}\text{ and } b_3\leq b_1
\\\nonumber&& 
\text{ and } c_1<\frac{C}{B}
\\\nonumber&& 
\textbf{or}  
\\\nonumber&& 
\sqrt{3} \sqrt{b_4^2}=b_2\text{ and } a_1>-a_3\text{ and } a_4>\frac{a_5^2+2 a_6 a_5+4 a_7 a_5+a_6^2+4 a_7^2+4 a_6 a_7}{a_1+a_3}
\\\nonumber&& 
\text{ and } c_2>-c_1
\text{ and } \mu _d^2>\frac{1}{2} \left(-a_5 v_{SM}^2-2 b_2 v_1^2-4 b_3 v_1^2\right)
\text{ and } v_1>b_1\text{ and } v_{SM}>b_1\text{ and } a_5<-a_6-2 a_7\text{ and } b_4<b_1
\\\nonumber&& 
\text{ and } c_1<\frac{C}{B}
\text{ and } -c_3<c_1\text{ and } a_6\leq \frac{a_7}{2}\text{ and } b_3\leq b_1
\\\nonumber&& 
\textbf{or}  
\\\nonumber&& 
a_1>-a_3\text{ and } a_4>\frac{a_5^2+2 a_6 a_5+4 a_7 a_5+a_6^2+4 a_7^2+4 a_6 a_7}{a_1+a_3}\text{ and } b_2>\sqrt{3} \sqrt{b_4^2}
\text{ and } c_1<\frac{C}{B}
\text{ and } -c_3<c_1
\\\nonumber&& 
\text{ and } b_3>b_1\text{ and } b_4>b_1\text{ and } c_2>-c_1\text{ and } \mu _d^2>\frac{1}{2} \left(-a_5 v_{SM}^2-2 b_2 v_1^2+4 b_3 v_1^2\right)
\text{ and } v_1>b_1
\\\nonumber&& 
\text{ and } v_{SM}>b_1\text{ and } a_5<-a_6-2 a_7
\text{ and } a_6<\frac{a_7}{2}
\\\nonumber&& 
\textbf{or}  
\\\nonumber&& 
a_1>-a_3\text{ and } a_4>\frac{a_5^2+2 a_6 a_5+4 a_7 a_5+a_6^2+4 a_7^2+4 a_6 a_7}{a_1+a_3}
\text{ and } b_2>\sqrt{3} \sqrt{b_4^2}\text{ and } b_3>b_1\text{ and } b_4>b_1\text{ and } c_2>-c_1
\\\nonumber&& 
\text{ and } \mu _d^2>\frac{1}{2} \left(-a_5 v_{SM}^2+2 a_6 v_{SM}^2-a_7 v_{SM}^2-2 b_2 v_1^2+4 b_3 v_1^2\right)\text{ and } v_1>b_1
\text{ and } v_{SM}>b_1\text{ and } a_6\geq \frac{a_7}{2}
\\\nonumber&& 
\text{ and } c_1<\frac{C}{B}
\text{ and } -c_3<c_1
\text{ and } a_5<-a_6-2 a_7
 \\\nonumber&& 
\textbf{or}  
\\\nonumber&& 
a_1>-a_3\text{ and } a_4>\frac{a_5^2+2 a_6 a_5+4 a_7 a_5+a_6^2+4 a_7^2+4 a_6 a_7}{a_1+a_3} 
\text{ and } b_2>\sqrt{3} \sqrt{b_4^2}\text{ and } b_3>b_1
\\\nonumber&& 
\text{ and } c_2>-c_1
\text{ and } \mu _d^2>\frac{1}{2} \left(-a_5 v_{SM}^2-2 b_2 v_1^2+4 b_3 v_1^2\right)
\\\nonumber&& 
\text{ and } v_1>b_1\text{ and } v_{SM}>b_1\text{ and } a_5<-a_6-2 a_7\text{ and } a_6<\frac{a_7}{2}\text{ and } b_4<b_1
\text{ and } c_1<\frac{C}{B} 
\text{ and } -c_3<c_1
\end{eqnarray*}

\begin{eqnarray*}
&& 
\textbf{or}  
\\\nonumber&& 
a_1>-a_3\text{ and } a_4>\frac{a_5^2+2 a_6 a_5+4 a_7 a_5+a_6^2+4 a_7^2+4 a_6 a_7}{a_1+a_3}
\text{ and } b_2>\sqrt{3} \sqrt{b_4^2}\text{ and } b_3>b_1 \text{ and } a_5<-a_6-2 
\\\nonumber&& 
\text{ and } c_2>-c_1\text{ and } \mu _d^2>\frac{1}{2} \left(-a_5 v_{SM}^2+2 a_6 v_{SM}^2-a_7 v_{SM}^2-2 b_2 v_1^2+4 b_3 v_1^2\right)
\text{ and } v_1>b_1\text{ and } v_{SM}>b_1
\\\nonumber&& 
\text{ and } c_1<\frac{C}{B}
\text{ and } -c_3<c_1
\text{ and } a_6 \geq \frac{a_7}{2} \text{ and } b_4<b_1
\\\nonumber&& 
\textbf{or}  
\\\nonumber&& 
a_1>-a_3\text{ and } a_4>\frac{a_5^2+2 a_6 a_5+4 a_7 a_5+a_6^2+4 a_7^2+4 a_6 a_7}{a_1+a_3}
\\\nonumber&& 
\text{ and } b_2>\sqrt{3} \sqrt{b_4^2}\text{ and } b_4>b_1\text{ and } c_2>-c_1
\text{ and } \mu _d^2>\frac{1}{2} \left(-a_5 v_{SM}^2-2 b_2 v_1^2-4 b_3 v_1^2\right)\text{ and } v_1>b_1\text{ and } v_{SM}>b_1
\\\nonumber&& 
\text{ and } a_5<-a_6-2 a_7\text{ and } a_6<\frac{a_7}{2}
\text{ and } c_1<\frac{C}{B} 
\text{ and } -c_3<c_1\text{ and } b_3\leq b_1
\\\nonumber&& 
\textbf{or}  
\\\nonumber&& 
a_1>-a_3\text{ and } a_4>\frac{a_5^2+2 a_6 a_5+4 a_7 a_5+a_6^2+4 a_7^2+4 a_6 a_7}{a_1+a_3}\text{ and } b_2>\sqrt{3} \sqrt{b_4^2}
\\\nonumber&& 
\text{ and } b_4>b_1\text{ and } c_2>-c_1\text{ and } \mu _d^2>\frac{1}{2} \left(-a_5 v_{SM}^2+2 a_6 v_{SM}^2-a_7 v_{SM}^2-2 b_2 v_1^2-4 b_3 v_1^2\right)
\\\nonumber&& 
\text{ and } v_1>b_1\text{ and } v_{SM}>b_1\text{ and } a_6\geq \frac{a_7}{2}\text{ and } a_5<-a_6-2 a_7
\text{ and } c_1<\frac{C}{B} 
\text{ and } -c_3<c_1\text{ and } b_3\leq b_1 
\\\nonumber&& 
\textbf{or}  
\\\nonumber&&
a_1>-a_3\text{ and } a_4>\frac{a_5^2+2 a_6 a_5+4 a_7 a_5+a_6^2+4 a_7^2+4 a_6 a_7}{a_1+a_3}\text{ and } b_2>\sqrt{3} \sqrt{b_4^2}
\\\nonumber&& 
\text{ and } c_2>-c_1\text{ and } \mu _d^2>\frac{1}{2} \left(-a_5 v_{SM}^2-2 b_2 v_1^2-4 b_3 v_1^2\right)\text{ and } v_1>b_1\text{ and } v_{SM}>b_1
\\\nonumber&& 
\text{ and } a_5<-a_6-2 a_7\text{ and } a_6<\frac{a_7}{2}\text{ and } b_4<b_1
\text{ and } c_1<\frac{C}{B}
\text{ and } -c_3<c_1\text{ and } b_3\leq b_1
\\\nonumber&& 
\textbf{or}  
\\\nonumber&&
 a_1>-a_3\text{ and } a_4>\frac{a_5^2+2 a_6 a_5+4 a_7 a_5+a_6^2+4 a_7^2+4 a_6 a_7}{a_1+a_3}
 \\\nonumber&& 
 \text{ and } b_2>\sqrt{3} \sqrt{b_4^2}\text{ and } c_2>-c_1 
 \text{ and } \mu _d^2>\frac{1}{2} \left(-a_5 v_{SM}^2+2 a_6 v_{SM}^2-a_7 v_{SM}^2-2 b_2 v_1^2-4 b_3 v_1^2\right)\text{ and } v_1>b_1
 \\\nonumber&& 
 \text{ and } v_{SM}>b_1\text{ and } a_6\geq \frac{a_7}{2}\text{ and } a_5<-a_6-2 a_7\text{ and } b_4<b_1
 \text{ and } c_1<\frac{C}{B}
  \text{ and } -c_3<c_1\text{ and } b_3\leq b_1
 \\\nonumber&& 
\textbf{or}  
\\\nonumber&&
a_1>-a_3\text{ and } a_6>\frac{a_7}{2}\text{ and } b_3>b_1
\\\nonumber&& 
\text{ and } c_2>-c_1\text{ and } \mu _d^2>\frac{1}{2} \left(-a_5 v_{SM}^2+2 a_6 v_{SM}^2-a_7 v_{SM}^2-2 b_2 v_1^2+4 b_3 v_1^2\right)
 \text{ and } v_1>b_1\text{ and } v_{SM}>b_1
 \\\nonumber&& 
 \text{ and } a_4<\frac{2 a_5^2 b_4^2+2 a_6^2 b_4^2+8 a_7^2 b_4^2+4 a_5 a_6 b_4^2+8 a_5 a_7 b_4^2+8 a_6 a_7 b_4^2}{-a_1 b_2^2-a_3 b_2^2+3 a_1 b_4^2+3 a_3 b_4^2}  \text{ and } b_4<b_1
 \text{ and } c_1<\frac{C}{B}  \text{ and } -c_3<c_1
 \\\nonumber&& 
 \text{ and } a_5<-a_6-2 a_7
 \text{ and } \frac{a_5^2+2 a_6 a_5+4 a_7 a_5+a_6^2+4 a_7^2+4 a_6 a_7}{a_1+a_3}<a_4\text{ and } b_2<\sqrt{3} \sqrt{b_4^2}\text{ and } -b_4<b_2
 \\\nonumber&& 
\textbf{or}  
\\\nonumber&& 
a_1>-a_3\text{ and } a_6>\frac{a_7}{2}\text{ and } c_2>-c_1
\\\nonumber&& 
\text{ and } \mu _d^2>\frac{1}{2} \left(-a_5 v_{SM}^2+2 a_6 v_{SM}^2-a_7 v_{SM}^2-2 b_2 v_1^2-4 b_3 v_1^2\right)
\text{ and } v_1>b_1\text{ and } v_{SM}>b_1
\\\nonumber&& 
\text{ and } a_4<\frac{2 a_5^2 b_4^2+2 a_6^2 b_4^2+8 a_7^2 b_4^2+4 a_5 a_6 b_4^2+8 a_5 a_7 b_4^2+8 a_6 a_7 b_4^2}{-a_1 b_2^2-a_3 b_2^2+3 a_1 b_4^2+3 a_3 b_4^2} \text{ and } c_1<\frac{C}{B}
\text{ and } -c_3<c_1\text{ and } b_3\leq b_1
\\\nonumber&& 
\text{ and } a_5<-a_6-2 a_7 
\text{ and } \frac{a_5^2+2 a_6 a_5+4 a_7 a_5+a_6^2+4 a_7^2+4 a_6 a_7}{a_1+a_3}<a_4
\text{ and } b_2<\sqrt{3} \sqrt{b_4^2}
\text{ and } -b_4<b_2\text{ and } b_4<b_1
 \end{eqnarray*}

\begin{eqnarray*}
\\\nonumber&& 
\textbf{or}  
\\\nonumber&& 
a_1>-a_3\text{ and } b_3>b_1\text{ and } b_4>b_1\text{ and } c_2>-c_1\text{ and } \mu _d^2>\frac{1}{2} \left(-a_5 v_{SM}^2-2 b_2 v_1^2+4 b_3 v_1^2\right)\text{ and } v_1>b_1\\\nonumber&& 
\text{ and } v_{SM}>b_1
\\\nonumber&& 
\text{ and } a_4<\frac{2 a_5^2 b_4^2+2 a_6^2 b_4^2+8 a_7^2 b_4^2+4 a_5 a_6 b_4^2+8 a_5 a_7 b_4^2+8 a_6 a_7 b_4^2}{-a_1 b_2^2-a_3 b_2^2+3 a_1 b_4^2+3 a_3 b_4^2} \text{ and } b_2<\sqrt{3} \sqrt{b_4^2}\
text{ and } b_4<b_2
\\\nonumber&& 
\text{ and } a_5<-a_6-2 a_7
\text{ and } a_6<\frac{a_7}{2}\text{ and } \frac{a_5^2+2 a_6 a_5+4 a_7 a_5+a_6^2+4 a_7^2+4 a_6 a_7}{a_1+a_3}<a_4 \text{ and } c_1<\frac{C}{B}
\text{ and } -c_3<c_1
 \\\nonumber&& 
\textbf{or}  
\\\nonumber&&
 a_1>-a_3\text{ and } b_3>b_1\text{ and } b_4>b_1
 \text{ and } c_2>-c_1\text{ and } \mu _d^2>\frac{1}{2} \left(-a_5 v_{SM}^2+2 a_6 v_{SM}^2-a_7 v_{SM}^2-2 b_2 v_1^2+4 b_3 v_1^2\right)\text{ and } v_1>b_1
 \\\nonumber&& 
 \text{ and } v_{SM}>b_1\text{ and } a_6\geq \frac{a_7}{2}
 \text{ and } a_4<\frac{2 a_5^2 b_4^2+2 a_6^2 b_4^2+8 a_7^2 b_4^2+4 a_5 a_6 b_4^2+8 a_5 a_7 b_4^2+8 a_6 a_7 b_4^2}{-a_1 b_2^2-a_3 b_2^2+3 a_1 b_4^2+3 a_3 b_4^2} \text{ and } c_1<\frac{C}{B}
 \text{ and } -c_3<c_1
 \\\nonumber&& 
 \text{ and } a_5<-a_6-2 a_7
 \text{ and } \frac{a_5^2+2 a_6 a_5+4 a_7 a_5+a_6^2+4 a_7^2+4 a_6 a_7}{a_1+a_3}<a_4\text{ and } b_2<\sqrt{3} \sqrt{b_4^2}
 \text{ and } b_4<b_2
  \\\nonumber&& 
\textbf{or}  
\\\nonumber&& 
a_1>-a_3\text{ and } b_3>b_1
\text{ and } c_2>-c_1\text{ and } \mu _d^2>\frac{1}{2} \left(-a_5 v_{SM}^2-2 b_2 v_1^2+4 b_3 v_1^2\right)
\\\nonumber&& 
\text{ and } v_1>b_1\text{ and } v_{SM}>b_1
\text{ and } a_4<\frac{2 a_5^2 b_4^2+2 a_6^2 b_4^2+8 a_7^2 b_4^2+4 a_5 a_6 b_4^2+8 a_5 a_7 b_4^2+8 a_6 a_7 b_4^2}{-a_1 b_2^2-a_3 b_2^2+3 a_1 b_4^2+3 a_3 b_4^2}
\\\nonumber&& 
\text{ and } a_5<-a_6-2 a_7
\text{ and } \frac{a_5^2+2 a_6 a_5+4 a_7 a_5+a_6^2+4 a_7^2+4 a_6 a_7}{a_1+a_3}<a_4\text{ and } b_2<\sqrt{3} \sqrt{b_4^2}
\\\nonumber&& 
\text{ and } -b_4<b_2\text{ and } b_4<b_1
\text{ and } c_1<\frac{C}{B}
\text{ and } -c_3<c_1\text{ and } a_6\leq \frac{a_7}{2}
\\\nonumber&& 
\textbf{or}  
\\\nonumber&& 
a_1>-a_3\text{ and } b_4>b_1\text{ and } c_2>-c_1
\text{ and } \mu _d^2>\frac{1}{2} \left(-a_5 v_{SM}^2-2 b_2 v_1^2-4 b_3 v_1^2\right)\text{ and } v_1>b_1 \text{ and } c_1<\frac{C}{B}
\text{ and } -c_3<c_1
\\\nonumber&& 
\text{ and } v_{SM}>b_1
\text{ and } a_4<\frac{2 a_5^2 b_4^2+2 a_6^2 b_4^2+8 a_7^2 b_4^2+4 a_5 a_6 b_4^2+8 a_5 a_7 b_4^2+8 a_6 a_7 b_4^2}{-a_1 b_2^2-a_3 b_2^2+3 a_1 b_4^2+3 a_3 b_4^2} \text{ and } b_3\leq b_1
\\\nonumber&& 
\text{ and } a_5<-a_6-2 a_7\text{ and } a_6<\frac{a_7}{2}
\text{ and } \frac{a_5^2+2 a_6 a_5+4 a_7 a_5+a_6^2+4 a_7^2+4 a_6 a_7}{a_1+a_3}<a_4 
\text{ and } b_2<\sqrt{3} \sqrt{b_4^2}\text{ and } b_4<b_2
\\\nonumber&& 
\textbf{or}  
\\\nonumber&&
 a_1>-a_3\text{ and } b_4>b_1\text{ and } c_2>-c_1
 \text{ and } \mu _d^2>\frac{1}{2} \left(-a_5 v_{SM}^2+2 a_6 v_{SM}^2-a_7 v_{SM}^2-2 b_2 v_1^2-4 b_3 v_1^2\right)
 \\\nonumber&& 
 \text{ and } v_1>b_1\text{ and } v_{SM}>b_1
 \text{ and } a_6\geq \frac{a_7}{2}
 \\\nonumber&& 
 \text{ and } a_4<\frac{2 a_5^2 b_4^2+2 a_6^2 b_4^2+8 a_7^2 b_4^2+4 a_5 a_6 b_4^2+8 a_5 a_7 b_4^2+8 a_6 a_7 b_4^2}{-a_1 b_2^2-a_3 b_2^2+3 a_1 b_4^2+3 a_3 b_4^2}
 \text{ and } a_5<-a_6-2 a_7
 \\\nonumber&& 
 \text{ and } \frac{a_5^2+2 a_6 a_5+4 a_7 a_5+a_6^2+4 a_7^2+4 a_6 a_7}{a_1+a_3}<a_4
 \text{ and } b_2<\sqrt{3} \sqrt{b_4^2}\text{ and } b_4<b_2 
 \text{ and } c_1<\frac{C}{B}
 \text{ and } -c_3<c_1\text{ and } b_3\leq b_1
\\\nonumber&& 
\textbf{or}  
\\\nonumber&& 
a_1>-a_3\text{ and } c_2>-c_1\text{ and } \mu _d^2>\frac{1}{2} \left(-a_5 v_{SM}^2-2 b_2 v_1^2-4 b_3 v_1^2\right)\text{ and } v_1>b_1
\\\nonumber&& 
\text{ and } v_{SM}>b_1
\text{ and } a_4<\frac{2 a_5^2 b_4^2+2 a_6^2 b_4^2+8 a_7^2 b_4^2+4 a_5 a_6 b_4^2+8 a_5 a_7 b_4^2+8 a_6 a_7 b_4^2}{-a_1 b_2^2-a_3 b_2^2+3 a_1 b_4^2+3 a_3 b_4^2}
\\\nonumber&& 
\text{ and } a_5<-a_6-2 a_7
\\\nonumber&& 
\text{ and } \frac{a_5^2+2 a_6 a_5+4 a_7 a_5+a_6^2+4 a_7^2+4 a_6 a_7}{a_1+a_3}<a_4\text{ and } b_2<\sqrt{3} \sqrt{b_4^2}\text{ and } -b_4<b_2
\text{ and } b_4<b_1
\\\nonumber&& 
\text{ and } c_1<\frac{C}{B}
\text{ and } -c_3<c_1\text{ and } a_6\leq \frac{a_7}{2}\text{ and } b_3\leq b_1
\end{eqnarray*}

\begin{center}
\fbox{{If $a5+a6+2 a7\leq 0$, $2 a1+2 a3\leq 0$ and $b2+b4\leq 0$, then:}}
\end{center}
 There is no solution.

\begin{center}
\fbox{{If $a5+a6+2 a7\leq 0$, $2 a1+2 a3\leq 0$ and $b2-b4\leq 0$, then:}}
\end{center}
There is no solution.
\\

where we made $A = 2 a_4 a_1 b_2^2+2 a_3 a_4 b_2^2-3 a_4 a_1^2 c_3+2 a_5^2 a_1 c_3+2 a_6^2 a_1 c_3+8 a_7^2 a_1 c_3-6 a_3 a_4 a_1 c_3+4 a_5 a_6 a_1 c_3+8 a_5 a_7 a_1 c_3+8 a_6 a_7 a_1 c_3+2 a_3 a_5^2 c_3+2 a_3 a_6^2 c_3+8 a_3 a_7^2 c_3-3 a_3^2 a_4 c_3+4 a_3 a_5 a_6 c_3+8 a_3 a_5 a_7 c_3+8 a_3 a_6 a_7 c_3$, $B = 3 a_4 a_1^2-2 a_5^2 a_1-2 a_6^2 a_1-8 a_7^2 a_1+6 a_3 a_4 a_1-4 a_5 a_6 a_1-8 a_5 a_7 a_1-8 a_6 a_7 a_1-2 a_3 a_5^2-2 a_3 a_6^2-8 a_3 a_7^2+3 a_3^2 a_4-4 a_3 a_5 a_6-8 a_3 a_5 a_7-8 a_3 a_6 a_7$ and $C = 2 a_4 a_1 b_2^2-6 a_4 a_1 b_4^2+2 a_3 a_4 b_2^2+4 a_5^2 b_4^2+4 a_6^2 b_4^2+16 a_7^2 b_4^2-6 a_3 a_4 b_4^2+8 a_5 a_6 b_4^2+16 a_5 a_7 b_4^2+16 a_6 a_7 b_4^2-3 a_4 a_1^2 c_3+2 a_5^2 a_1 c_3+2 a_6^2 a_1 c_3+8 a_7^2 a_1 c_3-6 a_3 a_4 a_1 c_3+4 a_5 a_6 a_1 c_3+8 a_5 a_7 a_1 c_3+8 a_6 a_7 a_1 c_3+2 a_3 a_5^2 c_3+2 a_3 a_6^2 c_3+8 a_3 a_7^2 c_3-3 a_3^2 a_4 c_3+4 a_3 a_5 a_6 c_3+8 a_3 a_5 a_7 c_3+8 a_3 a_6 a_7 c_3$

\begin{center}
\fbox{{If $0\leq 0$, $b2+b4\leq 0$ and $b2-b4\leq 0$, then: }}
\end{center}

\begin{eqnarray*}
\nonumber
&& b_4=b_1\text{ and } a_1>\frac{-a_3 c_1-a_3 c_3+b_2^2}{c_1+c_3}
\text{ and } a_5>-a_6-2 a_7\text{ and } b_3>b_1\text{ and } c_1>-c_3\text{ and } c_2>-c_1
\\\nonumber&& 
\text{ and } \mu _d^2>\frac{1}{2} \left(-a_5 v_{SM}^2-2 b_2 v_1^2+4 b_3 v_1^2\right)\text{ and } v_1>b_1
\text{ and } v_{SM}>b_1\text{ and } b_1<a_4
\text{ and } a_4<\frac{D}{E}\text{ and } a_6<\frac{a_7}{2}\text{ and } b_2\leq b_1
\\\nonumber&& 
\textbf{or}  
\\\nonumber&&
 b_4=b_1\text{ and } a_1>\frac{-a_3 c_1-a_3 c_3+b_2^2}{c_1+c_3}
 \text{ and } a_5>-a_6-2 a_7  \text{ and }  a_4<\frac{D}{E}\text{ and } b_2\leq b_1 \text{ and } v_1>b_1\text{ and } v_{SM}>b_1 \text{ and } b_1<a_4
 \\\nonumber&& 
 \text{ and } b_3>b_1\text{ and } c_1>-c_3\text{ and } c_2>-c_1\text{ and } \mu _d^2>\frac{1}{2} \left(-a_5 v_{SM}^2+2 a_6 v_{SM}^2-a_7 v_{SM}^2-2 b_2 v_1^2+4 b_3 v_1^2\right) \text{ and } a_6\geq \frac{a_7}{2}
  \\\nonumber&& 
\textbf{or}  
\\\nonumber&& 
b_4=b_1\text{ and } a_1>\frac{-a_3 c_1-a_3 c_3+b_2^2}{c_1+c_3}
\text{ and } a_5>-a_6-2 a_7\text{ and } c_1>-c_3\text{ and } c_2>-c_1 \text{ and } a_6<\frac{a_7}{2}
\text{ and } b_2\leq b_1
\\\nonumber&& 
\text{ and } \mu _d^2>\frac{1}{2} \left(-a_5 v_{SM}^2-2 b_2 v_1^2-4 b_3 v_1^2\right)
\text{ and } v_1>b_1\text{ and } v_{SM}>b_1
\text{ and } b_1<a_4\text{ and } a_4<\frac{D}{E} \text{ and } b_3\leq b_1
\\\nonumber&& 
\textbf{or}  
\\\nonumber&& b_4=b_1\text{ and } a_1>\frac{-a_3 c_1-a_3 c_3+b_2^2}{c_1+c_3}\text{ and } a_5>-a_6-2 a_7\text{ and } c_1>-c_3\text{ and } c_2>-c_1 \text{ and } a_4<\frac{D}{E}\text{ and } b_2\leq b_1\text{ and } b_3\leq b_1
\\\nonumber&& 
\text{ and } \mu _d^2>\frac{1}{2} \left(-a_5 v_{SM}^2+2 a_6 v_{SM}^2-a_7 v_{SM}^2-2 b_2 v_1^2-4 b_3 v_1^2\right)\text{ and } v_1>b_1 
\text{ and } v_{SM}>b_1\text{ and } a_6\geq \frac{a_7}{2}\text{ and } b_1<a_4
\\\nonumber&& 
\textbf{or}  
\\\nonumber&& 
 -\sqrt{3} \sqrt{b_4^2}=b_2\text{ and } a_1>\frac{-a_3 c_1-a_3 c_3+b_2^2-2 b_4 b_2+b_4^2}{c_1+c_3}\text{ and } a_5>-a_6-2 a_7
 \\\nonumber&& 
 \text{ and } a_6>\frac{a_7}{2}\text{ and } b_3>b_1\text{ and } b_4>b_1 
 \text{ and } c_1>-c_3\text{ and } c_2>-c_1
 \\\nonumber&& 
 \text{ and } \mu _d^2>\frac{1}{2} \left(-a_5 v_{SM}^2+2 a_6 v_{SM}^2-a_7 v_{SM}^2-2 b_2 v_1^2+4 b_3 v_1^2\right)\text{ and } v_1>b_1
 \text{ and } v_{SM}>b_1\text{ and } b_1<a_4
 \text{ and } a_4<\frac{F}{G}
\\\nonumber&& 
\textbf{or}  
\\\nonumber&& 
-\sqrt{3} \sqrt{b_4^2}=b_2\text{ and } a_1>\frac{-a_3 c_1-a_3 c_3+b_2^2-2 b_4 b_2+b_4^2}{c_1+c_3}\text{ and } a_5>-a_6-2 a_7\text{ and } a_6>\frac{a_7}{2}\text{ and } b_4>b_1
\\\nonumber&& 
\text{ and } \mu _d^2>\frac{1}{2} \left(-a_5 v_{SM}^2+2 a_6 v_{SM}^2-a_7 v_{SM}^2-2 b_2 v_1^2-4 b_3 v_1^2\right)\text{ and } v_1>b_1 \text{ and } c_1>-c_3\text{ and } c_2>-c_1
\\\nonumber&& 
\text{ and } v_{SM}>b_1\text{ and } b_1<a_4
\text{ and } a_4<\frac{F}{G}
\text{ and } b_3\leq b_1
 \end{eqnarray*}

\begin{eqnarray*}
 \\\nonumber&& 
\textbf{or}  
\\\nonumber&& -\sqrt{3} \sqrt{b_4^2}=b_2\text{ and } a_1>\frac{-a_3 c_1-a_3 c_3+b_2^2-2 b_4 b_2+b_4^2}{c_1+c_3}\text{ and } a_5>-a_6-2 a_7 \text{ and } b_3>b_1\text{ and } b_4>b_1\text{ and } c_1>-c_3
\\\nonumber&& 
\text{ and } \mu _d^2>\frac{1}{2} \left(-a_5 v_{SM}^2-2 b_2 v_1^2+4 b_3 v_1^2\right)\text{ and } v_1>b_1\text{ and } v_{SM}>b_1
\text{ and } b_1<a_4
\text{ and } a_4<\frac{F}{G}
\text{ and } a_6\leq \frac{a_7}{2} \text{ and } c_2>-c_1
\\\nonumber&& 
\textbf{or}  
\\\nonumber&&
 -\sqrt{3} \sqrt{b_4^2}=b_2\text{ and } a_1>\frac{-a_3 c_1-a_3 c_3+b_2^2-2 b_4 b_2+b_4^2}{c_1+c_3}\text{ and } a_5>-a_6-2 a_7
 \text{ and } b_4>b_1\text{ and } c_1>-c_3
 \\\nonumber&& 
 \text{ and } c_2>-c_1\text{ and } \mu _d^2>\frac{1}{2} \left(-a_5 v_{SM}^2-2 b_2 v_1^2-4 b_3 v_1^2\right)
 \text{ and } v_1>b_1\text{ and } v_{SM}>b_1\text{ and } b_1<a_4
 \text{ and } a_4<\frac{F}{G}
 \\\nonumber&& 
 \text{ and } a_6\leq \frac{a_7}{2}\text{ and } b_3\leq b_1
  \\\nonumber&& 
\textbf{or}  
\\\nonumber&&
-\sqrt{3} \sqrt{b_4^2}=b_2\text{ and } a_1>\frac{-a_3 c_1-a_3 c_3+b_2^2+2 b_4 b_2+b_4^2}{c_1+c_3}
\text{ and } a_4<\frac{F}{G}
\text{ and } b_4<b_1 
\\\nonumber&& 
\text{ and } a_5>-a_6-2 a_7\text{ and } a_6>\frac{a_7}{2}\text{ and } b_3>b_1\text{ and } c_1>-c_3
\\\nonumber&& 
\text{ and } c_2>-c_1\text{ and } \mu _d^2>\frac{1}{2} \left(-a_5 v_{SM}^2+2 a_6 v_{SM}^2-a_7 v_{SM}^2-2 b_2 v_1^2+4 b_3 v_1^2\right)\text{ and } v_1>b_1\text{ and } v_{SM}>b_1\text{ and } b_1<a_4
\\\nonumber&& 
\textbf{or}  
\\\nonumber&& 
 -\sqrt{3} \sqrt{b_4^2}=b_2\text{ and } a_1>\frac{-a_3 c_1-a_3 c_3+b_2^2+2 b_4 b_2+b_4^2}{c_1+c_3}\text{ and } a_5>-a_6-2 a_7\text{ and } a_6>\frac{a_7}{2}\text{ and } c_1>-c_3\text{ and } c_2>-c_1
 \\\nonumber&& 
 \text{ and } \mu _d^2>\frac{1}{2} \left(-a_5 v_{SM}^2+2 a_6 v_{SM}^2-a_7 v_{SM}^2-2 b_2 v_1^2-4 b_3 v_1^2\right)\text{ and } v_1>b_1\text{ and } v_{SM}>b_1\text{ and } b_1<a_4
 \\\nonumber&& 
 \text{ and } a_4<\frac{F}{G}
 \text{ and } b_4<b_1\text{ and } b_3\leq b_1
\\\nonumber&& 
\textbf{or}  
\\\nonumber&& 
-\sqrt{3} \sqrt{b_4^2}=b_2\text{ and } a_1>\frac{-a_3 c_1-a_3 c_3+b_2^2+2 b_4 b_2+b_4^2}{c_1+c_3}\text{ and } a_5>-a_6-2 a_7\text{ and } b_3>b_1
\\\nonumber&& 
\text{ and } c_1>-c_3\text{ and } c_2>-c_1\text{ and } \mu _d^2>\frac{1}{2} \left(-a_5 v_{SM}^2-2 b_2 v_1^2+4 b_3 v_1^2\right)\text{ and } v_1>b_1\text{ and } v_{SM}>b_1\text{ and } b_1<a_4
\\\nonumber&& 
\text{ and } a_4<\frac{F}{G}
\\\nonumber&& 
\text{ and } b_4<b_1\text{ and } a_6\leq \frac{a_7}{2}
\\\nonumber&& 
\textbf{or}  
\\\nonumber&& 
-\sqrt{3} \sqrt{b_4^2}=b_2\text{ and } a_1>\frac{-a_3 c_1-a_3 c_3+b_2^2+2 b_4 b_2+b_4^2}{c_1+c_3}\text{ and } a_5>-a_6-2 a_7
\\\nonumber&& 
\text{ and } c_1>-c_3\text{ and } c_2>-c_1\text{ and } \mu _d^2>\frac{1}{2} \left(-a_5 v_{SM}^2-2 b_2 v_1^2-4 b_3 v_1^2\right)
\text{ and } v_1>b_1\text{ and } v_{SM}>b_1\text{ and } b_1<a_4
\\\nonumber&& 
\text{ and } a_4<\frac{F}{G}
\text{ and } b_4<b_1\text{ and } a_6\leq \frac{a_7}{2}\text{ and } b_3\leq b_1 
\\\nonumber&& 
\textbf{or}  
\\\nonumber&& 
 a_1>\frac{-a_3 c_1-a_3 c_3+b_2^2-2 b_4 b_2+b_4^2}{c_1+c_3}\text{ and } a_5>-a_6-2 a_7\text{ and } b_3>b_1\text{ and } b_4>b_1\text{ and } c_1>-c_3\text{ and } c_2>-c_1
 \\\nonumber&& 
 \text{ and } \mu _d^2>\frac{1}{2} \left(-a_5 v_{SM}^2-2 b_2 v_1^2+4 b_3 v_1^2\right)
 \text{ and } v_1>b_1\text{ and } v_{SM}>b_1\text{ and } b_1<a_4
 \\\nonumber&& 
 \text{ and } a_4<\frac{F}{G}
 \text{ and } a_6<\frac{a_7}{2}\text{ and } b_2<-\sqrt{3} \sqrt{b_4^2}
 \end{eqnarray*}

\begin{eqnarray*}
 \nonumber&& 
\textbf{or}  
\\\nonumber&& 
a_1>\frac{-a_3 c_1-a_3 c_3+b_2^2-2 b_4 b_2+b_4^2}{c_1+c_3}\text{ and } a_5>-a_6-2 a_7\text{ and } b_3>b_1
\text{ and } b_4>b_1\text{ and } c_1>-c_3
\\\nonumber&& 
\text{ and } c_2>-c_1\text{ and } \mu _d^2>\frac{1}{2} \left(-a_5 v_{SM}^2+2 a_6 v_{SM}^2-a_7 v_{SM}^2-2 b_2 v_1^2+4 b_3 v_1^2\right)\text{ and } v_1>b_1
\text{ and } v_{SM}>b_1\text{ and } a_6\geq \frac{a_7}{2}
\\\nonumber&&
 \text{ and } b_1<a_4\text{ and } a_4<\frac{F}{G} 
 \text{ and } b_2<-\sqrt{3} \sqrt{b_4^2}
\\\nonumber&& 
\textbf{or}  
\\\nonumber&& 
a_1>\frac{-a_3 c_1-a_3 c_3+b_2^2-2 b_4 b_2+b_4^2}{c_1+c_3}\text{ and } a_5>-a_6-2 a_7\text{ and } b_4>b_1\text{ and } c_1>-c_3
\\\nonumber&& 
\text{ and } c_2>-c_1\text{ and } \mu _d^2>\frac{1}{2} \left(-a_5 v_{SM}^2-2 b_2 v_1^2-4 b_3 v_1^2\right)\text{ and } v_1>b_1\text{ and } v_{SM}>b_1\text{ and } b_1<a_4
\\\nonumber&& 
\text{ and } a_4<\frac{F}{G}
\text{ and } a_6<\frac{a_7}{2}\text{ and } b_2<-\sqrt{3} \sqrt{b_4^2}\text{ and } b_3\leq b_1
\\\nonumber&&
\textbf{or}  
\\\nonumber&& 
a_1>\frac{-a_3 c_1-a_3 c_3+b_2^2-2 b_4 b_2+b_4^2}{c_1+c_3}\text{ and } a_5>-a_6-2 a_7\text{ and } b_4>b_1\text{ and } c_1>-c_3
\\\nonumber&& 
\text{ and } c_2>-c_1\text{ and } \mu _d^2>\frac{1}{2} \left(-a_5 v_{SM}^2+2 a_6 v_{SM}^2-a_7 v_{SM}^2-2 b_2 v_1^2-4 b_3 v_1^2\right)\text{ and } v_1>b_1\text{ and } v_{SM}>b_1\text{ and } a_6\geq \frac{a_7}{2} 
\text{ and } b_1<a_4\text{ and } a_4<\frac{F}{G}
\\\nonumber&& 
\text{ and } b_2<-\sqrt{3} \sqrt{b_4^2}\text{ and } b_3\leq b_1
\\\nonumber&& 
\textbf{or}  
\\\nonumber&& 
 a_1>\frac{-a_3 c_1-a_3 c_3+b_2^2+2 b_4 b_2+b_4^2}{c_1+c_3}\text{ and } a_5>-a_6-2 a_7\text{ and } a_6>\frac{a_7}{2}
 \\\nonumber&& 
 \text{ and } b_3>b_1\text{ and } c_1>-c_3\text{ and } c_2>-c_1
 \text{ and } \mu _d^2>\frac{1}{2} \left(-a_5 v_{SM}^2+2 a_6 v_{SM}^2-a_7 v_{SM}^2-2 b_2 v_1^2+4 b_3 v_1^2\right)\text{ and } v_1>b_1\text{ and } v_{SM}>b_1
 \\\nonumber&& 
 \text{ and } b_1<a_4\text{ and } a_4<\frac{F}{G}
 \text{ and } b_2<3 b_4\text{ and } b_4<b_1
 \\\nonumber&& 
\textbf{or}  
\\\nonumber&& a_1>\frac{-a_3 c_1-a_3 c_3+b_2^2+2 b_4 b_2+b_4^2}{c_1+c_3}\text{ and } a_5>-a_6-2 a_7\text{ and } a_6>\frac{a_7}{2}\text{ and } c_1>-c_3
\\\nonumber&& 
\text{ and } c_2>-c_1\text{ and } \mu _d^2>\frac{1}{2} \left(-a_5 v_{SM}^2+2 a_6 v_{SM}^2-a_7 v_{SM}^2-2 b_2 v_1^2-4 b_3 v_1^2\right)
\text{ and } v_1>b_1\text{ and } v_{SM}>b_1\text{ and } b_1<a_4
\\\nonumber&& 
\text{ and } a_4<\frac{F}{G}
\text{ and } b_2<3 b_4\text{ and } b_4<b_1
\text{ and } b_3\leq b_1
\\\nonumber&& 
\textbf{or}  
\\\nonumber&& 
a_1>\frac{-a_3 c_1-a_3 c_3+b_2^2+2 b_4 b_2+b_4^2}{c_1+c_3}\text{ and } a_5>-a_6-2 a_7\text{ and } b_3>b_1\text{ and } c_1>-c_3
\\\nonumber&& 
\text{ and } c_2>-c_1\text{ and } \mu _d^2>\frac{1}{2} \left(-a_5 v_{SM}^2-2 b_2 v_1^2+4 b_3 v_1^2\right)\text{ and } v_1>b_1\text{ and } v_{SM}>b_1\text{ and } b_1<a_4
\text{ and } a_4<\frac{F}{G}
\\\nonumber&& 
\text{ and } b_2<3 b_4
\text{ and } b_4<b_1\text{ and } a_6\leq \frac{a_7}{2}
\\\nonumber&& 
\textbf{or}  
\\\nonumber&& 
a_1>\frac{-a_3 c_1-a_3 c_3+b_2^2+2 b_4 b_2+b_4^2}{c_1+c_3}\text{ and } a_5>-a_6-2 a_7\text{ and } c_1>-c_3\text{ and } c_2>-c_1
\\\nonumber&& 
\text{ and } \mu _d^2>\frac{1}{2} \left(-a_5 v_{SM}^2-2 b_2 v_1^2-4 b_3 v_1^2\right)
\\\nonumber&& 
\text{ and } v_1>b_1\text{ and } v_{SM}>b_1
\\\nonumber&& 
\text{ and } b_1<a_4\text{ and } a_4<\frac{F}{G}
\\\nonumber&& 
\text{ and } b_2<3 b_4
\\\nonumber&& 
\text{ and } b_4<b_1\text{ and } a_6\leq \frac{a_7}{2}\text{ and } b_3\leq b_1 
\end{eqnarray*}

\begin{eqnarray*}
\\\nonumber&& 
\textbf{or}  
\\\nonumber&& 
a_1>\frac{-a_3 c_1-a_3 c_3+b_2^2-2 b_4 b_2+b_4^2}{c_1+c_3}\text{ and } a_5>-a_6-2 a_7\text{ and } b_3>b_1\text{ and } b_4>b_1
\\\nonumber&& 
\text{ and } c_1>-c_3\text{ and } c_2>-c_1\text{ and } \mu _d^2>\frac{1}{2} \left(-a_5 v_{SM}^2-2 b_2 v_1^2+4 b_3 v_1^2\right)\text{ and } v_1>b_1\text{ and } v_{SM}>b_1
\text{ and } b_1<a_4\text{ and } a_4<\frac{F}{G}
\\\nonumber&& 
\text{ and } a_6<\frac{a_7}{2}\text{ and } -\sqrt{3} \sqrt{b_4^2}<b_2\text{ and } b_2\leq -b_4
\\\nonumber&& 
\textbf{or}  
\\\nonumber&& 
a_1>\frac{-a_3 c_1-a_3 c_3+b_2^2-2 b_4 b_2+b_4^2}{c_1+c_3}\text{ and } a_5>-a_6-2 a_7\text{ and } b_3>b_1
\\\nonumber&& 
\text{ and } b_4>b_1\text{ and } c_1>-c_3\text{ and } c_2>-c_1\text{ and } \mu _d^2>\frac{1}{2} \left(-a_5 v_{SM}^2+2 a_6 v_{SM}^2-a_7 v_{SM}^2-2 b_2 v_1^2+4 b_3 v_1^2\right)
\\\nonumber&& 
\text{ and } v_1>b_1\text{ and } v_{SM}>b_1
\text{ and } a_6\geq \frac{a_7}{2}
\\\nonumber&& 
\text{ and } b_1<a_4\text{ and } a_4<\frac{F}{G} 
\text{ and } -\sqrt{3} \sqrt{b_4^2}<b_2\text{ and } b_2\leq -b_4
\\\nonumber&& 
\textbf{or}  
\\\nonumber&& 
a_1>\frac{-a_3 c_1-a_3 c_3+b_2^2-2 b_4 b_2+b_4^2}{c_1+c_3}\text{ and } a_5>-a_6-2 a_7\text{ and } b_4>b_1\text{ and } c_1>-c_3
\\\nonumber&& 
\text{ and } c_2>-c_1\text{ and } \mu _d^2>\frac{1}{2} \left(-a_5 v_{SM}^2-2 b_2 v_1^2-4 b_3 v_1^2\right)\text{ and } v_1>b_1\text{ and } v_{SM}>b_1\text{ and } b_1<a_4
\\\nonumber&& 
\text{ and } a_4<\frac{F}{G}
\text{ and } a_6<\frac{a_7}{2}
\text{ and } -\sqrt{3} \sqrt{b_4^2}<b_2\text{ and } b_2\leq -b_4\text{ and } b_3\leq b_1
 \\\nonumber&& 
\textbf{or}  
\\\nonumber&& 
a_1>\frac{-a_3 c_1-a_3 c_3+b_2^2-2 b_4 b_2+b_4^2}{c_1+c_3}\text{ and } a_5>-a_6-2 a_7\text{ and } b_4>b_1\text{ and } c_1>-c_3
\\\nonumber&& 
\text{ and } c_2>-c_1\text{ and } \mu _d^2>\frac{1}{2} \left(-a_5 v_{SM}^2+2 a_6 v_{SM}^2-a_7 v_{SM}^2-2 b_2 v_1^2-4 b_3 v_1^2\right)
\\\nonumber&& 
\text{ and } v_1>b_1\text{ and } v_{SM}>b_1\text{ and } a_6\geq \frac{a_7}{2}\text{ and } b_1<a_4
\\\nonumber&& 
\text{ and } a_4<\frac{F}{G}
\\\nonumber&& 
\text{ and } -\sqrt{3} \sqrt{b_4^2}<b_2\text{ and } b_2\leq -b_4\text{ and } b_3\leq b_1 
\\\nonumber&& 
\textbf{or}  
\\\nonumber&&
a_1>\frac{-a_3 c_1-a_3 c_3+b_2^2+2 b_4 b_2+b_4^2}{c_1+c_3}\text{ and } a_5>-a_6-2 a_7\text{ and } b_3>b_1\text{ and } c_1>-c_3\text{ and } c_2>-c_1
\\\nonumber&& 
\text{ and } \mu _d^2>\frac{1}{2} \left(-a_5 v_{SM}^2-2 b_2 v_1^2+4 b_3 v_1^2\right)
\\\nonumber&& 
\text{ and } v_1>b_1\text{ and } v_{SM}>b_1
\\\nonumber&& 
\text{ and } b_1<a_4\text{ and } a_4<\frac{F}{G}
\\\nonumber&& 
\text{ and } a_6<\frac{a_7}{2}\text{ and } b_2<-\sqrt{3} \sqrt{b_4^2}\text{ and } b_4<b_1\text{ and } 3 b_4\leq b_2
\\\nonumber&& 
\textbf{or}  
\\\nonumber&&
 a_1>\frac{-a_3 c_1-a_3 c_3+b_2^2+2 b_4 b_2+b_4^2}{c_1+c_3}\text{ and } a_5>-a_6-2 a_7\text{ and } b_3>b_1\text{ and } c_1>-c_3
 \\\nonumber&& 
 \text{ and } c_2>-c_1\text{ and } \mu _d^2>\frac{1}{2} \left(-a_5 v_{SM}^2-2 b_2 v_1^2+4 b_3 v_1^2\right)
 \\\nonumber&& 
 \text{ and } v_1>b_1\text{ and } v_{SM}>b_1
 \\\nonumber&& 
 \text{ and } b_1<a_4\text{ and } a_4<\frac{F}{G}
 \\\nonumber&& 
 \text{ and } a_6<\frac{a_7}{2}\text{ and } b_4<b_1
 \text{ and } -\sqrt{3} \sqrt{b_4^2}<b_2\text{ and } b_2\leq b_4 
  \end{eqnarray*}

\begin{eqnarray*}
 \\\nonumber&& 
\textbf{or}  
\\\nonumber&& 
a_1>\frac{-a_3 c_1-a_3 c_3+b_2^2+2 b_4 b_2+b_4^2}{c_1+c_3}\text{ and } a_5>-a_6-2 a_7\text{ and } b_3>b_1\text{ and } c_1>-c_3
\\\nonumber&& 
\text{ and } c_2>-c_1\text{ and } \mu _d^2>\frac{1}{2} \left(-a_5 v_{SM}^2+2 a_6 v_{SM}^2-a_7 v_{SM}^2-2 b_2 v_1^2+4 b_3 v_1^2\right)\text{ and } v_1>b_1\text{ and } v_{SM}>b_1\text{ and } a_6\geq \frac{a_7}{2}
\\\nonumber&& 
\text{ and } b_1<a_4
\text{ and } a_4<\frac{F}{G} 
\text{ and } b_2<-\sqrt{3} \sqrt{b_4^2}\text{ and } b_4<b_1\text{ and } 3 b_4\leq b_2
\\\nonumber&& 
\textbf{or}  
\\\nonumber&& 
a_1>\frac{-a_3 c_1-a_3 c_3+b_2^2+2 b_4 b_2+b_4^2}{c_1+c_3}\text{ and } a_5>-a_6-2 a_7\text{ and } b_3>b_1
\\\nonumber&& 
\text{ and } c_1>-c_3\text{ and } c_2>-c_1\text{ and } \mu _d^2>\frac{1}{2} \left(-a_5 v_{SM}^2+2 a_6 v_{SM}^2-a_7 v_{SM}^2-2 b_2 v_1^2+4 b_3 v_1^2\right)
\\\nonumber&& 
\text{ and } v_1>b_1\text{ and } v_{SM}>b_1\text{ and } a_6\geq \frac{a_7}{2}\text{ and } b_1<a_4
\text{ and } a_4<\frac{F}{G}
\\\nonumber&& 
\text{ and } b_4<b_1
\text{ and } -\sqrt{3} \sqrt{b_4^2}<b_2\text{ and } b_2\leq b_4
 \\\nonumber&& 
\textbf{or}  
\\\nonumber&&
a_1>\frac{-a_3 c_1-a_3 c_3+b_2^2+2 b_4 b_2+b_4^2}{c_1+c_3}\text{ and } a_5>-a_6-2 a_7\text{ and } c_1>-c_3\text{ and } c_2>-c_1
\\\nonumber&& 
\text{ and } \mu _d^2>\frac{1}{2} \left(-a_5 v_{SM}^2-2 b_2 v_1^2-4 b_3 v_1^2\right)\text{ and } v_1>b_1\text{ and } v_{SM}>b_1\text{ and } b_1<a_4
\text{ and } a_4<\frac{F}{G}
\\\nonumber&& 
\text{ and } a_6<\frac{a_7}{2}
\text{ and } b_2<-\sqrt{3} \sqrt{b_4^2}\text{ and } b_4<b_1\text{ and } b_3\leq b_1\text{ and } 3 b_4\leq b_2
\\\nonumber&& 
\textbf{or}  
\\\nonumber&& 
a_1>\frac{-a_3 c_1-a_3 c_3+b_2^2+2 b_4 b_2+b_4^2}{c_1+c_3}\text{ and } a_5>-a_6-2 a_7\text{ and } c_1>-c_3\text{ and } c_2>-c_1
\\\nonumber&& 
\text{ and } \mu _d^2>\frac{1}{2} \left(-a_5 v_{SM}^2-2 b_2 v_1^2-4 b_3 v_1^2\right)\text{ and } v_1>b_1\text{ and } v_{SM}>b_1\text{ and } b_1<a_4
\\\nonumber&& 
\text{ and } a_4<\frac{F}{G}
\text{ and } a_6<\frac{a_7}{2}\text{ and } b_4<b_1\text{ and } -\sqrt{3} \sqrt{b_4^2}<b_2\text{ and } b_2\leq b_4\text{ and } b_3\leq b_1
 \\\nonumber&& 
\textbf{or}  
\\\nonumber&&
a_1>\frac{-a_3 c_1-a_3 c_3+b_2^2+2 b_4 b_2+b_4^2}{c_1+c_3}\text{ and } a_5>-a_6-2 a_7\text{ and } c_1>-c_3\text{ and } c_2>-c_1
\\\nonumber&& 
\text{ and } \mu _d^2>\frac{1}{2} \left(-a_5 v_{SM}^2+2 a_6 v_{SM}^2-a_7 v_{SM}^2-2 b_2 v_1^2-4 b_3 v_1^2\right)\text{ and } v_1>b_1\text{ and } v_{SM}>b_1\text{ and } a_6\geq \frac{a_7}{2} 
\\\nonumber&& 
\text{ and } b_2<-\sqrt{3} \sqrt{b_4^2}\text{ and } b_4<b_1\text{ and } b_3\leq b_1\text{ and } 3 b_4\leq b_2 \text{ and } b_1<a_4\text{ and } a_4<\frac{F}{G}
\\\nonumber&& 
\textbf{or}  
\\\nonumber&& 
a_1>\frac{-a_3 c_1-a_3 c_3+b_2^2+2 b_4 b_2+b_4^2}{c_1+c_3}\text{ and } a_5>-a_6-2 a_7\text{ and } c_1>-c_3\text{ and } c_2>-c_1
\\\nonumber&& 
\text{ and } \mu _d^2>\frac{1}{2} \left(-a_5 v_{SM}^2+2 a_6 v_{SM}^2-a_7 v_{SM}^2-2 b_2 v_1^2-4 b_3 v_1^2\right)\text{ and } v_1>b_1
\text{ and } v_{SM}>b_1\text{ and } a_6\geq \frac{a_7}{2}
\\\nonumber&& 
\text{ and } b_1<a_4\text{ and } a_4<\frac{F}{G} \text{ and } b_4<b_1\text{ and } -\sqrt{3} \sqrt{b_4^2}<b_2\text{ and } b_2\leq b_4\text{ and } b_3\leq b_1
\end{eqnarray*}

where we made: 

$D = 2 a_1 a_5^2 c_1+2 a_3 a_5^2 c_1+2 a_1 a_5^2 c_3+2 a_3 a_5^2 c_3+4 a_1 a_6 a_5 c_1+4 a_3 a_6 a_5 c_1+8 a_1 a_7 a_5 c_1+8 a_3 a_7 a_5 c_1+4 a_1 a_6 a_5 c_3+4 a_3 a_6 a_5 c_3+8 a_1 a_7 a_5 c_3+8 a_3 a_7 a_5 c_3+2 a_1 a_6^2 c_1+2 a_3 a_6^2 c_1+8 a_1 a_7^2 c_1+8 a_3 a_7^2 c_1+8 a_1 a_6 a_7 c_1+8 a_3 a_6 a_7 c_1+2 a_1 a_6^2 c_3+2 a_3 a_6^2 c_3+8 a_1 a_7^2 c_3+8 a_3 a_7^2 c_3+8 a_1 a_6 a_7 c_3+8 a_3 a_6 a_7 c_3$, $E = -2 a_1 b_2^2-2 a_3 b_2^2+3 a_1^2 c_1+3 a_1^2 c_3+6 a_3 a_1 c_1+6 a_3 a_1 c_3+3 a_3^2 c_1+3 a_3^2 c_3$, $F = 4 a_5^2 b_4^2+8 a_6 a_5 b_4^2+16 a_7 a_5 b_4^2+4 a_6^2 b_4^2+16 a_7^2 b_4^2+16 a_6 a_7 b_4^2+ 2 a_1 a_5^2 c_1+2 a_3 a_5^2 c_1+2 a_1 a_5^2 c_3+2 a_3 a_5^2 c_3+4 a_1 a_6 a_5 c_1+4 a_3 a_6 a_5 c_1+8 a_1 a_7 a_5 c_1+8 a_3 a_7 a_5 c_1+4 a_1 a_6 a_5 c_3+4 a_3 a_6 a_5 c_3+8 a_1 a_7 a_5 c_3+8 a_3 a_7 a_5 c_3+2 a_1 a_6^2 c_1+2 a_3 a_6^2 c_1+8 a_1 a_7^2 c_1+8 a_3 a_7^2 c_1+8 a_1 a_6 a_7 c_1+8 a_3 a_6 a_7 c_1+2 a_1 a_6^2 c_3+2 a_3 a_6^2 c_3+8 a_1 a_7^2 c_3+8 a_3 a_7^2 c_3+8 a_1 a_6 a_7 c_3+8 a_3 a_6 a_7 c_3$ and $G = -2 a_1 b_2^2+6 a_1 b_4^2-2 a_3 b_2^2+6 a_3 b_4^2+3 a_1^2 c_1+3 a_1^2 c_3+6 a_3 a_1 c_1+6 a_3 a_1 c_3+3 a_3^2 c_1+3 a_3^2 c_3$

\section{Custodial symmetry}
\label{sec:custodialsymmetry}

An accidental $\mathbb{Z}_2$ symmetry remains after the break of all the gauge and discrete symmetries in some extensions of the SM, which implies that in all interactions an inert neutral or charged scalar is always accompanied by a sterile neutrino and one known particle. For this reason, all the phenomenological consequences of the model appears through one or more loops. Multi-Higgs models are among the models that contribute to the oblique parameters, mainly to $\Delta \rho$ (or its equivalent the $T$ parameter)
\begin{equation}
\Delta\rho=\frac{m^2_W}{m^2_Z\hat{c}^2_W(M_Z)\hat{\rho}}-1\simeq \hat{\alpha}(M_Z) T,
\label{deltarho1}
\end{equation}
where $m_W$ and $m_Z$ are the masses of the SM weak vector bosons. This means that the parameter $T$ govern the size of weak isospin violating corrections to the relation between $m_W$ and $m_Z$. There are still at least two other oblique parameters $S$ and $U$ which characterizes the $q^2/M^2$ corrections, while $U$ requires both effects~\cite{Peskin:1990zt,Bertolini:1985ia,Grimus:2007if,Grimus}. These quantum effects arise because the vacuum polarization is sensitive to any field that couples with $W^\pm$ and/or $Z^0$. Notwithstanding, the $T$ parameter is usually the dominating one when the scalar masses are much larger than $m_Z$ \cite{Grimus}, which we expect to be true for these exotic particles. Therefore, we will not worry about the other oblique parameters in this work and shall only calculate the lowest order corrections to $\Delta \rho$.

Following the calculations presented in \cite{Grimus} for multi-Higgs-doublets models, $\Delta \rho$ is calculated as:
\begin{eqnarray*}
    \Delta \rho &=& \frac { g ^ { 2 } } { 64 \pi ^ { 2 } m _ { W } ^ { 2 } } \Bigg\{ \sum _ { a = 2 } ^ { n } \sum _ { b = 2 } ^ { m } \left| \left( U ^ { \dagger } V \right) _ { a b } \right| ^ { 2 } F \left( m _ { a } ^ { 2 } , \mu _ { b } ^ { 2 } \right)  
    - \sum _ { b = 2 } ^ { m - 1 } \sum _ { b ^ { \prime } = b + 1 } ^ { m } \left[ Im \left( V ^ { \dagger } V \right) _ { b b ^ { \prime } } \right] ^ { 2 } F \left( \mu _ { b } ^ { 2 } , \mu _ { b ^ { \prime } } ^ { 2 } \right) \nonumber \\&&
    - 2 \sum _ { a = 2 } ^ { n - 1 } \sum _ { a ^ { \prime } = a + 1 } ^ { n } \left| \left( U ^ { \dagger } U \right) _ { a a ^ { \prime } } \right| ^ { 2 } F \left( m _ { a } ^ { 2 } , m _ { a ^ { \prime } } ^ { 2 } \right)
    + 3 \sum _ { b = 2 } ^ { m } \left[  Im \left( V ^ { \dagger } V \right) _ { 1 b } \right] ^ { 2 } \left[ F \left( m _ { Z } ^ { 2 } , \mu _ { b } ^ { 2 } \right) - F \left( m _ { W } ^ { 2 } , \mu _ { b } ^ { 2 } \right) \right]  \label{eq:DeltaRhoGeral} \nonumber \\&&
    - 3 \left[ F \left( m _ { Z } ^ { 2 } , m _ { H } ^ { 2 } \right) - F \left( m _ { W } ^ { 2 } , m _ { H } ^ { 2 } \right) \right] \Bigg\}, 
\end{eqnarray*}
\begin{equation}
    F ( x , y ) \equiv \left\{ \begin{array} { l l } { \frac { x + y } { 2 } - \frac { x y } { x - y } \ln \frac { x } { y } } & { \text{, if }\, x \neq y, } \\ { 0 } & { \text{, if }\, x = y. } \end{array} \right.
\end{equation}
In the above equations, $m$ denotes the number of neutral mass eigenstates (8 in our case), $n$ denotes the number of charged mass eigenstates (3 in our case), $m_a$ are the masses of the charged scalars, and $\mu_b$ the masses of the neutral scalars. Also, $m_H$, $m_Z$ and $m_W$ are the masses of the SM Higgs, $Z$, and $W^\pm$ bosons, respectively.

$V$ and $U$ are matrices that relate the symmetry and mass eigenstates, in our case they are defined as
\begin{equation}
\left(
    \begin{array}{c}
         Re[S^0]+i Im[S^0]  \\
         \xi_1 \\
         \xi_2 \\
         \eta_1 + i \chi_1 \\
         \eta_2 + i \chi_2
    \end{array}
    \right)
    =V
    \left(
    \begin{array}{c}
         G^0_A  \\
         H \\
         H^0_1 \\
         H^0_2 \\
         h^0_1 \\
         h^0_2 \\
         A^0_1 \\
         A^0_2
    \end{array}
    \right),
\end{equation}
\begin{equation}
    V=
    \left(
\begin{array}{cccccccc}
 i & 1 & 0 & 0 & 0 & 0 & 0 & 0 \\
 0 & 0 & cos \theta & sin \theta & 0 & 0 & 0 & 0 \\
 0 & 0 & -sin \theta & cos \theta & 0 & 0 & 0 & 0 \\
 0 & 0 & 0 & 0 & \pm \frac{1}{\sqrt{2}} & \mp \frac{1}{\sqrt{2}} & \pm \frac{i}{\sqrt{2}} & \mp \frac{i}{\sqrt{2}} \\
 0 & 0 & 0 & 0 & \frac{1}{\sqrt{2}} & \frac{1}{\sqrt{2}} & \frac{i}{\sqrt{2}} & \frac{i}{\sqrt{2}} \\
\end{array}
\right).
\end{equation}
Here we used the rotation matrix $R_{E2}$ from Eq. \ref{eq:RotacaoCPpar3x3} for the mixing of the CP-even eigenstates (i.e., we are not using any of the solutions presented in Appendix \ref{sec:SolucoesCPpar3x3}), while the upper signs of $\pm$ and $\mp$ corresponds to the vacuum stability criteria from Eq. \ref{eq:solderivadas2} and the lower signs corresponds to the solution presented in Eq. \ref{eq:solderivadas3}.

As for the $U$ matrix we have:
\begin{equation}
\left(
    \begin{array}{c}
         S^+ \\
         D_1^+ \\
         D_2^+ \\
         \xi_1^+=0 \\
         \xi_2^+=0 \\
    \end{array}
    \right)
    =U
\left(
\begin{array}{c}
     G^+  \\
     H_1^+ \\
     H_2^+
\end{array}
\right)
    = 
    \left(
\begin{array}{ccc}
 1 & 0 & 0 \\
 0 & \pm \frac{1}{\sqrt{2}} & \mp \frac{1}{\sqrt{2}} \\
 0 & \frac{1}{\sqrt{2}} & \frac{1}{\sqrt{2}} \\
 0 & 0 & 0 \\
 0 & 0 & 0 \\
\end{array}
\right)
\left(
\begin{array}{c}
     G^+  \\
     H_1^+ \\
     H_2^+
\end{array}
\right). \label{eq:matrizU}
\end{equation}
Despite not having the fields $\xi_{1,2}^+$ in our model (that is why we have indicated them as equal to zero), we have to add them in Eq. \ref{eq:matrizU} so that the matrices $U$ and $V$ have the proper dimensions to multiply each other in Eq. \ref{eq:DeltaRhoGeral}. Nonetheless, the projection of these non-existent fields over the mass eigenstates is zero, not influencing the calculation of $\Delta \rho$. Also, it is worth mentioning that we set the Goldstone bosons as the first elements of our column matrices of mass eigenstates. That is because in Eq. \ref{eq:DeltaRhoGeral} the sums in $m$ and $n$ start at 2, this is meant to skip the Goldstone bosons in the calculation (see Ref. \cite{Grimus} for details).

Appyling Eq. \ref{eq:DeltaRhoGeral} to our model we find that in the third term the product $U^\dagger U$ gives the identity matrix, which in turn leads to $F(x,x)=0$ in the summation. Meanwhile, the fourth term of Eq. \ref{eq:DeltaRhoGeral} leads to $3 \left[ F \left( m _ { Z } ^ { 2 } , m _ { H } ^ { 2 } \right) - F \left( m _ { W } ^ { 2 } , m _ { H } ^ { 2 } \right) \right]$, which cancels the last term. In the end, when considering the vacuum stability solutions from Eqs. \ref{eq:solderivadas2} and \ref{eq:solderivadas3}, and that the masses of $H$, $H^0_1$ and $H^0_2$ are all different, we find
\begin{eqnarray}
    \Delta \rho=&\frac{g^2}{128 \pi ^2 M_W^2} \Bigg[ \frac{2 m_{A_1}^2 m_{h_1}^2 \log \left(\frac{m_{h_1}^2}{m_{A_1}^2}\right)}{m_{h_1}^2-m_{A_1}^2}+\frac{2 m_{A_2}^2 m_{h_2}^2 \log \left(\frac{m_{h_2}^2}{m_{A_2}^2}\right)}{m_{h_2}^2-m_{A_2}^2}+ \frac{2 m_H^2 m_{H_1^+}^2 \log \left(\frac{m_{H_1^+}^2}{m_H^2}\right)}{m_H^2-m_{H_1^+}^2}+\nonumber \\&\frac{2 m_H^2 m_{H_2^+}^2 \log \left(\frac{m_{H_2^+}^2}{m_H^2}\right)}{m_H^2-m_{H_2^+}^2} +2 m_H^2-m_{h_1}^2-m_{h_2}^2+m_{H_1^+}^2+m_{H_2^+}^2 -m_{A_1}^2 -m_{A_2}^2 \Bigg]. 
\end{eqnarray}
Being the same result for each vacuum stability condition.

\section{Conclusions}\label{sec:conclusions}

We have presented the copositivity criteria for the scalar sector of our $S_3 \otimes \mathbb{Z}_2$ model, which ensures that it is bounded from below. We have also shown the scalar mass eigenstates, and identified the SM Higgs boson among these. With the results we found one can further explore the model, where the mass eigenstates allows the calculation of the interaction terms necessary for phenomenological studies, while the copositivity criteria can have their consistency verified once phenomenological results impose future limits on the model parameters. The $\Delta \rho$ parameter adds another constraint to the model, imposing limits on the scalar masses. However, the complicated result (even when considering $m_H=m_{H_1^0}=m_{H_2^0}$), and the several possibilities to be considered from the copositivity criteria and the mass eigenstates solutions that we presented here, leaves us with too many possible routes to pursue in imposing limits over the model parameters at the moment. The same goes for the unitarity condition for the scalar potential, which we could have followed the procedure described in \cite{Jurciukonis:2018skr}. However, these conditioons depend on the mass eigenstate solutions used. Given that we have 104  possible solutions for the eigenstates of our scalar sector, such unitarity analysis becomes nearly unfeasible. Therefore, it seems wise to leave these as they are until other phenomenological studies bring different constraints on the model parameters, which can give us a direction to follow among the many possibilities here presented.

Some possible phenomenological studies are: the electron and muon anomalous magnetic dipole moments \cite{DeConto:2016ith}, the neutron electric dipole moment \cite{DeConto:2014fza,DeConto:2016osh}, and Flavor-Changing Neutral Currents \cite{Machado:2013jca}. All these observables are well suited for studies in multi-Higgs models. Also, the works presented in \cite{Dias:2012bh} and \cite{Fortes:2017ndr} can be revisited, to check whether the values for the model parameters agree with the copositivity criteria here presented.

Models with a $S_3$ symmetry have been studied in a variety of previous works. Among these, some are: the model's scalar potential, including its mass eigenstates and self-couplings \cite{Kubo:2004ps,Barradas-Guevara:2014yoa,Das:2014fea}, the quark sector of these models \cite{Canales:2013cga}, and the neutrino masses and their mixing \cite{Canales:2012dr}. Despite some of these topics overlapping our work, we feel that what we present here is relevant. In our model two scalar singlets are added, which further increases the scalar sector and its complexity, leading to different copositivity conditions, mass eigenstates and Yukawa sectors. Therefore, our results are not the same as the ones shown in the works just cited.

\section*{Acknowledgements}
G. De Conto and A. C. B. Machado would like to thank Coordena\c{c}\~{a}o de Aperfei\c{c}oamento de Pessoal de N\'{i}vel Superior (CAPES) for financial support.

\appendix

\section{Diagonalization solutions for the $3\times3$ CP-even mass matrix}\label{sec:SolucoesCPpar3x3}

In this appendix we show all the solutions for the $3\times3$ CP-even mass matrix discussed in Sec. \ref{sec:CPpar3x3}, except the ones where $v_1=v_2=0$. In the first subsection we consider the vacuum stability condition from Eq. \ref{eq:solderivadas2}, and in the second the condition from Eq. \ref{eq:solderivadas3}. All solutions are presented in the same format: first we show the relations that the parameters must obey, then the diagonalized matrix with the masses squared, and finally the orthogonal diagonalization matrix.

\subsection[Using $v_2= -v_1$]{Using $v_2= -v_1$, $\mu_\zeta^2= \frac{1}{2} \left(-b_1 v_{SM}^2-4 v_1^2 (c_1+c_2)+\mu_{12}^2\right)$, and $\mu_{SM}=-a_4 v_{SM}^2-b_1 v_1^2$.}

\begin{itemize}

\item Solution 1: $\left\{a_4\to \frac{2 v_1^2 (c_1+c_2)}{v_{SM}^2},b_1\to 0,sin \theta \to -\sqrt{1-cos \theta ^2},\mu_{12}^2\to 4 v_1^2 (c_1+c_2)\right\}$\subitem $ diag(m^2_H, m^2_{H_1^0}, m^2_{H_2^0})=\left(
\begin{array}{ccc}
4 v_1^2 (c_1+c_2) & 0 & 0 \\
0 & 4 v_1^2 (c_1+c_2) & 0 \\
0 & 0 & 4 v_1^2 (c_1+c_2) \\
\end{array}
\right)$\subitem $R_{E2}=\left(
\begin{array}{ccc}
1 & 0 & 0 \\
0 & cos \theta  & -\sqrt{1-cos \theta ^2} \\
0 & \sqrt{1-cos \theta ^2} & cos \theta  \\
\end{array}
\right)$

\item Solution 2: $\left\{a_4\to \frac{2 v_1^2 (c_1+c_2)}{v_{SM}^2},b_1\to 0,sin \theta \to \sqrt{1-cos \theta ^2},\mu_{12}^2\to 4 v_1^2 (c_1+c_2)\right\}$\subitem $ diag(m^2_H, m^2_{H_1^0}, m^2_{H_2^0})=\left(
\begin{array}{ccc}
4 v_1^2 (c_1+c_2) & 0 & 0 \\
0 & 4 v_1^2 (c_1+c_2) & 0 \\
0 & 0 & 4 v_1^2 (c_1+c_2) \\
\end{array}
\right)$\subitem $R_{E2}=\left(
\begin{array}{ccc}
1 & 0 & 0 \\
0 & cos \theta  & \sqrt{1-cos \theta ^2} \\
0 & -\sqrt{1-cos \theta ^2} & cos \theta  \\
\end{array}
\right)$

\item Solution 3: $\left\{b_1\to 0,cos \theta \to -\frac{1}{\sqrt{2}},sin \theta \to -\frac{1}{\sqrt{2}},\mu_{12}^2\to 2 a_4 v_{SM}^2\right\}$\subitem $ diag(m^2_H, m^2_{H_1^0}, m^2_{H_2^0})=\left(
\begin{array}{ccc}
2 a_4 v_{SM}^2 & 0 & 0 \\
0 & 4 v_1^2 (c_1+c_2) & 0 \\
0 & 0 & 2 a_4 v_{SM}^2 \\
\end{array}
\right)$\subitem $R_{E2}=\left(
\begin{array}{ccc}
1 & 0 & 0 \\
0 & -\frac{1}{\sqrt{2}} & -\frac{1}{\sqrt{2}} \\
0 & \frac{1}{\sqrt{2}} & -\frac{1}{\sqrt{2}} \\
\end{array}
\right)$

\item Solution 4: $\left\{b_1\to 0,cos \theta \to \frac{1}{\sqrt{2}},sin \theta \to -\frac{1}{\sqrt{2}},\mu_{12}^2\to 2 a_4 v_{SM}^2\right\}$\subitem $ diag(m^2_H, m^2_{H_1^0}, m^2_{H_2^0})=\left(
\begin{array}{ccc}
2 a_4 v_{SM}^2 & 0 & 0 \\
0 & 2 a_4 v_{SM}^2 & 0 \\
0 & 0 & 4 v_1^2 (c_1+c_2) \\
\end{array}
\right)$\subitem $R_{E2}=\left(
\begin{array}{ccc}
1 & 0 & 0 \\
0 & \frac{1}{\sqrt{2}} & -\frac{1}{\sqrt{2}} \\
0 & \frac{1}{\sqrt{2}} & \frac{1}{\sqrt{2}} \\
\end{array}
\right)$

\item Solution 5: $\left\{b_1\to 0,cos \theta \to -\frac{1}{\sqrt{2}},sin \theta \to \frac{1}{\sqrt{2}},\mu_{12}^2\to 2 a_4 v_{SM}^2\right\}$\subitem $ diag(m^2_H, m^2_{H_1^0}, m^2_{H_2^0})=\left(
\begin{array}{ccc}
2 a_4 v_{SM}^2 & 0 & 0 \\
0 & 2 a_4 v_{SM}^2 & 0 \\
0 & 0 & 4 v_1^2 (c_1+c_2) \\
\end{array}
\right)$\subitem $R_{E2}=\left(
\begin{array}{ccc}
1 & 0 & 0 \\
0 & -\frac{1}{\sqrt{2}} & \frac{1}{\sqrt{2}} \\
0 & -\frac{1}{\sqrt{2}} & -\frac{1}{\sqrt{2}} \\
\end{array}
\right)$

\item Solution 6: $\left\{b_1\to 0,cos \theta \to \frac{1}{\sqrt{2}},sin \theta \to \frac{1}{\sqrt{2}},\mu_{12}^2\to 2 a_4 v_{SM}^2\right\}$\subitem $ diag(m^2_H, m^2_{H_1^0}, m^2_{H_2^0})=\left(
\begin{array}{ccc}
2 a_4 v_{SM}^2 & 0 & 0 \\
0 & 4 v_1^2 (c_1+c_2) & 0 \\
0 & 0 & 2 a_4 v_{SM}^2 \\
\end{array}
\right)$\subitem $R_{E2}=\left(
\begin{array}{ccc}
1 & 0 & 0 \\
0 & \frac{1}{\sqrt{2}} & \frac{1}{\sqrt{2}} \\
0 & -\frac{1}{\sqrt{2}} & \frac{1}{\sqrt{2}} \\
\end{array}
\right)$

\item Solution 7: $\left\{a_4\to \frac{2 v_1^2 (c_1+c_2)}{v_{SM}^2},b_1\to 0,cos \theta \to -\frac{1}{\sqrt{2}},sin \theta \to -\frac{1}{\sqrt{2}},\mu_{12}^2\to 4 v_1^2 (c_1+c_2)\right\}$\subitem $ diag(m^2_H, m^2_{H_1^0}, m^2_{H_2^0})=\left(
\begin{array}{ccc}
4 v_1^2 (c_1+c_2) & 0 & 0 \\
0 & 4 v_1^2 (c_1+c_2) & 0 \\
0 & 0 & 4 v_1^2 (c_1+c_2) \\
\end{array}
\right)$\subitem $R_{E2}=\left(
\begin{array}{ccc}
1 & 0 & 0 \\
0 & -\frac{1}{\sqrt{2}} & -\frac{1}{\sqrt{2}} \\
0 & \frac{1}{\sqrt{2}} & -\frac{1}{\sqrt{2}} \\
\end{array}
\right)$

\item Solution 8: $\left\{a_4\to \frac{2 v_1^2 (c_1+c_2)}{v_{SM}^2},b_1\to 0,cos \theta \to \frac{1}{\sqrt{2}},sin \theta \to -\frac{1}{\sqrt{2}},\mu_{12}^2\to 4 v_1^2 (c_1+c_2)\right\}$\subitem $ diag(m^2_H, m^2_{H_1^0}, m^2_{H_2^0})=\left(
\begin{array}{ccc}
4 v_1^2 (c_1+c_2) & 0 & 0 \\
0 & 4 v_1^2 (c_1+c_2) & 0 \\
0 & 0 & 4 v_1^2 (c_1+c_2) \\
\end{array}
\right)$\subitem $R_{E2}=\left(
\begin{array}{ccc}
1 & 0 & 0 \\
0 & \frac{1}{\sqrt{2}} & -\frac{1}{\sqrt{2}} \\
0 & \frac{1}{\sqrt{2}} & \frac{1}{\sqrt{2}} \\
\end{array}
\right)$

\item Solution 9: $\left\{a_4\to \frac{2 v_1^2 (c_1+c_2)}{v_{SM}^2},b_1\to 0,cos \theta \to -\frac{1}{\sqrt{2}},sin \theta \to \frac{1}{\sqrt{2}},\mu_{12}^2\to 4 v_1^2 (c_1+c_2)\right\}$\subitem $ diag(m^2_H, m^2_{H_1^0}, m^2_{H_2^0})=\left(
\begin{array}{ccc}
4 v_1^2 (c_1+c_2) & 0 & 0 \\
0 & 4 v_1^2 (c_1+c_2) & 0 \\
0 & 0 & 4 v_1^2 (c_1+c_2) \\
\end{array}
\right)$\subitem $R_{E2}=\left(
\begin{array}{ccc}
1 & 0 & 0 \\
0 & -\frac{1}{\sqrt{2}} & \frac{1}{\sqrt{2}} \\
0 & -\frac{1}{\sqrt{2}} & -\frac{1}{\sqrt{2}} \\
\end{array}
\right)$

\item Solution 10: $\left\{a_4\to \frac{2 v_1^2 (c_1+c_2)}{v_{SM}^2},b_1\to 0,cos \theta \to \frac{1}{\sqrt{2}},sin \theta \to \frac{1}{\sqrt{2}},\mu_{12}^2\to 4 v_1^2 (c_1+c_2)\right\}$\subitem $ diag(m^2_H, m^2_{H_1^0}, m^2_{H_2^0})=\left(
\begin{array}{ccc}
4 v_1^2 (c_1+c_2) & 0 & 0 \\
0 & 4 v_1^2 (c_1+c_2) & 0 \\
0 & 0 & 4 v_1^2 (c_1+c_2) \\
\end{array}
\right)$\subitem $R_{E2}=\left(
\begin{array}{ccc}
1 & 0 & 0 \\
0 & \frac{1}{\sqrt{2}} & \frac{1}{\sqrt{2}} \\
0 & -\frac{1}{\sqrt{2}} & \frac{1}{\sqrt{2}} \\
\end{array}
\right)$

\end{itemize}

\subsection[Using $v_2= v_1$]{Using $v_2= v_1$, $\mu_\zeta^2= \frac{1}{2} \left(-b_1 v_{SM}^2-4 v_1^2 (c_1+c_2)-\mu_{12}^2\right)$ and $\mu_{SM}=-a_4 v_{SM}^2-b_1 v_1^2$.}

\begin{itemize}

\item Solution 1: $\left\{a_4\to \frac{2 v_1^2 (c_1+c_2)}{v_{SM}^2},b_1\to 0,sin \theta \to -\sqrt{1-cos \theta ^2},\mu_{12}^2\to -4 v_1^2 (c_1+c_2)\right\}$\subitem $ diag(m^2_H, m^2_{H_1^0}, m^2_{H_2^0})=\left(
\begin{array}{ccc}
4 v_1^2 (c_1+c_2) & 0 & 0 \\
0 & 4 v_1^2 (c_1+c_2) & 0 \\
0 & 0 & 4 v_1^2 (c_1+c_2) \\
\end{array}
\right)$\subitem $R_{E2}=\left(
\begin{array}{ccc}
1 & 0 & 0 \\
0 & cos \theta  & -\sqrt{1-cos \theta ^2} \\
0 & \sqrt{1-cos \theta ^2} & cos \theta  \\
\end{array}
\right)$

\item Solution 2: $\left\{a_4\to \frac{2 v_1^2 (c_1+c_2)}{v_{SM}^2},b_1\to 0,sin \theta \to \sqrt{1-cos \theta ^2},\mu_{12}^2\to -4 v_1^2 (c_1+c_2)\right\}$\subitem $ diag(m^2_H, m^2_{H_1^0}, m^2_{H_2^0})=\left(
\begin{array}{ccc}
4 v_1^2 (c_1+c_2) & 0 & 0 \\
0 & 4 v_1^2 (c_1+c_2) & 0 \\
0 & 0 & 4 v_1^2 (c_1+c_2) \\
\end{array}
\right)$\subitem $R_{E2}=\left(
\begin{array}{ccc}
1 & 0 & 0 \\
0 & cos \theta  & \sqrt{1-cos \theta ^2} \\
0 & -\sqrt{1-cos \theta ^2} & cos \theta  \\
\end{array}
\right)$

\item Solution 3: $\left\{b_1\to 0,cos \theta \to -\frac{1}{\sqrt{2}},sin \theta \to -\frac{1}{\sqrt{2}},\mu_{12}^2\to -2 a_4 v_{SM}^2\right\}$\subitem $ diag(m^2_H, m^2_{H_1^0}, m^2_{H_2^0})=\left(
\begin{array}{ccc}
2 a_4 v_{SM}^2 & 0 & 0 \\
0 & 2 a_4 v_{SM}^2 & 0 \\
0 & 0 & 4 v_1^2 (c_1+c_2) \\
\end{array}
\right)$\subitem $R_{E2}=\left(
\begin{array}{ccc}
1 & 0 & 0 \\
0 & -\frac{1}{\sqrt{2}} & -\frac{1}{\sqrt{2}} \\
0 & \frac{1}{\sqrt{2}} & -\frac{1}{\sqrt{2}} \\
\end{array}
\right)$

\item Solution 4: $\left\{b_1\to 0,cos \theta \to \frac{1}{\sqrt{2}},sin \theta \to -\frac{1}{\sqrt{2}},\mu_{12}^2\to -2 a_4 v_{SM}^2\right\}$\subitem $ diag(m^2_H, m^2_{H_1^0}, m^2_{H_2^0})=\left(
\begin{array}{ccc}
2 a_4 v_{SM}^2 & 0 & 0 \\
0 & 4 v_1^2 (c_1+c_2) & 0 \\
0 & 0 & 2 a_4 v_{SM}^2 \\
\end{array}
\right)$\subitem $R_{E2}=\left(
\begin{array}{ccc}
1 & 0 & 0 \\
0 & \frac{1}{\sqrt{2}} & -\frac{1}{\sqrt{2}} \\
0 & \frac{1}{\sqrt{2}} & \frac{1}{\sqrt{2}} \\
\end{array}
\right)$

\item Solution 5: $\left\{b_1\to 0,cos \theta \to -\frac{1}{\sqrt{2}},sin \theta \to \frac{1}{\sqrt{2}},\mu_{12}^2\to -2 a_4 v_{SM}^2\right\}$\subitem $ diag(m^2_H, m^2_{H_1^0}, m^2_{H_2^0})=\left(
\begin{array}{ccc}
2 a_4 v_{SM}^2 & 0 & 0 \\
0 & 4 v_1^2 (c_1+c_2) & 0 \\
0 & 0 & 2 a_4 v_{SM}^2 \\
\end{array}
\right)$\subitem $R_{E2}=\left(
\begin{array}{ccc}
1 & 0 & 0 \\
0 & -\frac{1}{\sqrt{2}} & \frac{1}{\sqrt{2}} \\
0 & -\frac{1}{\sqrt{2}} & -\frac{1}{\sqrt{2}} \\
\end{array}
\right)$

\item Solution 6: $\left\{b_1\to 0,cos \theta \to \frac{1}{\sqrt{2}},sin \theta \to \frac{1}{\sqrt{2}},\mu_{12}^2\to -2 a_4 v_{SM}^2\right\}$\subitem $ diag(m^2_H, m^2_{H_1^0}, m^2_{H_2^0})=\left(
\begin{array}{ccc}
2 a_4 v_{SM}^2 & 0 & 0 \\
0 & 2 a_4 v_{SM}^2 & 0 \\
0 & 0 & 4 v_1^2 (c_1+c_2) \\
\end{array}
\right)$\subitem $R_{E2}=\left(
\begin{array}{ccc}
1 & 0 & 0 \\
0 & \frac{1}{\sqrt{2}} & \frac{1}{\sqrt{2}} \\
0 & -\frac{1}{\sqrt{2}} & \frac{1}{\sqrt{2}} \\
\end{array}
\right)$

\item Solution 7: $\left\{a_4\to \frac{2 v_1^2 (c_1+c_2)}{v_{SM}^2},b_1\to 0,cos \theta \to -\frac{1}{\sqrt{2}},sin \theta \to -\frac{1}{\sqrt{2}},\mu_{12}^2\to -4 v_1^2 (c_1+c_2)\right\}$\subitem $ diag(m^2_H, m^2_{H_1^0}, m^2_{H_2^0})=\left(
\begin{array}{ccc}
4 v_1^2 (c_1+c_2) & 0 & 0 \\
0 & 4 v_1^2 (c_1+c_2) & 0 \\
0 & 0 & 4 v_1^2 (c_1+c_2) \\
\end{array}
\right)$\subitem $R_{E2}=\left(
\begin{array}{ccc}
1 & 0 & 0 \\
0 & -\frac{1}{\sqrt{2}} & -\frac{1}{\sqrt{2}} \\
0 & \frac{1}{\sqrt{2}} & -\frac{1}{\sqrt{2}} \\
\end{array}
\right)$

\item Solution 8: $\left\{a_4\to \frac{2 v_1^2 (c_1+c_2)}{v_{SM}^2},b_1\to 0,cos \theta \to \frac{1}{\sqrt{2}},sin \theta \to -\frac{1}{\sqrt{2}},\mu_{12}^2\to -4 v_1^2 (c_1+c_2)\right\}$\subitem $ diag(m^2_H, m^2_{H_1^0}, m^2_{H_2^0})=\left(
\begin{array}{ccc}
4 v_1^2 (c_1+c_2) & 0 & 0 \\
0 & 4 v_1^2 (c_1+c_2) & 0 \\
0 & 0 & 4 v_1^2 (c_1+c_2) \\
\end{array}
\right)$\subitem $R_{E2}=\left(
\begin{array}{ccc}
1 & 0 & 0 \\
0 & \frac{1}{\sqrt{2}} & -\frac{1}{\sqrt{2}} \\
0 & \frac{1}{\sqrt{2}} & \frac{1}{\sqrt{2}} \\
\end{array}
\right)$

\item Solution 9: $\left\{a_4\to \frac{2 v_1^2 (c_1+c_2)}{v_{SM}^2},b_1\to 0,cos \theta \to -\frac{1}{\sqrt{2}},sin \theta \to \frac{1}{\sqrt{2}},\mu_{12}^2\to -4 v_1^2 (c_1+c_2)\right\}$\subitem $ diag(m^2_H, m^2_{H_1^0}, m^2_{H_2^0})=\left(
\begin{array}{ccc}
4 v_1^2 (c_1+c_2) & 0 & 0 \\
0 & 4 v_1^2 (c_1+c_2) & 0 \\
0 & 0 & 4 v_1^2 (c_1+c_2) \\
\end{array}
\right)$\subitem $R_{E2}=\left(
\begin{array}{ccc}
1 & 0 & 0 \\
0 & -\frac{1}{\sqrt{2}} & \frac{1}{\sqrt{2}} \\
0 & -\frac{1}{\sqrt{2}} & -\frac{1}{\sqrt{2}} \\
\end{array}
\right)$

\item Solution 10: $\left\{a_4\to \frac{2 v_1^2 (c_1+c_2)}{v_{SM}^2},b_1\to 0,cos \theta \to \frac{1}{\sqrt{2}},sin \theta \to \frac{1}{\sqrt{2}},\mu_{12}^2\to -4 v_1^2 (c_1+c_2)\right\}$\subitem $ diag(m^2_H, m^2_{H_1^0}, m^2_{H_2^0})=\left(
\begin{array}{ccc}
4 v_1^2 (c_1+c_2) & 0 & 0 \\
0 & 4 v_1^2 (c_1+c_2) & 0 \\
0 & 0 & 4 v_1^2 (c_1+c_2) \\
\end{array}
\right)$\subitem $R_{E2}=\left(
\begin{array}{ccc}
1 & 0 & 0 \\
0 & \frac{1}{\sqrt{2}} & \frac{1}{\sqrt{2}} \\
0 & -\frac{1}{\sqrt{2}} & \frac{1}{\sqrt{2}} \\
\end{array}
\right)$

\end{itemize}

\end{document}